\documentclass[12pt,letterpaper,pdftex]{article} 

\usepackage{algorithmic}
\usepackage{algorithm}
\usepackage{hyphenat}
\usepackage{lastpage}

\title{Monte Carlo Sensitivity Coefficients and Analytical Benchmarks for Unresolved Resonance Probability Tables}
\author{Brian C. Kiedrowski}

\usepackage{amsmath}
\usepackage{amsthm, amssymb}
\usepackage[margin=1in]{geometry}
\usepackage{setspace}
\usepackage{multirow}
\usepackage{textcomp}
\usepackage[pdftex]{graphicx}
\usepackage{sectsty}
\usepackage{subfig}
\usepackage{hyperref}
\usepackage{bm}

\newcommand{\dir}{\hat{\mathbf{\Omega}}}

\newcommand{\dho}[2]{ \dfrac{ \partial {#1} }{ \partial {#2} } }

\newcommand{\adj}{\Psi^\dagger}
\newcommand{\iso}[2]{${}^{\scriptsize \textrm{#2}}${#1}}

\sectionfont{\normalsize}
\subsectionfont{\normalsize}
\subsubsectionfont{\normalsize}

\begin{document}


\doublespacing

%
%
%
%

\begin{center}
\noindent {\bf \large Monte Carlo Sensitivity Coefficients and Analytical Benchmarks for Unresolved Resonance Probability Tables}

\vspace{0.2in}

\noindent Brian C. Kiedrowski \\
Department of Nuclear Engineering and Radiological Sciences, University of Michigan, 2355 Bonisteel Boulevard, Ann Arbor, MI 48109

\vspace{0.2in}


\vspace{0.5in}

\noindent
\textbf{Abstract} \\
\end{center}
The Monte Carlo differential operator sampling method is applied to the computation of sensitivity coefficients of unresolved resonance probability table cross sections. Three new analytical benchmarks for verifying unresolved resonance treatments and sensitivity coefficient computations are developed. The method and its research-code implementation are verified against these benchmarks and agreement is observed. Numerical results for unresolved resonance sensitivity coefficients are obtained for the Big Ten benchmark and a simplified Molten Chloride Fast Reactor model. Energy-integrated eigenvalue sensitivity coefficients for the unresolved resonance range agree with MCNP6.2 calculations of these two models.


\section{Introduction} \label{Sec:Introduction}
%

This paper presents a Monte Carlo method to compute sensitivity coefficients of cross sections in unresolved resonance probability tables~\cite{Levitt_MonteCarloProbabilityTableMethod_NSE49_1972}, analytical benchmarks, and results showing agreement. 

Neutron cross sections have a complicated energy dependence because of nuclear resonances. These resonances need to be well represented to obtain accurate estimates of engineering quantities of interest. For many heavy nuclides, the resonances are too close together for currently available experimental data to resolve each resonance individually. Statistical models that estimate the distribution of resonance spacings and widths are used to describe for the resonance behavior in this unresolved region, and probability tables are constructed by nuclear data processing codes using random sampling approaches~\cite{Muir_NJOYManual_LA-UR-12-27079_2012}. 


Monte Carlo calculations subsequently use these probability tables by randomly sampling cross sections from them, typically during the neutron random walks. Sampling the random cross sections as opposed to simply using their expected values allows for the effect of self-shielding of the unresolved resonances to be accurately accounted for. The impact may be significant in many applications including fast spectrum reactors, neutron detection in nuclear nonproliferation and safeguards, criticality safety, and experiments designed to assess and improve the accuracy of specific nuclear cross sections. For instance, the impact on the effective multiplication factor $k$ in fast reactors may be as high as a few hundred pcm ($1~\text{pcm} = 10^{-5}$ or ten parts per million), large enough to have a significant impact on the design. Additionally, Monte Carlo calculations are increasingly used to generate application-specific multigroup cross section sets for fast deterministic transport solvers. Including the unresolved resonance self-shielding effects when generating these multigroup cross sections is important for accurate deterministic transport calculations.

In addition to estimating the expected value of some response or quantity of interest, it is often desirable for establishing safety limits to have a measure of its uncertainty as well. An efficient approach for uncertainty quantification involves the use of a first-order Taylor series expansion of the variation of model parameters. This approach requires the estimation of first derivatives of responses with respect to the nuclear cross sections and is often cast in terms of the sensitivity coefficient,
\begin{align}
  S_{R,x} = \frac{ \partial \left( \log R \right) }{ \partial \left( \log x \right) } = \frac{R}{x} \dho{R}{x} .
\end{align}
The sensitivity coefficient is the derivative of the logarithm of the response $R$ with respect to the logarithm of the model parameter $x$. More commonly, this is stated as the expected relative differential change in the response with respect to some relative differential change of a model parameter. The model parameter $x$ relevant in this paper is a cross section or multiplicative factor on the cross section within the resonance probability table.

There are established approaches for computing sensitivity coefficients in Monte Carlo calculations~\cite{Rief_ReviewMonteCarloPerturbationTechniques_NSE92_1986}. The simplest method is to perform direct perturbations using a finite difference approximation (usually a central difference) involving two or more calculations where the parameter is perturbed by a small amount $\Delta x$. While this approach is useful for validating a sensitivity coefficient produced using other methods, it often impractical as it requires running numerous particle histories to statistically resolve $\Delta R$ for a sufficiently small $\Delta x$. 

More efficient approaches are based on perturbation theory. One approach is the differential operator sampling method that produces an estimator for the sensitivity coefficient directly. Another is the correlated sampling technique that estimates the sensitivity coefficient using a small change in a response $\Delta R$ caused by some small, but finite change in response parameter $\Delta x$. Another class of perturbation-based methods employs computing adjoint weighting functions within the integrals of the mathematical expressions from perturbation theory. These integrals can be evaluated either explicitly during post processing using the computed adjoint functions or implicitly during Monte Carlo sampling.

In this article, the differential operator sampling method is applied to devise a scheme for computing sensitivity coefficients for the unresolved resonance cross sections stored in the probability tables. This method is implemented in a continuous-energy Monte Carlo research code called Shuriken written in C++. A version of Shuriken specifically for testing the unresolved resonance physics along with the analytical benchmarks and results included in this article are in available in a software repository\footnote{Shuriken software repository: https://github.com/bckiedrowski/shuriken-urr}.

To the best knowledge of the author, there are currently no analytical benchmarks that verify the probability table sampling implementation. To address this gap, this article presents three new benchmarks. Analytical solutions of leakage and collision rates are obtained for 1-D slab geometry, fixed-source problems that use continuous-energy cross sections sampled from probability table data. The benchmark solutions are consistent with the sampling implementation in the production MCNP code, specifically version 6.2~\cite{Werner_MCNP6.2ReleaseNotes_LA-UR-18-20802_2018}, and the research code Shuriken, which uses the same algorithm. Derivatives of the analytical solutions are taken to obtain expressions that benchmark the sensitivity coefficients. The results of these comparisons show agreement, suggesting the method and its implementation in Shuriken are correct.

Unfortunately, the paper cannot apply these sensitivity coefficients to uncertainty quantification because covariance data for unresolved resonance probability tables does not yet exist. Furthermore, the Evaluated Nuclear Data File (ENDF) format~\cite{ENDF6FormatsManual_2009} does not have a method for representing the data and, to best knowledge of the author, nor do any of the formats of derivative processed files such as A Compact ENDF (ACE). The Generalized Nuclear Database Structure (GNDS) format should have the necessary capabilities to support this. The author hopes that providing a method for computing the sensitivity coefficients motivates the creation of the necessary formats, covariance data, and the routines in the nuclear data processing codes to put them into a readily usable format for Monte Carlo transport solvers.

This article is divided into three major sections. Section~\ref{Sec:Methodology} provides an overview of unresolved resonance probability tables, details the method for the computation of sensitivity coefficients to the probability table parameters, and gives information on the Shuriken implementation. Section~\ref{Sec:Verification} gives derivations of three analytical benchmarks of increasing complexity using $^{90}$Zr data. This section provides numerical comparisons with these benchmark results. Finally, Sec.~\ref{Sec:Results} provides numerical results of unresolved resonance probability table sensitivity coefficients for $^{238}$U and $^{235}$U for the Big Ten benchmark~\cite{Hansen_BigTen_NSE72_1979} and a Molten Chloride Fast Reactor model~\cite{Mausloff_MCFR_NED379_2021}. Agreement between Shuriken and MCNP6.2 is observed for the integral quantities and eigenvalue sensitivity coefficients integrated over the unresolved resonance range.

\section{Methodology} \label{Sec:Methodology}

The computation of sensitivity coefficients of unresolved resonance parameters is done using differential operator sampling with a fission source correction. Section~\ref{Sec:Methodology_URProbabilityTables} explains the unresolved resonance data and the sampling and interpolation schemes that need to be considered. Section~\ref{Sec:Methodology_DifferentialOperatorSampling} then summarizes the theory of the differential operator sampling technique, providing a generic form of the estimators including partial derivatives of cross sections with respect to the unresolved resonance probability table data. Section~\ref{Sec:Methodology_URSensitivityCoefficients} derives specific forms of these partial derivatives. Finally, Sec.~\ref{Sec:Methodology_Implementation} provides the algorithm and details implementation requirements.

\subsection{Unresolved Resonance Probability Tables} \label{Sec:Methodology_URProbabilityTables}

Monte Carlo neutron transport codes usually represent the unresolved resonances using the probability table method. While different codes have slightly different implementations, usually in the interpolation schemes, the general features are largely the same and the resultant differences typically range from being insignificant to minor. 

The probability tables represent the resonances as a set of cross section bands on a grid of incident neutron energies. At each band, a cross section value or multiplicative factor applied to a mean-value cross section is given along with a probability that the cross section is within that band. The relevant nuclear reactions for which unresolved resonances occur are elastic scattering, fission, and radiative capture (n,$\gamma$). Sometimes derived data such as the total cross section, the sum of the individual reactions in the unresolved resonance table plus the contributions from other reactions, or neutron heating values are provided. The band probabilities are shared among all of the reactions, but could differ between energy grid points.

An implementation for the unresolved resonance cross section is as follows: When a neutron of energy $E$ enters a region, the code checks which nuclides are present within that region and then for any nuclide for which unresolved resonance cross sections have not already been computed, the code samples the probability tables for each nuclide. Each probability table sampling begins by first finding the energy bin on the unresolved resonance probability tables with bounds $E_g \le E < E_{g-1}$; the usual convention of descending energy with ascending group index is adopted here. Next, a single uniform random number $[0,1)$ is sampled that is then used to determine the band $k_g$ and $k_{g-1}$ based upon the cumulative probability distribution provided at both energy grid points $E_g$ and $E_{g-1}$. The subscripts on the band index emphasize that the probabilities for each band can vary with energy.

For the case where the cross sections for reaction $x$ and band $k$, $\sigma_{x,g,k_g}$ and $\sigma_{x,g-1,k_{g-1}}$, are provided, the value of the sampled cross section is determined from an interpolation. The most common interpolation scheme is linear interpolation
\begin{align} \label{Eqn:Methodology_LinearInterpolation_XSValues}
  \sigma_x(E) = ( 1 - \rho ) \sigma_{x,g,k_g}  + \rho \sigma_{x,g-1,k_{g-1}},
\end{align}
where $\rho$ is the energy interpolation factor ranging from $[0,1)$. For linear interpolation in energy,
\begin{align} \label{Eqn:Methodology_LinearInterpolationFactor}
  \rho(E) = \frac{ E - E_g }{ E_{g-1} - E_g } , \quad E_g \le E < E_{g-1}.
\end{align}
Sometimes logarithmic interpolation is employed. The cross section using logarithmic interpolation is
\begin{align} \label{Eqn:Methodology_LogInterpolation_XSValues}
  \sigma_x(E) = \sigma_{x,g,k_g} \left(  \frac{\sigma_{x,g-1,k_{g-1}}}{\sigma_{x,g,k_g}} \right)^\rho .
\end{align}
The logarithmic energy interpolation factor is
\begin{align} \label{Eqn:Methodology_LogInterpolationFactor}
  \rho(E) = \frac{ \log \left( \dfrac{E}{E_g} \right) }{  \log \left( \dfrac{E_{g-1}}{E_g} \right) } , \quad E_g \le E < E_{g-1}.
\end{align}
Semi-logarithmic interpolations are also possible by mixing and matching the interpolation schemes for the cross sections and energies.

Some implementations, such as the one found in MCNP6.2, directly use the value of the cross section as given in Eqs.~\eqref{Eqn:Methodology_LinearInterpolation_XSValues} or \eqref{Eqn:Methodology_LogInterpolation_XSValues}. Other implementations, such as the one employed by Shift, perform a second interpolation of the cross sections between bands $k_g$ and $k_g+1$ based on the uniform random number that was sampled, leading to a bilinear (or bilogarithmic) interpolation across energy and band. This paper uses the simple linear interpolation and fixed band values. In practice, the difference in calculated values is usually small.

The other case for the data on the probability tables is multiplicative factors. These are given by $\kappa_{x,g,k_g}$ and $\kappa_{x,g-1,k_{g-1}}$ for the respective energy group and band. The factors at each energy grid point are then interpolated in the same manner as in the case of linear-linear interpolation with Eq.~\eqref{Eqn:Methodology_LinearInterpolation_XSValues}, except the cross sections are replaced with the factors, to obtain $\kappa_x(E)$. The computed cross section is then determined by taking the product of the interpolated factor and the mean-value cross section:
\begin{align} \label{Eqn:Methodology_XSFromUnresolvedFactor}
  \sigma_x(E) = \kappa_x(E) \overline{\sigma}_x(E).
\end{align}
The mean-value cross section $\overline{\sigma}_x(E)$ is obtained from the standard energy-cross section grid in the same manner as the code uses to determine any other cross section outside the unresolved resonance region (e.g., linear-linear interpolation on pointwise cross section data). An advantage of the multiplicative factor representation is the factors and the mean-value cross sections can be on different energy grids, where usually the unresolved resonance grid is coarser than the mean-value cross section grid.

It is vital that is that the randomly sampled cross sections preserve balance and sum to the total to ensure the Monte Carlo sampling has well-formed reaction probabilities for collision sampling and consistent reaction multipliers for estimators. This is done in MCNP6.2 by using the sampled random number to compute the elastic, fission, and radiative capture cross sections and then computing the total cross section from
\begin{align} \label{Eqn:Methodology_TotalXSBalance}
  \sigma_t(E) = \sigma_{el}(E) + \sigma_f(E) + \sigma_\gamma(E) + \sigma_{in}(E) + \sigma_{oa}(E),
\end{align}
where $\sigma_{el}, \sigma_f$, and $\sigma_\gamma$ are the respective randomly sampled cross sections and $\sigma_{in}$ and $\sigma_{oa}$ are the inelastic scattering and other non-radiative capture absorption, e.g. (n,p), (n,$\alpha$), etc. reaction cross sections that lack resonances and are therefore determined from the standard energy-cross section grid. Note that the ACE format provides unresolved resonance data for the total cross sections, but these are not used by MCNP6.2 and the total cross section is determined from the balance relation in Eq.~\eqref{Eqn:Methodology_TotalXSBalance}.

Once the cross sections are determined, they must be preserved for the duration that the particle and any copies produced from variance reduction techniques (e.g., particle splitting) have the initial incident energy $E$. In this manner, should the particle leave the current region and at some point during the trajectory enter a different one with the same nuclide, the sampled cross sections are the same. Also, any copies that would be put into the particle bank need information stored so that they would use the same cross sections upon being pulled from particle bank. 

There are two options that may be used to accomplish this task, both involving caching some information. The first is to cache the sampled cross section values for each nuclide in the problem as they are sampled from the unresolved resonance probability table.  The alternative is to cache the random number used to sample each nuclide and recompute the cross section as needed. This information also needs to be stored in the particle bank when copies from variance reduction are made along with the rest of the particle state. The cache is flushed and the cross sections values are resampled using a different random number only when the particle's energy changes following a collision. The advantage of caching the cross sections for each nuclide is that it is more computationally efficient, avoiding binary search operations, but requires more memory usage than storing the respective random numbers.

The other important consideration is that the correlation between different cross section datasets of the same nuclide but different temperatures needs to be preserved in the manner that Doppler broadening is applied consistently. This is usually accomplished by storing the random number used to sample the cross sections of that nuclide at one temperature and using that same random number to compute new cross sections when the particle encounters a region with the same nuclide at a different temperature.

\subsection{Differential Operator Sampling} \label{Sec:Methodology_DifferentialOperatorSampling}

Differential operator sampling provides an estimator for a derivative of a response $R$ with respect to a model parameter $x$ that is made during a standard Monte Carlo history $\xi$. A general abstract description of the response $R$ is given as the following expectation:
\begin{align}  \label{Eqn:Methodology_AbstractResponse}
  R = \int_\Xi r(\xi) p(\xi) d\xi.
\end{align}
The integral here is actually a formal integral operator that denotes taking the ensemble of the set of all possible random histories $\Xi$ where each history is denoted by the variable $\xi$. In the integrand, $r(\xi)$ is the score made for random history $\xi$ and $p(\xi)$ is the probability of that random history. This integral is estimated with Monte Carlo by simulating random histories and taking the sample mean of the score $r$ in each history.

To obtain an estimator of the sensitivity coefficient with respect to parameter $x$, apply the operator $x \hspace{0.1em} \partial/\partial x$ to Eq.~\eqref{Eqn:Methodology_AbstractResponse}. An assumption is made that the differential perturbation of $x$ is such that it does not change the space of all possible random walks $\Xi$ (a contrary example is adding material to a voided region) so that the operator can be brought into the integral. After some rearrangement, the derivative is
\begin{align} \label{Eqn:Methodology_AbstractResponseDerivative}
  x \dho{R}{x} = \int_\Xi \left( x \dho{r}{x} + r(\xi) \frac{x}{p(\xi)} \dho{p}{x} \right) p(\xi) d\xi .
\end{align}
The term in parentheses is the score made during each history. The sensitivity coefficient is then obtained by dividing this by the response $R$.

The first term in the score in Eq.~\eqref{Eqn:Methodology_AbstractResponseDerivative} is the partial derivative of the score with respect to the model parameter $x$. This is called the direct effect because it is accounts for how the perturbation immediately changes the interaction causing the response. For a given history, the score is sum of the contributions $r_m$ over all steps $m$ in the random walk,
\begin{align} \label{Eqn:Methodology_HistoryScore}
  r(\xi) = \sum_m r_m(\xi) .
\end{align}
Applying $x \hspace{0.1em} \partial/\partial x$ gives
\begin{align} \label{Eqn:Methodology_HistoryScoreDerivative}
  x \dho{r}{x} = \sum_m x \dho{r_m}{x} .
\end{align}
This expression implies that at each step in the random walk a contribution for the derivative of the contribution must be added to an accumulator for that history. For a track-length estimator of a reaction rate with macroscopic cross section $\Sigma_y$ for a particle traveling a random distance $s$, this is
\begin{align} \label{Eqn:Methodology_HistoryScoreDerivative_TrackLength}
  x \dho{r}{x} = \sum_m x \dho{\Sigma_y}{x} s.
\end{align}
The partial derivative with respect to the cross section is specific to the type of sensitivity coefficient and is discussed in detail in the next section.

The second term within the parentheses in Eq.~\eqref{Eqn:Methodology_AbstractResponseDerivative} is the score times the indirect effect. The indirect effect accounts for how the perturbation changes the random walk within the history. The probability for a given history $p(\xi)$ can be expressed as the product of the probabilities or densities of each step. For example, if a history terminates upon absorption after $M$ steps,
\begin{align} \label{Eqn:Methodology_HistoryProbability}
  p(\xi) = Q T_1 C_1 T_2 C_2 \cdots T_M C_M.
\end{align}
Here $Q$ is the source probability density. $T_m$ is the transition or streaming probability density for the $m$th step. The transition begins following a source emission or collision and ends with a neutron either undergoing a collision or exiting the problem. The transition step may be further decomposed into multiple segments,
\begin{align}
  T_m = T_{m,1} T_{m,2} \cdots T_{m,J} ,
\end{align}
where each segment $j < J$ denotes free flight across a single region without a collision. The probability density function for the $j$th segment is
\begin{align} \label{Eqn:Methodology_HistoryTransitionProbability}
  T_{m,j} = \Sigma_t e^{-\Sigma_t s} + \delta( s - b ) e^{-\Sigma_t b}, \quad 0 \le s \le b.
\end{align}
Here $s$ denotes the random distance travelled during the segment, $\Sigma_t$ is the total cross section during the segment, and $b$ is the distance to the edge of the region, which may be an internal interface or exterior boundary. (The $m$ subscript for the random walk step is nominally present, but excluded here for clarity.) The first term is the particle colliding between $0 \le s < b$ and the second term containing the Dirac delta function at $x = b$ denotes the particle reaching the edge of the region. 

$C_m$ is the probability for the $m$th collision with reaction $y$,
\begin{align} \label{Eqn:Methodology_HistoryCollisionProbability}
  C_m = \frac{\Sigma_y}{\Sigma_t} .
\end{align}

A history that terminates with leakage is identical except that Eq.~\eqref{Eqn:Methodology_HistoryProbability} ends with $T_M$ and does not include $C_M$. To keep notation simple here, the form is given for an analog simulation (no variance reduction) and no within-history multiplication. Within an eigenvalue calculation, multiplication from fission is handled by banking the resulting secondaries as a source term in the subsequent cycle and other multiplying reactions such as (n,2n) can be handled by multiplying the particle statistical weight by the multiplicity of the reaction. This equation can be extended to include those, but it clutters the notation and is unnecessary for the topic of this paper. 

Applying the operator $x \hspace{0.1em} \partial/\partial x$ to Eq.~\eqref{Eqn:Methodology_HistoryProbability} gives
\begin{align} \label{Eqn:Methodology_HistoryProbabilityDerivative}
  \frac{x}{p} \dho{p}{x} = \frac{x}{Q}\dho{Q}{x} + \sum_{m=1}^M \frac{x}{T_m}\dho{T_m}{x} + \frac{x}{C_m}\dho{C_m}{x} .
\end{align}
The first term gives the effect of the perturbation of the source, and is discussed after the others. The terms in the summation provide the indirect effects on the transition and collision steps in the random walk.

Inserting Eq.~\eqref{Eqn:Methodology_HistoryTransitionProbability} into the term for the derivative of the transition step and writing the resulting expression separately for the case where a collision occurs within the segment, $0 \le s < b$, or does not, $s = b$, gives
\begin{align} \label{Eqn:Methodology_HistoryTransitionProbabilityDerivative}
 \frac{x}{T_m}\dho{T_m}{x} = \left\{ \begin{array}{l r}
  x \dho{\Sigma_t}{x} \left( \dfrac{1}{\Sigma_t} -  s \right) , & \quad 0 \le s < b, \vspace{0.1em} \\
  - x \dho{\Sigma_t}{x} b , & s = b. \\ \end{array} \right.
\end{align}
The partial derivative of the total cross section is specific to the model parameter $x$.

The analogous collision derivative term for reaction $y$ is obtained by inserting Eq.~\eqref{Eqn:Methodology_HistoryCollisionProbability}. After expanding, this gives
\begin{align} \label{Eqn:Methodology_HistoryCollisionProbabilityDerivative}
  \frac{x}{C_m}\dho{C_m}{x} = \frac{x}{\Sigma_y} \dho{\Sigma_y}{x} - \frac{x}{\Sigma_t} \dho{\Sigma_t}{x} .
\end{align}
Note that the second term with a $1/\Sigma_t$ in this collision derivative is equal and opposite to the first term of the transition derivative when a collision occurs, so they cancel. Since the $1/\Sigma_t$ term is also absent when the neutron streams through the transition segment without collision, then it is always absent for either event.

In a fixed source calculation, the derivative of the source term in Eq.~\eqref{Eqn:Methodology_HistoryProbabilityDerivative} is usually zero since the emission of source neutrons is typically, but not always, independent of most model parameters. In an eigenvalue calculation, however, the fission source depends on the random histories in the previous cycles and the term contains contributions from how the perturbation impacted the transport and fission neutron production in those cycles. Furthermore, between cycles the fission neutron population is renormalized to retain a fixed average population and the effect of the perturbation must also be propagated through this renormalization. 

Accounting for the perturbation in the fission source can be done in a couple mathematically equivalent ways. The first stores contributions in the current cycle and and applies adjoint weighting functions, either computed directly with Monte Carlo in a future cycle or serving as a precomputed multiplier based on the location of the fission production site. 

The other approach, which is used in this paper, is to account for the cumulative effect of the perturbation across the previous cycles using differential operator sampling. This is done each cycle using Eq.~\eqref{Eqn:Methodology_HistoryProbabilityDerivative}, including the accumulated perturbed source effect from the previous cycles, the indirect effect contributions during transport, using Eqs.~\eqref{Eqn:Methodology_HistoryTransitionProbabilityDerivative} and \eqref{Eqn:Methodology_HistoryCollisionProbabilityDerivative}, plus the direct effect contribution capturing the impact on the production of a fission neutrons during each collision. For this direct effect, the expected number of fission neutrons produced each collision is determined by a collision estimator as
\begin{align} \label{Eqn:Methodology_FissionProductionResponseFunction}
  r_f = \frac{1}{k} \frac{ \nu \Sigma_f }{ \Sigma_t }.
\end{align}
The effect of perturbation on the production rate of each neutron, assuming $x$ is a cross section, is
\begin{align} \label{Eqn:Methodology_FissionProductionResponseFunctionDerivative}
  \frac{x}{r_f} \dho{r_f}{x} = \frac{x}{\Sigma_f} \dho{\Sigma_f}{x} - \frac{x}{\Sigma_t} \dho{\Sigma_t}{x} .
\end{align}
This gives the impact on the perturbation of the fission source particles produced. This equation is analogous to Eq.~\eqref{Eqn:Methodology_HistoryScoreDerivative_TrackLength}, except that it is for a collision estimator and not a track-length estimator.

The sum of all the various effects (perturbed source, indirect, and direct for fission production) is stored for each fission neutron production site, which is given as $w_{f,j}$. Between cycles the effect of the perturbation for each source particle is renormalized such that the total change in the fission source weight is zero, such that the initial perturbed source weight in the following cycle is
\begin{align} \label{Eqn:Methodology_SourceDerivative}
  \frac{x}{Q}\dho{Q}{x} = w_0 - \dfrac{ \displaystyle\sum_j n_j w_{f,j} }{ \displaystyle\sum_j n_j } .
\end{align}
Here $w_0$ is the fission source particle weight for the current cycle, $j$ is an index for fission producing events in the previous cycle, $n_j$ is the number of fission neutrons banked in the $j$th event, and $w_{f,j}$ is the cumulative indirect and direct effect banked in the previous cycle for that fission event. This process of accumulating this perturbed source effect requires several cycles to reach an equilibrium similar to the standard fission source convergence. The accumulation however, can begin in the inactive cycles and so long as a few more (typically 5-10) cycles are skipped past when the fission source would be converged, then the perturbed source effect can be adequately determined.

\subsection{Unresolved Resonance Sensitivity Coefficients} \label{Sec:Methodology_URSensitivityCoefficients}

The previous section provides the equations for the scoring functions necessary for computing response sensitivities with respect to cross sections using differential operator sampling. Applying these to the unresolved resonance table requires that the partial derivatives of the cross sections in Eqs.~\eqref{Eqn:Methodology_HistoryScoreDerivative_TrackLength}, \eqref{Eqn:Methodology_HistoryTransitionProbabilityDerivative}, \eqref{Eqn:Methodology_HistoryCollisionProbabilityDerivative}, and \eqref{Eqn:Methodology_FissionProductionResponseFunctionDerivative} be explicitly evaluated. For each interpolation scheme, there are two cases that need to be considered. The first is where the unresolved resonance data is given as cross section values and the second is where they are given as factors.

The model parameter $x$ is taken to be some provided reaction cross section for some nuclide $j$ that is on the unresolved probability table, $\sigma_{x,g,k_g}^j$, where the reaction type is either total, elastic scattering, fission, or radiative capture and $j$ is a superscript index for the nuclide. The total macroscopic cross section for a material containing multiple nuclides is the sum of the product of atomic densities and their respective microscopic total cross sections:
\begin{align} \label{Eqn:Methodology_TotalXS}
  \Sigma_t(E) = \sum_j N^j \sigma_t^j(E).
\end{align}
Inserting Eq.~\eqref{Eqn:Methodology_TotalXSBalance} for the microscopic total cross section gives
\begin{align} \label{Eqn:Methodology_TotalXSExpanded}
  \Sigma_t(E) = \sum_j N^j \left[ \sigma_{el}^j(E) + \sigma_f^j(E) + \sigma_\gamma^j(E) + \sigma_{in}^j(E) + \sigma_{oa}^j(E)  \right]
\end{align}
Taking the partial derivative with respect to the unresolved resonance microscopic cross section gives
\begin{align} \label{Eqn:Methodology_TotalXSDerivative_Step1}
  x \dho{\Sigma_t}{x} = \sigma_{x,g,k_g}^j \dho{\Sigma_t}{\sigma_{x,g,k_g}^j} =  N^j \sigma_{x,g,k_g}^j \dho{ \sigma_x^j(E) }{ \sigma_{x,g,k_g}^j } .
\end{align}

The next step is to evaluate the partial derivative of the microscopic reaction cross section $\sigma_x^j(E)$. If a linear interpolation scheme is used for the cross section, this microscopic reaction cross section is given by Eq.~\eqref{Eqn:Methodology_LinearInterpolation_XSValues}. Inspecting this equation, there are a few cases that need to be considered depending on the energy of the incident neutron. If the neutron energy is within the range $E_g \le E < E_{g-1}$, where $g$ is taken to be the energy index of the cross section on the unresolved resonance probability table, then the model parameter cross section is on the lower grid point of the range. Likewise, if $E_{g+1} \le E < E_g$, then the model parameter cross section is on the upper grid point. Finally, if it is neither of those, then the reaction cross section does not depend upon the model parameter cross section. 

For linear interpolation in the cross section, Eq.~\eqref{Eqn:Methodology_TotalXSDerivative_Step1} becomes
\begin{align}  \label{Eqn:Methodology_TotalXSDerivative_Linear}
  x \dho{\Sigma_t}{x} = N^j \sigma_{x,g,k_g} \dho{ \sigma_x^j(E) }{ \sigma_{x,g,k_g}^j } = \left\{ \begin{array}{l r}
  N^j \sigma_{x,g,k_g}^j ( 1 - \rho ) 	& \quad E_g \le E < E_{g-1}, \\
  N^j \sigma_{x,g,k_g}^j \rho			& \quad E_{g+1} \le E < E_g, \\
  0,									& \quad \text{otherwise}, \\ \end{array} \right.
\end{align}
where $\rho$ is the interpolation factor within the grid element containing energy $E$ from Eqs.~\eqref{Eqn:Methodology_LinearInterpolationFactor} or \eqref{Eqn:Methodology_LogInterpolationFactor} for linear and logarithmic interpolation in energy respectively. 

For logarithmic interpolation in the cross section, the corresponding expression is
\begin{align}  \label{Eqn:Methodology_TotalXSDerivative_Logarithmic}
  x \dho{\Sigma_t}{x} = N^j \sigma_{x,g,k_g} \dho{ \sigma_x^j(E) }{ \sigma_{x,g,k_g}^j } = \left\{ \begin{array}{l r}
  N^j \sigma_{x,g,k_g}^j ( 1 - \rho ) \left( \dfrac{ \sigma_{x,g-1,k_{g-1}}^j }{ \sigma_{x,g,k_g}^j } \right)^\rho 	& \quad E_g \le E < E_{g-1}, \vspace{0.2em} \\
  N^j \sigma_{x,g,k_g}^j \rho \left( \dfrac{ \sigma_{x,g,k_g}^j }{ \sigma_{x,g+1,k_{g+1}}^j } \right)^{1 -\rho} 	& \quad E_{g+1} \le E < E_g, \vspace{0.2em} \\
  0,																												& \quad \text{otherwise}.	 \\ 
  \end{array} \right.
\end{align}

Equation~\eqref{Eqn:Methodology_HistoryCollisionProbabilityDerivative} for the collision probability derivative requires the derivative of the macroscopic cross section for the reaction $y$ that occurs in the simulation. Likewise, Eq.~\eqref{Eqn:Methodology_FissionProductionResponseFunctionDerivative} requires the derivative for the macroscopic fission cross section. The result of this is otherwise identical to Eqs.~\eqref{Eqn:Methodology_TotalXSDerivative_Linear} or \eqref{Eqn:Methodology_TotalXSDerivative_Logarithmic} except that they are zero if the reaction in the derivative does not match the perturbed cross section.

The unresolved resonance data may also be represented as energy-dependent multiplicative factors $\kappa_x(E)$ as in Eq.~\eqref{Eqn:Methodology_XSFromUnresolvedFactor}. The expression for the total cross section in Eq.~\eqref{Eqn:Methodology_TotalXSExpanded} in this case is
\begin{align}
  \Sigma_t(E) = \sum_j N^j \left[ \kappa_{el}^j(E) \overline{\sigma}_{el}^j(E) + \kappa_{f}^j(E) \overline{\sigma}_f^j(E) + \kappa_{\gamma}^j(E) \overline{\sigma}_\gamma^j(E) + \sigma_{in}^j(E) + \sigma_{oa}^j(E)  \right] .
\end{align}
The partial derivative of the macroscopic total cross section in Eq.~\eqref{Eqn:Methodology_TotalXS} with respect to a multiplicative factor for a particular reaction and nuclide $j$ is
\begin{align} \label{Eqn:Methodology_TotalXSDerivativeFactors_Step1}
  x \dho{\Sigma_t}{x} = \kappa_{x,g,k_g}^j \dho{\Sigma_t}{\kappa_{x,g,k_g}^j} =  N^j \kappa_{x,g,k_g}^j \overline{\sigma}_x^j(E) \dho{ \kappa_x^j(E) }{ \kappa_{x,g,k_g}^j } .
\end{align}
Equation~\eqref{Eqn:Methodology_TotalXSDerivative_Linear} for linear interpolation of the multiplicative factor and Eq.~\eqref{Eqn:Methodology_TotalXSDerivative_Logarithmic} for the corresponding logarithmic interpolation are instead
\begin{align}  \label{Eqn:Methodology_TotalXSDerivativeFactors_Linear}
  x \dho{\Sigma_t}{x} = \left\{ \begin{array}{l r}
  N^j \kappa_{x,g,k_g}^j \overline{\sigma}_x^j(E) ( 1 - \rho ) 	& \quad E_g \le E < E_{g-1}, \\
  N^j \kappa_{x,g,k_g}^j \overline{\sigma}_x^j(E) \rho			& \quad E_{g+1} \le E < E_g, \\
  0,															& \quad \text{otherwise}, \\ \end{array} \right.
\end{align}
and
\begin{align}  \label{Eqn:Methodology_TotalXSDerivativeFactors_Logarithmic}
  x \dho{\Sigma_t}{x} = \left\{ \begin{array}{l r}
  N^j \kappa_{x,g,k_g}^j \overline{\sigma}_x^j(E) ( 1 - \rho ) \left( \dfrac{ \kappa_{x,g-1,k_{g-1}}^j }{ \kappa_{x,g,k_g}^j } \right)^\rho 	& \quad E_g \le E < E_{g-1}, \vspace{0.2em} \\
  N^j \kappa_{x,g,k_g}^j \overline{\sigma}_x^j(E) \rho \left( \dfrac{ \kappa_{x,g,k_g}^j }{ \kappa_{x,g+1,k_{g+1}}^j } \right)^{1 -\rho} 	& \quad E_{g+1} \le E < E_g, \vspace{0.2em} \\
  0,																												& \quad \text{otherwise},	 \\ 
  \end{array} \right.
\end{align}
respectively. Based on these equations, the only difference in the estimator between linear and logarithmic interpolation is an extra factor of the ratio of the unresolved resonance probability data to either the power of $\rho$ or $1-\rho$. These results can be applied to the partial derivatives with respect to other macroscopic cross sections in the same manner as the linear interpolation case.

Note that the forms of the partial derivatives are the same as the ones obtained for sensitivity coefficients with respect to pointwise cross sections. Usually, however, sensitivity coefficients are tabulated over an energy range. For the most common case of linear interpolation, the sensitivity coefficients follow the linear summation rule: the sum of the sensitivity coefficients is equal to the sensitivity coefficient of the sum. This is straightforward to show since the factors of $1 - \rho$ and $\rho$ in Eqs.~\eqref{Eqn:Methodology_TotalXSDerivative_Linear} and \eqref{Eqn:Methodology_TotalXSDerivativeFactors_Linear} sum to one and equals the result obtained for the case of the sensitivity integrated over the entire bin. Therefore, the sensitivity coefficient of a cross section is taken over the entire unresolved resonance energy range is, for the case of linear interpolation, equal to the sum of the sensitivity coefficients over all energy grid points and cross section bands. 

The net result of this is that the current methods of computing sensitivity coefficients containing all or part of the unresolved resonance range are consistent with this more detailed approach. Differences are expected, however, if logarithmic interpolation is used for the unresolved resonance probability table data. A survey of evaluated nuclear data files in the ENDF/B-VII.1 library~\cite{Chadwick_ENDF7.1-NDS112_2011} shows that linear interpolation is the predominant scheme, suggesting that logarithmic interpolation is rarely prescribed. Therefore, any observed differences should be rare.

\subsection{Implementation} \label{Sec:Methodology_Implementation}

The method for computing unresolved resonance sensitivity coefficients is implemented in a research Monte Carlo code named Shuriken that supports continuous-energy physics using nuclear data read from ACE files. Shuriken supports history-based transport, analog physics (no variance reduction), basic constructive solid geometry, estimators for surface currents, fluxes, and reaction rates, in both fixed source and $k$ eigenvalue calculation modes. The focus of this section is primarily on eigenvalue calculations since fixed source calculations are a simpler version. The purpose of this section is to document what was done to provide the results in this paper by providing enough information so that another developer could implement the method in their code.

In the eigenvalue calculations, sensitivity coefficients for the $k$ eigenvalue and reaction rate ratios are supported. (A sensitivity coefficient with respect to an arbitrary estimate in a $k$ eigenvalue calculation does not make sense since the eigenfunction is only unique to within an arbitrary constant.) The sensitivity coefficient of a reaction rate ratio is obtained as difference of the sensitivity coefficients of the numerator and denominator. Let $R = R_1/R_2$, the sensitivity coefficient of this ratio is then
\begin{align}
  S_{R,x} = \frac{x}{R} \dho{R}{x} = \frac{x}{R_1} \dho{R_1}{x} - \frac{x}{R_2} \dho{R_2}{x} .
\end{align}

In the problem input, the user defines the relevant unresolved probability data of interest along with the relevant estimators to serve as the responses that the sensitivity coefficients are computed for. In the implementation used in this paper, the user specifies a vector of unresolved resonance data to have sensitivities computed with respect to a vector of responses. The user specifies the nuclide, reaction, and geometric cells for each element of the unresolved resonance data. For each element, the code produces an array dimensioned by the number of energy grid points times the number of probability bands on the unresolved resonance table (the ACE format only permits the data to have the same number of bands). Each element in this array contains memory to store indirect effect contributions and another array for direct effect contributions dimensioned by the number of responses.

At the beginning of a particle history, a source particle is taken from the fission source bank, which includes a pointer to an array of perturbed weights for each sensitivity coefficient for the perturbed fission source effect of this source particle. Equation~\eqref{Eqn:Methodology_SourceDerivative} is used to compute the starting perturbed source weight for the history that is used to initialize the accumulator for the indirect effect. The accumulator for the direct effect is initialized to zero. (In a fixed-source calculation, the indirect effect is initialized to zero since there is no perturbed source effect from previous cycles.)

A random distance to collision is sampled and the distance to the next event $s$ is determined as the minimum of this distance to collision or the distance to the boundary or interface. This distance $s$ is used to accumulate an estimator for the indirect effect using the negative term of Eq.~\eqref{Eqn:Methodology_HistoryTransitionProbabilityDerivative} for each perturbation. [As mentioned, when a collision occurs this cancels with an equal and opposite term in Eq.~\eqref{Eqn:Methodology_HistoryCollisionProbabilityDerivative}.] For unresolved resonance probability data, the indirect effect score contribution is
\begin{align}  \label{Eqn:Methodology_TransitionDerivativeScore_Linear}
  I_{T_m,x,g,k_g}^j = \left\{ \begin{array}{l r}
  -w N^j \sigma_{x,g,k_g}^j ( 1 - \rho ) ,	 	& \quad E_g \le E < E_{g-1}, \\
  -w N^j \sigma_{x,g,k_g}^j \rho s,				& \quad E_{g+1} \le E < E_g, \\
  0,									& \quad \text{otherwise}, \\ \end{array} \right.
\end{align}
for linear interpolation; here, $w$ is the particle statistical weight at the time of collision. The case for multiplicative factors with follows where the cross section $\sigma_{x,g,k_g}^j$ is replaced by the factor $\kappa_{x,g,k_g}^j \overline{\sigma}_x^j(E)$. An additional factor of the ratio to the $\rho$ or $1-\rho$ power is made per Eqs.~\eqref{Eqn:Methodology_TotalXSDerivative_Logarithmic} or \eqref{Eqn:Methodology_TotalXSDerivativeFactors_Logarithmic} in the case of logarithmic interpolation depending on the format of the data. To emphasize, there are two scores that get made per particle trajectory. The first case in the piecewise function in Eq.~\eqref{Eqn:Methodology_TransitionDerivativeScore_Linear} applies to the data at lower energy grid point. The second case is for the upper energy grid point. This is the case for all of the estimators that are subsequently discussed in this section. 

In active cycles, a contribution to the direct effect track-length estimator is made using Eq.~\eqref{Eqn:Methodology_HistoryScoreDerivative_TrackLength} for the combinations of all estimators and perturbations. The form is otherwise identical to Eq.~\eqref{Eqn:Methodology_TransitionDerivativeScore_Linear}.

If a collision occurs in an eigenvalue calculation, then a random number $n_j$ fission neutrons for the source term in the following cycle are banked based on the expected value given by Eq.~\eqref{Eqn:Methodology_FissionProductionResponseFunction}. If $n_j > 0$, then information about the perturbed fission source weight $w_{f,j}$ is stored for this fission source event $j$ for each perturbation. This $w_{f,j}$ is the accumulated indirect effect up to this point plus a potential contribution from the direct effect from the derivative of the collision estimate of fission production given in Eq.~\eqref{Eqn:Methodology_FissionProductionResponseFunction}. The direct effect score contribution to the perturbed source weight $w_{f,j}$ is
\begin{align}  \label{Eqn:Methodology_FissionProduction_Linear}
  F_{x,g,k_g}^j = \left\{ \begin{array}{l r}
  w N^j \sigma_{x,g,k_g}^j ( 1 - \rho ) \left( \dfrac{\delta_{xf}}{\Sigma_f} - \dfrac{1}{\Sigma_t} \right),	& \quad E_g \le E < E_{g-1}, \\
  w N^j \sigma_{x,g,k_g}^j \rho \left( \dfrac{\delta_{xf}}{\Sigma_f} - \dfrac{1}{\Sigma_t}         \right),	& \quad E_{g+1} \le E < E_g, \\
  0,																										& \quad \text{otherwise}, \\ \end{array} \right.
\end{align}
where $\delta_{xf}$ is one if reaction $x$ is fission and zero otherwise. Note that the cross sections are that for the entire material and not for just for the nuclide causing fission. An important point here is that information about the perturbed source is stored for all of the unresolved resonance data under consideration because of the accumulated indirect effect, whereas Eq.~\eqref{Eqn:Methodology_FissionProduction_Linear} only applies to the specific collision event.

A pointer or index for this information is stored within the fission source bank to prevent unnecessary duplication of data. Additionally, for each perturbation a score of $n_j w_{f,j}$ is made to an accumulator that is used to renormalize the fission source at the beginning of each history in the following cycle with Eq.~\eqref{Eqn:Methodology_SourceDerivative}.

Following the potential production of fission source neutrons, the outgoing reaction $y$ of the neutron with nuclide $j$ is sampled from the cross section data. If this reaction $y$ and nuclide $j$ matches the reaction of the perturbation, then for the linear interpolation case, the contribution based on the first term of Eq.~\eqref{Eqn:Methodology_HistoryCollisionProbabilityDerivative} is 
\begin{align}  \label{Eqn:Methodology_CollisionDerivativeScore_Linear}
  I_{C_m,x,g,k_g}^j = \left\{ \begin{array}{l r}
  w (1 - \rho),	 	& \quad E_g \le E < E_{g-1}, \\
  w \rho,			& \quad E_{g+1} \le E < E_g, \\
  0,				& \quad \text{otherwise}. \\ \end{array} \right.
\end{align}
Note that this score contribution appears to be missing a factor of $1/\Sigma_y^j$. This is factor is accounted for in the random walk process itself by virtue that it is the probability of reaction $y$ with nuclide $j$ occurring.

The estimators in this section require information that is normally not readily available at the time of scoring. The following is required: the relevant reaction cross section on the lower and upper bound, $\sigma_{x,g,k_g}^j$ and $\sigma_{x,g-1,k_{g-1}}^j$ when cross sections are provided and $\kappa_{x,g,k_g}^j \overline{\sigma}_x^j(E)$ and $\kappa_{x,g-1,k_{g-1}}^j \overline{\sigma}_x^j(E)$ when multiplicative factors are provided, respectively; the interpolation factor $\rho$; and the indices of the energy range $g$ and the band for lower and upper bounds, $k_g$ and $k_{g-1}$ respectively, for scoring in the appropriate bin for the band. 

There are two options for handling this data, having the classic memory versus computational time tradeoff. The memory efficient approach is to recompute this data as needed based upon the stored random number. In practice, this requires significant computational expense from searching the data. The computationally efficient approach, and the one employed in the code referenced in this paper is to cache the relevant data during the sampling and computation of the unresolved resonance cross sections. 

One drawback to the current implementation is that there is significant memory requirement for the perturbed source effect, as information is retained for all previous cycles as opposed to a fixed number of this. This is mathematically correct provided enough inactive cycles have been run to reach stationarity in the localized perturbed source following the convergence of the fission source. While this simplifies the implementation, eventually there will be nonzero contributions accrued for every unresolved resonance energy grid point and probability band and this informations need to be stored for every fission neutron producing event. This limits the practical batch size that can be stored in memory. The author recommends that a developer making a production implementation be mindful of this and adapt a strategy of limiting the number of cycles for the perturbed source effect. Should this be done, a large portion of the data for the perturbed source effect will be zero, and sparse storage schemes could be employed.

\section{Analytical Benchmark Verification} \label{Sec:Verification}
This section presents three analytical benchmark problems for the unresolved probability tables. The benchmarks are fixed-source problems in 1-D slab geometry using real continuous-energy nuclear data. These benchmarks can be employed both as a test of the implementation of the unresolved resonance probability table sampling and, for the purposes of this article specifically, the computation of sensitivity coefficients of the cross sections specified on the probability tables.

A survey of nuclear data relevant to nuclear reactor applications (and specifically fast reactors where the unresolved resonance range is expected to be significant) was conducted, and the ENDF/B-VII.1 $^{90}$Zr ACE files at room temperature (293.6~K) have attractive features that make them ideal for an analytical benchmark. 

In terms of application relevance, zirconium  has a low overall capture cross section and good structural properties making it useful as a fuel cladding or constituent in a fuel matrix. 

Properties that make the $^{90}$Zr data ideal for an analytical benchmark are that the prescribed energy grid for the probability tables aligns with the energy grid for the cross sections, which is often not the case for nuclides such as $^{238}$U. Because of this alignment, only a single energy discretization needs to be considered in constructing the solutions. Another useful property is that there are only two reactions in this energy range: elastic scattering and radiative $(\textrm{n},\gamma)$ capture, and no fission or contributions from inelastic scattering or other absorptions. Finally, the unresolved resonance data is also given as cross sections, as opposed to multiplicative factors, and the prescribed interpolation between energy grid points is linear-linear. In principle, any other nuclear data file with these properties could also be used with the following benchmark solutions.

There are $G$ incident neutron energy grid points within the unresolved resonance range where the cross section probability tables are provided. As in the previous section, the convention of using descending energies where the top energy is defined by $E_0$ and the bottom $E_G$ is adopted. The energy range with index $g$ includes all energies such that $E_{g} \le E < E_{g-1}$, including the lower energy bound and excluding of the upper energy. 

For the purposes of these benchmarks, zirconium is assumed to be pure $^{90}$Zr with an atomic density of $N = 4.3675 \times 10^{-2}$~b$^{-1}\cdot$cm$^{-1}$, which is typical of zirconium metal. The microscopic elastic scattering and radiative capture cross sections along with their associated probabilities are provided in Table~\ref{Table_Verification_90Zr_URRCrossSectionData}. There are 13 energy grid points in the data ranging from 53.5~keV to 100~keV. At the lower end of the unresolved range there is an energy bin from 53.5~keV to 54.0~keV. The remaining energy bins increase by 5~keV intervals up to 99~keV. The remaining two bins from 99~keV to 100~keV involve a large bin covering almost all of the energy range up to 1~eV below 100~keV and another very small bin to 100~keV; this small bin ensures a smooth transition into the unresolved resonance range, but since it is extremely narrow and the cross sections are effectively identical on each side, it is neglected and only 12 grid points are used. The unresolved resonance tables at each energy grid point have 16 probability bands. Each probability table has a different spacing of cumulative probabilities, and this fact must be considered when constructing the analytical solutions.

Three different responses from fixed-source calculations are considered in this section, each having sensitivity coefficients computed for responses that are significant and meaningful to that particular quantity. The general layout for this section is that each analytic solution is derived and then results for the equivalent Shuriken Monte Carlo calculation are reported and the degree of agreement is discussed. These results show that the implementation of the unresolved resonance probability table sampling is correct (or at least consistent with the provided implementation), the equations for the differential operator sampling sensitivity coefficients were derived correctly, and that the computation of the indirect and direct effects are implemented correctly.

The first of the three cases, detailed in Sec.~\ref{Sec:Verification_UncollidedTransmittance}, considers the leakage of uncollided neutrons (or a purely absorbing problem), for which analytical solutions can readily be obtained. A reference solution for the sensitivity coefficients of this leakage rate with respect to the total cross sections (the sum of the sensitivity coefficients for elastic scattering plus radiative capture) provided in the unresolved resonance tables are computed. This simple case establishes the computation of indirect effects in Eqs.~\eqref{Eqn:Methodology_TransitionDerivativeScore_Linear} and~\eqref{Eqn:Methodology_CollisionDerivativeScore_Linear} are done correctly.

The second case, discussed in Sec.~\ref{Sec:Verification_FirstCollisionCapture}, considers the response of neutrons being captured on their first collision. This is essentially equivalent to a purely absorbing problem, except that the response involves a reaction cross section, leading to a more complicated solution. Sensitivity coefficients with respect to the radiative capture and elastic scattering cross sections given in the unresolved resonance probability tables are computed. In addition to the expression being more complicated, the sensitivity coefficient for the radiative capture cross section tests the implementation of the direct effect.

The third case, given in Sec.~\ref{Sec:Verification_OnceCollidedTransmittance}, considers the response being the leakage of neutrons that have had exactly one collision. Modified scattering physics is used to make the solution tractable. The purpose of this benchmark is to establish that unresolved resonance cross section sampling following a collision is done correctly and that the sensitivity coefficients can handle more than an uncollided response.

As shown for case, the Monte Carlo results generally agree with the reference solutions. This provides evidence that the equations derived in the previous section and their implementation is correct. This and the limitations of these benchmarks are discussed in Sec.~\ref{Sec:Verification_Discussion}.

\begin{table}[t!] \tiny
\caption{Unresolved Resonance Data for $^{90}$Zr at 293.6~K from ENDF/B-VII.1 Processed by NJOY}
\begin{center}
\begin{tabular}{|l|c|c|c|c|c|c|c|c|c|} \hline
  $E$ (keV)  &  \multicolumn{3}{|c|}{53.5} &       \multicolumn{3}{|c|}{54.0} &       \multicolumn{3}{|c|}{59.0}  \\ \hline
Band $k$ 	& $\sigma_{s,k}$ (b)	& $\sigma_{\gamma,k}$ (b) 	& $p_{k}$		
			& $\sigma_{s,k}$ (b)	& $\sigma_{\gamma,k}$ (b)	& $p_{k}$		
			& $\sigma_{s,k}$ (b)	& $\sigma_{\gamma,k}$ (b)	& $p_{k}$		\\ \hline
  1 &   1.0015e-06 & 3.9687e-04 & 1.5625e-05 & 1.2251e-06 & 9.6691e-04 & 1.3125e-03 & 9.9801e-07 & 5.1658e-04 & 2.1094e-04 \\
  2 &   1.1854e-06 & 6.1470e-04 & 4.6094e-04 & 1.7389e-03 & 4.9979e-03 & 2.6812e-03 & 1.0390e-06 & 1.2455e-03 & 1.3000e-03 \\
  3 &   2.7649e-01 & 4.2641e-03 & 1.1316e-02 & 8.2760e-01 & 4.6443e-03 & 1.6761e-02 & 2.9736e-01 & 4.4895e-03 & 8.9938e-03 \\
  4 &   1.9634e+00 & 4.7909e-03 & 3.2094e-02 & 2.5949e+00 & 3.6598e-03 & 3.6297e-02 & 1.5853e+00 & 4.2088e-03 & 2.2658e-02 \\
  5 &   3.7732e+00 & 2.8310e-03 & 5.4078e-02 & 4.1598e+00 & 2.6054e-03 & 6.9805e-02 & 3.3021e+00 & 3.4658e-03 & 4.5559e-02 \\
  6 &   4.9900e+00 & 1.6768e-03 & 1.3471e-01 & 5.1072e+00 & 1.8127e-03 & 1.1324e-01 & 4.5420e+00 & 2.2358e-03 & 7.6086e-02 \\
  7 &   5.7063e+00 & 1.4092e-03 & 1.8533e-01 & 5.6399e+00 & 1.5591e-03 & 1.1971e-01 & 5.3223e+00 & 1.8017e-03 & 1.4240e-01 \\
  8 &   6.1685e+00 & 2.6604e-03 & 1.5177e-01 & 5.8934e+00 & 1.5847e-03 & 8.1573e-02 & 5.9074e+00 & 1.8959e-03 & 2.1041e-01 \\
  9 &   6.7023e+00 & 4.5730e-03 & 1.2146e-01 & 6.2628e+00 & 2.8397e-03 & 1.5076e-01 & 6.5428e+00 & 4.1062e-03 & 1.5972e-01 \\
 10 &   7.6021e+00 & 8.1334e-03 & 1.0852e-01 & 6.9717e+00 & 5.4599e-03 & 1.4828e-01 & 7.5577e+00 & 8.4072e-03 & 1.2122e-01 \\
 11 &   1.0234e+01 & 1.8853e-02 & 1.0204e-01 & 8.8966e+00 & 1.3020e-02 & 1.3609e-01 & 1.0217e+01 & 1.6672e-02 & 1.0547e-01 \\
 12 &   1.9774e+01 & 3.7937e-02 & 5.4930e-02 & 1.5835e+01 & 3.1984e-02 & 6.8822e-02 & 1.9197e+01 & 3.2770e-02 & 5.9692e-02 \\
 13 &   3.9799e+01 & 7.0792e-02 & 2.5867e-02 & 3.5348e+01 & 6.5917e-02 & 3.6155e-02 & 3.7745e+01 & 6.1906e-02 & 2.8952e-02 \\
 14 &   6.0311e+01 & 1.5651e-01 & 1.1920e-02 & 5.9014e+01 & 1.4220e-01 & 1.3112e-02 & 5.9604e+01 & 1.4794e-01 & 1.2626e-02 \\
 15 &   9.3155e+01 & 2.3798e-01 & 3.9984e-03 & 9.3536e+01 & 2.2616e-01 & 4.0531e-03 & 8.9057e+01 & 1.8227e-01 & 3.3219e-03 \\
 16 &   1.0772e+02 & 1.6797e-01 & 1.4969e-03 & 1.0842e+02 & 1.6840e-01 & 1.3546e-03 & 1.0131e+02 & 1.4719e-01 & 1.3859e-03 \\ \hline
  $E$ (keV)  &  \multicolumn{3}{|c|}{64.0} &       \multicolumn{3}{|c|}{69.0} &       \multicolumn{3}{|c|}{74.0}  \\ \hline
Band $k$ 	& $\sigma_{s,k}$ (b)	& $\sigma_{\gamma,k}$ (b) 	& $p_{k}$		
			& $\sigma_{s,k}$ (b)	& $\sigma_{\gamma,k}$ (b)	& $p_{k}$		
			& $\sigma_{s,k}$ (b)	& $\sigma_{\gamma,k}$ (b)	& $p_{k}$		\\ \hline
  1 &   3.2226e-06 & 1.1791e-03 & 1.6187e-03 & 1.7563e-05 & 2.2014e-03 & 2.7672e-03 & 1.5742e-06 & 8.0556e-04 & 1.0891e-03 \\
  2 &   8.4717e-02 & 4.5135e-03 & 4.0156e-03 & 3.0330e-01 & 5.1479e-03 & 5.6984e-03 & 1.5181e-04 & 3.2993e-03 & 2.0188e-03 \\
  3 &   9.2825e-01 & 4.5031e-03 & 1.5298e-02 & 1.3017e+00 & 4.2478e-03 & 1.5811e-02 & 6.5329e-01 & 4.5050e-03 & 1.1970e-02 \\
  4 &   2.7294e+00 & 3.8757e-03 & 4.1387e-02 & 3.0753e+00 & 3.5586e-03 & 4.4548e-02 & 2.1525e+00 & 3.9612e-03 & 2.7244e-02 \\
  5 &   4.3453e+00 & 2.2663e-03 & 7.6112e-02 & 4.4345e+00 & 2.1952e-03 & 6.6017e-02 & 3.9766e+00 & 2.5309e-03 & 7.0242e-02 \\
  6 &   5.2782e+00 & 1.8343e-03 & 1.5239e-01 & 5.2422e+00 & 1.7244e-03 & 1.3052e-01 & 5.1367e+00 & 1.7770e-03 & 1.2550e-01 \\
  7 &   5.8013e+00 & 1.7533e-03 & 1.3780e-01 & 5.7132e+00 & 1.5648e-03 & 1.0178e-01 & 5.7606e+00 & 1.6428e-03 & 1.7791e-01 \\
  8 &   6.1983e+00 & 2.6016e-03 & 1.2885e-01 & 5.9786e+00 & 2.0245e-03 & 9.5342e-02 & 6.2916e+00 & 2.8255e-03 & 1.6863e-01 \\
  9 &   6.7087e+00 & 4.5736e-03 & 1.1477e-01 & 6.3278e+00 & 2.9744e-03 & 1.1765e-01 & 7.0015e+00 & 5.3752e-03 & 1.2130e-01 \\
 10 &   7.6856e+00 & 8.7563e-03 & 1.2001e-01 & 6.9548e+00 & 5.4664e-03 & 1.3385e-01 & 8.1405e+00 & 1.0314e-02 & 9.5981e-02 \\
 11 &   1.0785e+01 & 1.7180e-02 & 1.1089e-01 & 8.8082e+00 & 1.2452e-02 & 1.4647e-01 & 1.1822e+01 & 1.8939e-02 & 1.0987e-01 \\
 12 &   2.1960e+01 & 3.6627e-02 & 5.8445e-02 & 1.6328e+01 & 2.6713e-02 & 8.5653e-02 & 2.4012e+01 & 3.5136e-02 & 5.1761e-02 \\
 13 &   4.0768e+01 & 6.2057e-02 & 2.5400e-02 & 3.4406e+01 & 4.8175e-02 & 3.6648e-02 & 3.9020e+01 & 5.5448e-02 & 2.3409e-02 \\
 14 &   6.5699e+01 & 1.6266e-01 & 8.8797e-03 & 5.8054e+01 & 1.2422e-01 & 1.2556e-02 & 6.2316e+01 & 1.2244e-01 & 9.1641e-03 \\
 15 &   8.4962e+01 & 1.4270e-01 & 2.7656e-03 & 7.8902e+01 & 1.2963e-01 & 2.6875e-03 & 7.5915e+01 & 9.9826e-02 & 2.4234e-03 \\
 16 &   9.4619e+01 & 1.3707e-01 & 1.3610e-03 & 8.7495e+01 & 1.1431e-01 & 1.9984e-03 & 8.3880e+01 & 1.0683e-01 & 1.4907e-03 \\ \hline
  $E$ (keV)  &  \multicolumn{3}{|c|}{79.0} &       \multicolumn{3}{|c|}{84.0} &       \multicolumn{3}{|c|}{89.0}  \\ \hline
Band $k$ 	& $\sigma_{s,k}$ (b)	& $\sigma_{\gamma,k}$ (b) 	& $p_{k}$		
			& $\sigma_{s,k}$ (b)	& $\sigma_{\gamma,k}$ (b)	& $p_{k}$		
			& $\sigma_{s,k}$ (b)	& $\sigma_{\gamma,k}$ (b)	& $p_{k}$		\\ \hline
  1 &   2.3288e-06 & 8.7044e-04 & 1.1875e-03 & 1.0117e-06 & 7.4938e-04 & 7.8750e-04 & 2.6618e-04 & 2.5107e-03 & 2.7062e-03 \\
  2 &   4.0687e-06 & 2.1522e-03 & 1.1562e-03 & 2.4771e-06 & 1.8826e-03 & 1.3078e-03 & 5.0324e-01 & 3.7969e-03 & 7.8516e-03 \\
  3 &   5.5878e-01 & 4.0103e-03 & 1.1780e-02 & 5.8466e-01 & 4.4962e-03 & 1.1672e-02 & 2.0863e+00 & 3.7970e-03 & 3.1059e-02 \\
  4 &   1.9696e+00 & 4.0135e-03 & 2.3458e-02 & 2.3198e+00 & 3.5093e-03 & 3.2233e-02 & 3.7487e+00 & 2.7513e-03 & 4.4200e-02 \\
  5 &   3.6493e+00 & 2.6237e-03 & 5.5123e-02 & 3.9918e+00 & 2.5713e-03 & 5.6453e-02 & 4.6753e+00 & 1.9233e-03 & 6.1550e-02 \\
  6 &   4.9549e+00 & 1.8515e-03 & 1.3802e-01 & 5.1212e+00 & 1.7696e-03 & 1.3362e-01 & 5.2627e+00 & 1.6863e-03 & 1.0671e-01 \\
  7 &   5.6729e+00 & 1.6751e-03 & 1.6975e-01 & 5.7685e+00 & 1.5536e-03 & 1.6228e-01 & 5.6789e+00 & 1.6131e-03 & 1.0718e-01 \\
  8 &   6.2055e+00 & 2.7008e-03 & 1.6384e-01 & 6.2763e+00 & 2.6387e-03 & 1.6468e-01 & 5.9423e+00 & 1.9421e-03 & 6.6983e-02 \\
  9 &   6.8995e+00 & 5.0101e-03 & 1.2437e-01 & 6.9988e+00 & 5.1951e-03 & 1.3614e-01 & 6.2466e+00 & 2.7770e-03 & 1.1118e-01 \\
 10 &   8.1676e+00 & 1.0175e-02 & 1.1528e-01 & 8.2446e+00 & 9.9528e-03 & 9.9184e-02 & 6.8481e+00 & 4.6997e-03 & 1.3342e-01 \\
 11 &   1.2343e+01 & 1.8820e-02 & 1.0852e-01 & 1.1859e+01 & 1.7643e-02 & 1.0663e-01 & 8.7282e+00 & 1.1151e-02 & 1.6889e-01 \\
 12 &   2.3735e+01 & 3.1954e-02 & 4.7191e-02 & 2.2475e+01 & 2.9549e-02 & 5.3222e-02 & 1.5752e+01 & 2.1491e-02 & 9.3975e-02 \\
 13 &   3.6891e+01 & 5.0720e-02 & 2.5564e-02 & 3.6048e+01 & 4.8780e-02 & 2.7402e-02 & 2.9797e+01 & 3.3837e-02 & 4.3406e-02 \\
 14 &   6.0627e+01 & 1.0090e-01 & 1.2133e-02 & 5.7818e+01 & 9.6771e-02 & 1.0319e-02 & 4.9005e+01 & 7.7229e-02 & 1.4350e-02 \\
 15 &   7.3424e+01 & 7.7556e-02 & 2.0875e-03 & 6.9993e+01 & 7.8967e-02 & 3.6875e-03 & 6.3056e+01 & 7.2694e-02 & 4.7359e-03 \\
 16 &   8.6975e+01 & 1.1059e-01 & 5.3580e-04 & 8.8185e+01 & 1.0001e-01 & 3.7800e-04 & 7.2835e+01 & 7.0098e-02 & 1.8109e-03 \\ \hline
  $E$ (keV)  &  \multicolumn{3}{|c|}{94.0} &       \multicolumn{3}{|c|}{99.0} &       \multicolumn{3}{|c|}{100.0}  \\ \hline
Band $k$ 	& $\sigma_{s,k}$ (b)	& $\sigma_{\gamma,k}$ (b) 	& $p_{k}$		
			& $\sigma_{s,k}$ (b)	& $\sigma_{\gamma,k}$ (b)	& $p_{k}$		
			& $\sigma_{s,k}$ (b)	& $\sigma_{\gamma,k}$ (b)	& $p_{k}$		\\ \hline
  1 &   9.9817e-07 & 4.6185e-04 & 6.4063e-05 & 9.9130e-07 & 3.6552e-04 & 8.7500e-05 & 1.6554e-02 & 2.9001e-03 & 3.1234e-03 \\
  2 &   4.9320e-06 & 1.7766e-03 & 1.8656e-03 & 1.0980e-06 & 1.3407e-03 & 1.5594e-03 & 5.1654e-01 & 3.4423e-03 & 6.3656e-03 \\
  3 &   5.9691e-01 & 3.9790e-03 & 9.6297e-03 & 6.0589e-01 & 4.0979e-03 & 1.0705e-02 & 2.2319e+00 & 3.7857e-03 & 3.6392e-02 \\
  4 &   2.0892e+00 & 3.4760e-03 & 2.4109e-02 & 1.9985e+00 & 3.8804e-03 & 2.0891e-02 & 4.0495e+00 & 2.4340e-03 & 5.3967e-02 \\
  5 &   3.7665e+00 & 2.8012e-03 & 5.0667e-02 & 3.5637e+00 & 2.7693e-03 & 4.5509e-02 & 4.9593e+00 & 1.7486e-03 & 7.9911e-02 \\
  6 &   4.9418e+00 & 1.6963e-03 & 1.1112e-01 & 4.7784e+00 & 1.6255e-03 & 1.0207e-01 & 5.4441e+00 & 1.5444e-03 & 9.0811e-02 \\
  7 &   5.6092e+00 & 1.5368e-03 & 1.5987e-01 & 5.5649e+00 & 1.6364e-03 & 2.0084e-01 & 5.7232e+00 & 1.5183e-03 & 7.4145e-02 \\
  8 &   6.0935e+00 & 2.2340e-03 & 1.4337e-01 & 6.1497e+00 & 2.6640e-03 & 1.4630e-01 & 5.9492e+00 & 1.9865e-03 & 7.4277e-02 \\
  9 &   6.6055e+00 & 3.7941e-03 & 1.1325e-01 & 6.8460e+00 & 4.7992e-03 & 1.3814e-01 & 6.2868e+00 & 2.8011e-03 & 1.1668e-01 \\
 10 &   7.4433e+00 & 6.9140e-03 & 1.2459e-01 & 8.1801e+00 & 8.8582e-03 & 1.1723e-01 & 7.0484e+00 & 4.9809e-03 & 1.6333e-01 \\
 11 &   1.0000e+01 & 1.3487e-02 & 1.3083e-01 & 1.1996e+01 & 1.5788e-02 & 1.1258e-01 & 9.6600e+00 & 1.1919e-02 & 1.6703e-01 \\
 12 &   1.7965e+01 & 2.3011e-02 & 7.2963e-02 & 2.1080e+01 & 2.4082e-02 & 5.3548e-02 & 1.8570e+01 & 2.2235e-02 & 8.2780e-02 \\
 13 &   3.0438e+01 & 3.5273e-02 & 3.6708e-02 & 3.2165e+01 & 3.7894e-02 & 3.2952e-02 & 3.1563e+01 & 3.7439e-02 & 3.3094e-02 \\
 14 &   4.9070e+01 & 7.0938e-02 & 1.5078e-02 & 4.9264e+01 & 6.5617e-02 & 1.0375e-02 & 4.9750e+01 & 6.6062e-02 & 1.2558e-02 \\
 15 &   6.0636e+01 & 6.1442e-02 & 3.4015e-03 & 5.6994e+01 & 5.8523e-02 & 3.8656e-03 & 5.8033e+01 & 5.2787e-02 & 3.1657e-03 \\
 16 &   6.7556e+01 & 6.9233e-02 & 2.4750e-03 & 6.3282e+01 & 5.8553e-02 & 3.3437e-03 & 6.5873e+01 & 6.8787e-02 & 2.3687e-03 \\ \hline
\end{tabular}
\end{center}
\label{Table_Verification_90Zr_URRCrossSectionData}
\end{table}%

\clearpage

\subsection{Uncollided Transmittance Sensitivity} \label{Sec:Verification_UncollidedTransmittance}
The first test problem consists of a monodirectional beam with a prescribed energy spectrum normally incident upon the slab of thickness~$L$. For simplicity the energy spectrum of the source is constant over the entirely unresolved resonance energy range such that the probability of a particular source neutron being within an energy range $g$ is
\begin{align} \label{Eqn:Verification_UniformEnergyBinWidth}
  w_g = \frac{E_g - E_{g-1}}{E_0 - E_G} .
\end{align}

The uncollided beam intensity is simply an exponential attenuated by the cross section $\Sigma_t(E)$. Within an energy range, the total cross section is determined from a linear interpolation in energy and randomly sampled cross sections at endpoints from the prescribed probability tables. 

Because the endpoints have different band probabilities $p_{k_g}$ and $p_{k_{g-1}}$, a unionized grid of cumulative sampling probabilities of each band within every energy range $g$ must be constructed to obtain a consistent expression for the linear interpolation. This grid is built by first building the cumulative probabiliites for each endpoint and taking the union. Each range of cumulative probabilities is assigned an index $\ell$ and mapped to a pair of band indices $k_g$ and $k_{g-1}$ that give the cross sections at the end points. The probability of selecting a particular pair of endpoints is then denoted by $p_{g,\ell}$, which is obtained by taking the difference of the cumulative probabilities on the unionized grid.

An example of this unionized grid is displayed in Table~\ref{Table_Verification_UnionizedGridExample}. The data is for a neutron with an energy within range $54 \le E < 59$ keV on $^{90}$Zr. The table can be viewed on two sides. The left side of the table consists of the first three columns giving the cumulative probabilities for the endpoints at 54 and 59 keV. The right side of the table are the remaining columns giving the unionized index $\ell$, the cumulative probability, and probability of that range on the unionized table, and the two corresponding non-unionized indices for the left and right sides. For the most part, there is an alternating pattern, but the pattern changes at $\ell = 16$ and then again at $\ell = 30$. 

\begin{table}[t!] \small
\caption{Example of Unionized Probability Grid for $54 \le E < 59$ keV}
\begin{center}
\begin{tabular}{|c|c|c|c|c|c|c|c|} \hline
  \multicolumn{3}{|c|}{ Endpoint Data } & \multicolumn{5}{|c|}{ Unionized Grid } \\ \hline
  $k$	& $c_{k_{g}}$ 	& $c_{k_{g-1}}$		& $\ell$	& $c_{g,\ell}$	& $p_{g,\ell}$	& $k_g$		& $k_{g-1}$		\\ \hline
	1	&	1.313e-03	& 2.109e-04			& 1			& 2.109e-04		& 2.109e-04		& 1			& 1				\\
	2	&	3.994e-03	& 1.511e-03			& 2			& 1.313e-03		& 1.102e-03		& 1			& 2				\\
	3	&	2.075e-02	& 1.050e-02			& 3			& 1.511e-03		& 1.984e-04		& 2			& 2				\\
	4	&	5.705e-02	& 3.316e-02			& 4			& 3.994e-03		& 2.483e-03		& 2			& 3				\\
	5	&	1.269e-01	& 7.872e-02			& 5			& 1.050e-02		& 6.511e-03		& 3			& 3				\\
	6	&	2.401e-01	& 1.548e-01			& 6			& 2.075e-02		& 1.025e-02		& 3			& 4				\\
	7	&	3.598e-01	& 2.972e-01			& 7			& 3.316e-02		& 1.241e-02		& 4			& 4				\\
	8	&	4.414e-01	& 5.076e-01			& 8			& 5.705e-02		& 2.389e-02		& 4			& 5				\\
	9	&	5.921e-01	& 6.673e-01			& 9			& 7.872e-02		& 2.167e-02		& 5			& 5				\\
	10	&	7.404e-01	& 7.886e-01			& 10		& 1.269e-01		& 4.813e-02		& 5			& 6				\\
	11	&	8.765e-01	& 8.940e-01			& 11		& 1.548e-01		& 2.795e-02		& 6			& 6				\\
	12	&	9.453e-01	& 9.537e-01			& 12		& 2.401e-01		& 8.529e-02		& 6			& 7				\\
	13	&	9.815e-01	& 9.827e-01			& 13		& 2.972e-01		& 5.711e-02		& 7			& 7				\\
	14	&	9.946e-01	& 9.953e-01			& 14		& 3.598e-01		& 6.260e-02		& 7			& 8				\\
	15	&	9.987e-01	& 9.986e-01			& 15		& 4.414e-01		& 8.157e-02		& 8			& 8				\\
	16	&	1.000e+00	& 1.000e+00			& 16		& 5.076e-01		& 6.624e-02		& 9			& 8				\\
		&				&					& 17		& 5.921e-01		& 8.452e-02		& 9			& 9 			\\
		&				&					& 18		& 6.673e-01		& 7.520e-02		& 10		& 9 			\\
		&				&					& 19		& 7.404e-01		& 7.308e-02		& 10		& 10			\\
		&				&					& 20		& 7.886e-01		& 4.814e-02		& 11		& 10			\\
		&				&					& 21		& 8.765e-01		& 8.795e-02		& 11		& 11			\\
		&				&					& 22		& 8.940e-01		& 1.752e-02		& 12		& 11			\\
		&				&					& 23		& 9.453e-01		& 5.130e-02		& 12		& 12			\\
		&				&					& 24		& 9.537e-01		& 8.388e-03		& 13		& 12			\\
		&				&					& 25		& 9.815e-01		& 2.777e-02		& 13		& 13			\\
		&				&					& 26		& 9.827e-01		& 1.185e-03		& 14		& 13			\\
		&				&					& 27		& 9.946e-01		& 1.193e-02		& 14		& 14			\\
		&				&					& 28		& 9.953e-01		& 6.990e-04		& 15		& 14			\\
		&				&					& 29		& 9.986e-01		& 3.322e-03		& 15		& 15			\\
		&				&					& 30		& 9.987e-01		& 3.216e-05		& 15		& 16			\\
		&				&					& 31		& 1.000e+00		& 1.348e-03		& 16		& 16			\\ \hline
\end{tabular}
\end{center}
\label{Table_Verification_UnionizedGridExample}
\end{table}%

The expected transmittance probability or current for monodirectional neutrons uniformly distributed within this energy range is then
\begin{align} \label{Eqn:Verification_Transmittance_Step1}
  J_0 =  \frac{1}{E_0 - E_G} \sum_{g=1}^G \sum_\ell p_{g,\ell} \int_{E_g}^{E_{g-1}} \exp \left[ -\Sigma_{t,\ell}(E) L \right] dE .
\end{align}
Since the interpolation scheme is linear-linear, the total cross section in the unresolved region for the particular probability band $\ell$ is obtained from Eq.~\eqref{Eqn:Methodology_LinearInterpolation_XSValues} as
\begin{align} \label{Eqn:Verification_LinearInterpolationTotalXS}
  \Sigma_{t,\ell}(E) = N \left[ ( 1 - \rho ) \sigma_{t,g,\ell} + \rho \sigma_{t,g-1,\ell} \right] , \quad E_{g} \le E < E_{g-1},
\end{align}
where $\rho$ is the linear interpolation factor from Eq.~\eqref{Eqn:Methodology_LinearInterpolationFactor} and the $\sigma_{t,g,\ell}$ are given in the probability table. Contrasting with Eq.~\eqref{Eqn:Methodology_LinearInterpolation_XSValues}, the band indices $k_g$ and $k_{g-1}$ for the respective lower and upper energy grid points are replaced with the unionized index $\ell$. 

The integral for the current given in Eq.~\eqref{Eqn:Verification_Transmittance_Step1} can be evaluated by performing a change of variables from energy to the interpolation fraction via
\begin{align} \label{Eqn:Verification_TransformationEnergyToInterpolationFactor}
  dE = ( E_{g-1} - E_g ) d\rho ,
\end{align}
giving
\begin{align} \label{Eqn:Verification_Transmittance_Step2}
  J_0 = \sum_{g=1}^G \sum_\ell w_g p_{g,\ell} \int_0^1 \exp \left[ -N L \left(  \sigma_{t,g,\ell} + \rho ( \sigma_{t,g-1,\ell} - \sigma_{t,g,\ell} ) \right) \right] d\rho.
\end{align}
Carrying out the integral gives
\begin{align} \label{Eqn:Verification_Transmittance_IntegralResult}
  f_{g,\ell} = \int_0^1 \exp \left[ -N L \left(  \sigma_{t,g,\ell} + \rho ( \sigma_{t,g-1,\ell} - \sigma_{t,g,\ell} ) \right) \right] d\rho = \frac{ e^{-\tau_{g,\ell}} - e^{-\tau_{g-1,\ell}} }{ \tau_{g-1,\ell} - \tau_{g,\ell} } ,
\end{align}
where $\tau$ is the optical distance at an energy grid point:
\begin{align} \label{Eqn:Verification_OpticalDistance}
  \tau_{g,\ell} = N L \sigma_{t,g,\ell} .
\end{align}
The total transmittance is the sum of the transmittances for each energy bin $g$ weighted by the relative bin width or probability of a source neutron within energy range $g$. The total transmittance is then written as
\begin{align} \label{Eqn:Verification_Transmittance}
  J_0 = \sum_{g=1}^G \sum_\ell w_g p_{g,\ell} f_{g,\ell} .
\end{align}

The sensitivity coefficient of the transmittance with respect to each provided probability table total cross section on the grid points $\sigma_{t,g,k_g}$ (here again $k_g$ is an index for the grid endpoints, not the unionized grid for the energy range) may be computed by taking the logarithmic derivative of Eq.~\eqref{Eqn:Verification_Transmittance}. To complicate matters, the summation over the probability bands is over the unionized grid for the energy range, whereas the parameters are given at on the grid endpoints, each having a different structure. Therefore, multiple values of $\ell$ on the unionized grid for the energy range may correspond to the same value at the energy grid point $k$. Furthermore, these grid points reside on the upper or lower edges of the energy range, which results in a different form for the derivative. Finally, for cross sections specified at the top $(g = 0)$ and bottom $(g = G)$ of the unresolved resonance range, the values are only used on a single side as opposed to both.

To address these notational complexities, an indicator function $1^\pm_{g,k_g,\ell}$ is used that is either one or zero. The value is one if the energy range $g$ specified is between $1 \le g \le G$ (this handles the edges of the unresolved resonance range) and if the unionized grid index $\ell$ corresponds to the index $k_g$ on either the upper ($+$ superscript) or lower ($-$ superscript) energy grid point of the energy range:
\begin{align} \label{Eqn:Verification_Transmittance_IndicatorMappingFunction}
  1^\pm_{g,k_g,\ell} = 
  \left\{ \begin{array}{l l}
    1, & \quad 1 \le g \le G, \ \ell \in k_g \text{ for upper $+$ or lower $-$ grid point,} \\
    0, & \quad \text{otherwise} . \\
  \end{array} \right.
\end{align}

The sensitivity coefficient of the transmittance with respect to a grid total cross section is then
\begin{align} \label{Eqn:Verification_Transmittance_URTotalXSSensitivity}
  S_{J_0,\sigma_{t,g,k_g}} = \frac{w_g N \sigma_{t,g,k_g} L}{J_0}  \sum_\ell 
    1_{g,k_g,\ell}^-        p_{g,\ell}   \left( \frac{ f_{g,\ell} - e^{-\tau_{g,\ell}}   }{ \tau_{g-1,\ell} - \tau_{g,\ell} } \right)
  + 1_{g+1,k_{g+1},\ell}^+  p_{g+1,\ell} \left( \frac{ e^{-\tau_{g,\ell}} - f_{g+1,\ell} }{ \tau_{g,\ell} - \tau_{g+1,\ell} } \right) .
\end{align}
Considering an energy grid point $g$, the first term in the summation corresponds to the upper energy range $E_g \le E < E_{g-1}$ where $E_g$ is on its lower bound. The second term therefore corresponds to the lower energy range $E_{g+1} \le E < E_{g}$ where $E_g$ is on its upper bound. In this manner, the data on each interior energy grid point contributes to the neutron transport in the neighboring ranges. The case of the top or bottom of the unresolved resonance range, $g = 0$ or $g = G$ respectively, is handled by the value of the first index of the indicator function. At the top of the unresolved resonance range, $1^-_{0,k,\ell} = 0$ meaning that there is no energy range where $g = 0$ resides on the lower edge. Likewise, at the bottom of the unresolved resonance range, $1^+_{G+1,k,\ell} = 0$ meaning there is no energy range where $g = G$ resides on the upper edge.

The sensitivity coefficient for the total cross section over the entire unresolved resonance range can be found by taking the sum of the sensitivity coefficients for each energy grid point $g$ and probability table band $k$:
\begin{align} \label{Eqn:Verification_Transmittance_TotalXSSensitivity_Sum}
  S_{J_0,\sigma_t} = \sum_{g=1}^G \sum_k  S_{J_0,\sigma_{t,g,k}} .
\end{align} 

Equation~\eqref{Eqn:Verification_Transmittance_URTotalXSSensitivity} is evaluated using the provided $^{90}$Zr unresolved resonance cross sections to obtain reference solution for the sensitivity coefficients of the uncollided current or transmission to the total cross section for a slab with a thickness of $L = 10$~cm. The numerical results for the sensitivity coefficient of each individual datum in the unresolved resonance table is provided in Table~\ref{Table_Verification_BenchmarkUncollidedTransmittance_SensitivityTotalXS}. Sum totals over all the probability bands at each energy grid point are given along with the grand total over all bands and energy grid points. The reference solution shows that for this specific problem, only probability bands 3 through 12 have a significant impact on the result.

Numerical results were obtained using a Shuriken Monte Carlo calculation with $10^9$ particle histories. These results are given in Table~\ref{Table_Verification_MonteCarloUncollidedTransmittance_SensitivityTotalXS}. The uncertainties for the sensitivity coefficient for each individual cross section value are not displayed because it would be difficult to fit into a table meaningfully, but the significant values have uncertainties that are on the order of few tenths of a percent. The uncertainty is provided for the overall current and the energy-integrated sensitivity coefficient, which has a 1-$\sigma$ uncertainty within 0.016\%. The agreement of these overall values is within 1.5$\sigma$ of the respective analytic reference solution, which is within the generally accepted value of 2$\sigma$ (a 95\% confidence interval). There is also general agreement in the individual sensitivity coefficients that are significant in magnitude. 

Note that the Monte Carlo results are zero for probability bands 14 through 16. This is because there were no scores to those specific estimators. In other words, no uncollided particles leaked out of the slab when these bands were selected. This is expected for two reasons. First, and most importantly, the cross section values in Table~\ref{Table_Verification_90Zr_URRCrossSectionData} are large enough such that when it is multiplied by the number density and thickness and exponentiated, the leakage probability is extremely small. Second, the probability of sampling these bands is each less than 1\%, meaning that there are relatively fewer chances to leak out. Besides, as mentioned, the reference solutions show that the sensitivity coefficients are on the order of $10^{-10}$ or small in all but a few instances. 

\begin{table}[t!] \small
\caption{Benchmark Sensitivity Coefficients for the Uncollided Transmission Probability with Respect to the Total Cross Section for Slab Thickness $L = 10$~cm}
\begin{center}
\begin{tabular}{|l|c|c|c|c|c|c|} \hline
  $E$ (keV)  &       53.5 &       54.0 &       59.0 &       64.0 &       69.0 &       74.0 \\ \hline
  Band = 1 & -1.774e-10 & -3.980e-07 & -6.232e-08 & -1.091e-06 & -3.494e-06 & -5.017e-07 \\
  2 & -8.102e-09 & -5.439e-06 & -9.248e-07 & -1.920e-04 & -8.469e-04 & -3.901e-06 \\
  3 & -7.708e-05 & -2.951e-03 & -1.310e-03 & -5.276e-03 & -6.610e-03 & -3.331e-03 \\
  4 & -7.563e-04 & -9.472e-03 & -1.001e-02 & -1.934e-02 & -2.052e-02 & -1.279e-02 \\
  5 & -1.128e-03 & -1.488e-02 & -2.042e-02 & -2.853e-02 & -2.442e-02 & -2.827e-02 \\
  6 & -2.203e-03 & -1.951e-02 & -2.727e-02 & -4.581e-02 & -3.949e-02 & -3.959e-02 \\
  7 & -2.501e-03 & -1.808e-02 & -4.242e-02 & -3.594e-02 & -2.738e-02 & -4.759e-02 \\
  8 & -1.796e-03 & -1.149e-02 & -5.352e-02 & -3.046e-02 & -2.391e-02 & -3.894e-02 \\
  9 & -1.220e-03 & -1.925e-02 & -3.412e-02 & -2.347e-02 & -2.692e-02 & -2.273e-02 \\
 10 & -8.106e-04 & -1.534e-02 & -1.899e-02 & -1.789e-02 & -2.523e-02 & -1.226e-02 \\
 11 & -3.563e-04 & -7.579e-03 & -7.229e-03 & -6.319e-03 & -1.506e-02 & -4.391e-03 \\
 12 & -8.620e-06 & -4.010e-04 & -2.145e-04 & -8.568e-05 & -9.669e-04 & -4.034e-05 \\
 13 & -1.661e-09 & -8.872e-07 & -2.781e-07 & -2.184e-08 & -1.161e-06 & -3.207e-08 \\
 14 & -8.530e-14 & -3.369e-11 & -4.020e-11 & -3.705e-13 & -1.221e-10 & -7.526e-13 \\
 15 & -1.259e-17 & -7.881e-16 & -9.360e-17 & -1.482e-17 & -1.282e-15 & -4.030e-15 \\
 16 & -5.018e-23 & -4.146e-22 & -1.097e-20 & -1.673e-19 & -6.215e-18 & -2.720e-17 \\ \hline
 Sum   & -1.086e-02 & -1.190e-01 & -2.155e-01 & -2.133e-01 & -2.114e-01 & -2.099e-01 \\ \hline
   $E$ (keV)    &       79.0 &       84.0 &       89.0 &       94.0 &       99.0 &      100.0 \\ \hline
  Band = 1 & -5.916e-07 & -3.373e-07 & -4.198e-06 & -1.693e-08 & -1.098e-08 & -3.314e-06 \\
  2 & -1.410e-06 & -1.407e-06 & -1.799e-03 & -1.884e-06 & -7.161e-07 & -1.487e-04 \\
  3 & -2.935e-03 & -2.986e-03 & -1.462e-02 & -2.539e-03 & -1.690e-03 & -1.722e-03 \\
  4 & -1.080e-02 & -1.553e-02 & -1.831e-02 & -1.156e-02 & -5.879e-03 & -2.143e-03 \\
  5 & -2.302e-02 & -2.272e-02 & -2.076e-02 & -2.146e-02 & -1.161e-02 & -2.663e-03 \\
  6 & -4.460e-02 & -4.235e-02 & -3.223e-02 & -3.638e-02 & -2.057e-02 & -2.576e-03 \\
  7 & -4.568e-02 & -4.357e-02 & -2.901e-02 & -4.459e-02 & -3.346e-02 & -2.036e-03 \\
  8 & -3.829e-02 & -3.827e-02 & -1.671e-02 & -3.497e-02 & -2.088e-02 & -1.904e-03 \\
  9 & -2.390e-02 & -2.530e-02 & -2.580e-02 & -2.368e-02 & -1.610e-02 & -2.724e-03 \\
 10 & -1.525e-02 & -1.235e-02 & -2.595e-02 & -1.985e-02 & -8.986e-03 & -3.005e-03 \\
 11 & -3.636e-03 & -4.085e-03 & -1.846e-02 & -9.530e-03 & -2.846e-03 & -1.345e-03 \\
 12 & -2.131e-05 & -7.427e-05 & -1.204e-03 & -4.076e-04 & -5.723e-05 & -4.358e-05 \\
 13 & -1.154e-07 & -1.632e-07 & -4.068e-06 & -1.593e-06 & -3.339e-07 & -6.494e-08 \\
 14 & -4.232e-11 & -9.105e-12 & -9.075e-10 & -1.560e-09 & -7.805e-11 & -4.994e-11 \\
 15 & -1.155e-15 & -2.040e-14 & -5.454e-13 & -4.102e-13 & -2.296e-12 & -9.803e-14 \\
 16 & -8.144e-18 & -6.415e-18 & -2.144e-15 & -2.161e-14 & -6.304e-14 & -4.203e-15 \\ \hline
 Sum   & -2.081e-01 & -2.072e-01 & -2.049e-01 & -2.050e-01 & -1.221e-01 & -2.031e-02 \\ \hline
 \multicolumn{7}{|l|}{ Energy Integrated Sensitivity = -1.94750} \\ \hline
 \multicolumn{7}{|l|}{ Transmission Probability = 8.22421e-2} \\ \hline
\end{tabular}
\end{center}
\label{Table_Verification_BenchmarkUncollidedTransmittance_SensitivityTotalXS}
\end{table}%

\begin{table}[t!] \small
\caption{Shuriken Monte Carlo Computed Sensitivity Coefficients for the Uncollided Transmission Probability with Respect to the Total Cross Section for Slab Thickness $L = 10$~cm}
\begin{center}
\begin{tabular}{|l|c|c|c|c|c|c|} \hline
  $E$ (keV)  &       53.5 &       54.0 &       59.0 &       64.0 &       69.0 &       74.0 \\ \hline
  Band = 1 & -1.762e-10 & -3.996e-07 & -6.249e-08 &-1.091e-06 &-3.487e-06 & -5.022e-07 \\
  2	& -7.977e-09 & -5.445e-06 & -9.243e-07 & -1.919e-04 & -8.481e-04 & -3.886e-06 \\
  3	& -7.690e-05 & -2.953e-03 & -1.310e-03 & -5.287e-03 & -6.620e-03 & -3.325e-03 \\
  4	& -7.509e-04 & -9.468e-03 & -1.002e-02 & -1.932e-02 & -2.051e-02 & -1.279e-02 \\
  5	& -1.126e-03 & -1.490e-02 & -2.043e-02 & -2.852e-02 & -2.441e-02 & -2.827e-02 \\
  6	& -2.198e-03 & -1.950e-02 & -2.724e-02 & -4.578e-02 & -3.944e-02 & -3.959e-02 \\
  7	& -2.503e-03 & -1.807e-02 & -4.241e-02 & -3.589e-02 & -2.739e-02 & -4.761e-02 \\
  8	& -1.796e-03 & -1.152e-02 & -5.347e-02 & -3.043e-02 & -2.389e-02 & -3.901e-02 \\
  9	& -1.221e-03 & -1.926e-02 & -3.413e-02 & -2.349e-02 & -2.693e-02 & -2.273e-02 \\
 10	& -7.994e-04 & -1.534e-02 & -1.901e-02 & -1.789e-02 & -2.522e-02 & -1.230e-02 \\
 11	& -3.585e-04 & -7.564e-03 & -7.224e-03 & -6.347e-03 & -1.504e-02 & -4.375e-03 \\
 12	& -7.595e-06 & -3.988e-04 & -2.224e-04 & -8.729e-05 & -9.609e-04 & -4.045e-05 \\
 13	&  0.000e+00 & -1.008e-06 & -3.395e-07 & -1.722e-07 & -1.873e-06 & -4.769e-08 \\
 14	&  0.000e+00 &  0.000e+00 &  0.000e+00 &  0.000e+00 &  0.000e+00 &  0.000e+00 \\
 15	&  0.000e+00 &  0.000e+00 &  0.000e+00 &  0.000e+00 &  0.000e+00 &  0.000e+00 \\
 16	&  0.000e+00 &  0.000e+00 &  0.000e+00 &  0.000e+00 &  0.000e+00 &  0.000e+00 \\ \hline
Sum	& -1.084e-02 & -1.190e-01 & -2.155e-01 & -2.132e-01 & -2.113e-01 & -2.100e-01 \\ \hline
   $E$ (keV)    &       79.0 &       84.0 &       89.0 &       94.0 &       99.0 &      100.0 \\ \hline
  Band = 1 & -5.925e-07 & -3.372e-07 & -4.198e-06 & -1.708e-08 & -1.091e-08 & -3.334e-06 \\
  2 & -1.408e-06 & -1.401e-06 & -1.798e-03 & -1.883e-06 & -7.166e-07 & -1.480e-04 \\
  3 & -2.932e-03 & -2.985e-03 & -1.463e-02 & -2.539e-03 & -1.691e-03 & -1.728e-03 \\
  4 & -1.081e-02 & -1.553e-02 & -1.828e-02 & -1.157e-02 & -5.886e-03 & -2.149e-03 \\
  5 & -2.303e-02 & -2.270e-02 & -2.077e-02 & -2.148e-02 & -1.161e-02 & -2.670e-03 \\
  6 & -4.465e-02 & -4.235e-02 & -3.222e-02 & -3.639e-02 & -2.057e-02 & -2.577e-03 \\
  7 & -4.565e-02 & -4.350e-02 & -2.902e-02 & -4.458e-02 & -3.350e-02 & -2.043e-03 \\
  8 & -3.825e-02 & -3.822e-02 & -1.671e-02 & -3.503e-02 & -2.090e-02 & -1.904e-03 \\
  9 & -2.394e-02 & -2.533e-02 & -2.579e-02 & -2.369e-02 & -1.611e-02 & -2.726e-03 \\
 10 & -1.527e-02 & -1.234e-02 & -2.595e-02 & -1.985e-02 & -8.979e-03 & -2.976e-03 \\
 11 & -3.628e-03 & -4.072e-03 & -1.846e-02 & -9.519e-03 & -2.847e-03 & -1.334e-03 \\
 12 & -2.237e-05 & -7.538e-05 & -1.204e-03 & -4.190e-04 & -5.967e-05 & -4.416e-05 \\
 13 & -3.824e-07 & -2.834e-08 & -4.015e-06 & -1.472e-06 & -3.147e-07 & -1.482e-07 \\
 14 &  0.000e+00 &  0.000e+00 &  0.000e+00 &  0.000e+00 &  0.000e+00 &  0.000e+00 \\
 15 &  0.000e+00 &  0.000e+00 &  0.000e+00 &  0.000e+00 &  0.000e+00 &  0.000e+00 \\
 16 &  0.000e+00 &  0.000e+00 &  0.000e+00 &  0.000e+00 &  0.000e+00 &  0.000e+00 \\ \hline
Sum & -2.082e-01 & -2.071e-01 & -2.048e-01 & -2.051e-01 & -1.222e-01 & -2.030e-02 \\ \hline
\multicolumn{7}{|l|}{ Energy Integrated Sensitivity = -1.947481e+00 $\pm$ 3.103477e-04} \\ \hline
\multicolumn{7}{|l|}{ Transmission Probability = 8.222908e-02 $\pm$ 8.687201e-06} \\ \hline
\end{tabular}
\end{center}
\label{Table_Verification_MonteCarloUncollidedTransmittance_SensitivityTotalXS}
\end{table}%

\clearpage

\subsection{Analytic Benchmark for First-Collision Radiative Capture Sensitivities} \label{Sec:Verification_FirstCollisionCapture}
The analytic uncollided transmittance sensitivity benchmark is relatively simple, but is rather limited in the portions of the method verified. Only the total cross section can be meaningfully tested, and the transmittance does not involve any direct effect. A slightly more complicated version of this benchmark can be constructed by computing the sensitivity coefficient to the probability that a neutron experiences a particular reaction type on its first collision. Using the same problem setup as the transmittance benchmark, the probability that neutrons undergo radiative capture $(\text{n},\gamma)$ on their first collision is
\begin{align}
  R_\gamma =  \frac{1}{E_0 - E_G} \sum_{g=1}^G \sum_\ell p_{g,\ell} \int_{E_g}^{E_{g-1}} \frac{\Sigma_{\gamma,\ell}(E)}{\Sigma_{t,\ell}(E)} \bigg( 1 - \exp \left[ -\Sigma_{t,\ell}(E) L \right] \bigg) dE .
\end{align}
The radiative capture $(\text{n},\gamma)$ cross section is $\Sigma_\gamma(E)$, which like the total cross section in Eq.~\eqref{Eqn:Verification_LinearInterpolationTotalXS}, is given by a linear interpolation within an energy range $E_g \le E < E_{g-1}$ using the same interpolation factor $\rho$. 

As with the transmittance case, the integral can be evaluated by transforming from energy to interpolation factor using the relationship in Eq.~\eqref{Eqn:Verification_TransformationEnergyToInterpolationFactor} to give an expression of the form
\begin{align}
  R_\gamma = \sum_{g=1}^G \sum_\ell w_g p_{g,\ell} f_{g,\ell} 
\end{align}
with
\begin{align}
  f_{g,\ell} &= \int_0^1 
  \frac{\sigma_{\gamma,g,\ell} + \rho ( \sigma_{\gamma,g-1,\ell} - \sigma_{\gamma,g,\ell} )}
       {\sigma_{t,g,\ell}      + \rho ( \sigma_{t,g-1,\ell}      - \sigma_{t,g,\ell} )} 
  \bigg( 1 - \exp \bigg[ -N L \left(  \sigma_{t,g,\ell} + \rho ( \sigma_{t,g-1,\ell} - \sigma_{t,g,\ell} ) \right) \bigg] \bigg) d\rho \nonumber \\
  &= \frac{ h_{g,\ell} }{  ( \tau_{g,\ell} - \tau_{g-1,\ell} )^2 }, 
\end{align}
where
\begin{align}
   h_{g,\ell} 
   &= ( \tau_{\gamma,g,\ell} - \tau_{\gamma,g-1,\ell} ) \bigg[ \tau_{g,\ell} - \tau_{g-1,\ell} + e^{-\tau_{g,\ell}} - e^{-\tau_{g-1,\ell}} \bigg] \nonumber \\
   &+ ( \tau_{\gamma,g-1,\ell} \tau_{g,\ell} - \tau_{\gamma,g,\ell} \tau_{g-1,\ell} ) \bigg[ \ln\left( \frac{\tau_{g,\ell}}{\tau_{g-1,\ell}} \right) + \text{E}_1(\tau_{g,\ell}) - \text{E}_1(\tau_{g-1,\ell}) \bigg] .
\end{align}
The quantity $\tau_{\gamma,g,\ell}$ is the optical distance [Eq.~\eqref{Eqn:Verification_OpticalDistance}] reduced by the radiative capture probability,
\begin{align}
  \tau_{\gamma,g,\ell} = N L \sigma_{\gamma,g,\ell} .
\end{align}
The $\text{E}_n(x)$ function is the exponential integral defined by
\begin{align}
  \text{E}_n(x) = \int_1^\infty \frac{e^{-xt}}{t^n} dt.
\end{align}

Two sensitivity coefficients are computed. The first is the sensitivity to the capture rate with respect to the radiative capture $(\text{n},\gamma)$ cross section, and the second is with respect to the elastic scattering cross section. Note that when taking the derivatives, the total cross section is the sum of the radiative capture and elastic scattering cross sections.

The sensitivity coefficient for the capture rate with respect to an unresolved resonance probability table capture cross section is
\begin{align}
  S_{R_\gamma,\sigma_{\gamma,g,k_g}} = \frac{w_g N \sigma_{\gamma,g,k_g} L}{R_\gamma}  &\sum_\ell 
    1_{g,k_g,\ell}^-    p_{g,\ell}   \left[ \frac{ h_{\gamma,g,\ell}^- - 2 ( \tau_{g,\ell} - \tau_{g-1,\ell} ) f_{g,\ell} }{ ( \tau_{g,\ell} - \tau_{g-1,\ell} )^2 } \right] \nonumber \\
  &+ 1_{g+1,k_{g+1},\ell}^+  p_{g+1,\ell}\left[ \frac{ h_{\gamma,g+1,\ell}^+ + 2 ( \tau_{g+1,\ell} - \tau_{g,\ell} ) f_{g+1,\ell} }{ ( \tau_{g+1,\ell} - \tau_{g,\ell} )^2 } \right] ,
\end{align}
where $h_{\gamma,g,\ell}^\pm$ are derivatives of $h_{g,\ell}$ with respect to the capture cross section located at the upper ($+$ superscript) or lower ($-$ superscript) respective edges at the energy range; these are
\begin{subequations}
\begin{align}
  h_{\gamma,g,\ell}^- 
  &= \tau_{g,\ell} - \tau_{g-1,\ell} + e^{-\tau_{g,\ell}} - e^{-\tau_{g-1,\ell}} 
   + ( \tau_{\gamma,g,\ell}   - \tau_{\gamma,g-1,\ell} ) \left( 1 - e^{-\tau_{g,\ell}} \right) \nonumber \\
  &+ ( \tau_{\gamma,g-1,\ell} - \tau_{g-1,\ell} ) \bigg[ \ln\left( \frac{\tau_{g,\ell}}{\tau_{g-1,\ell}} \right) + \text{E}_1(\tau_{g,\ell}) - \text{E}_1(\tau_{g-1,\ell}) \bigg] \nonumber \\
  &+ ( \tau_{\gamma,g-1,\ell} \tau_{g,\ell} - \tau_{\gamma,g,\ell} \tau_{g-1,\ell}  ) \bigg[ \frac{1 - e^{-\tau_{g,\ell}}}{ \tau_{g,\ell} }\bigg] , \\
  h_{\gamma,g,\ell}^+
  &=  \tau_{g-1,\ell} - \tau_{g,\ell} + e^{-\tau_{g-1,\ell}} - e^{-\tau_{g,\ell}} 
   + ( \tau_{\gamma,g-1,\ell}   - \tau_{\gamma,g,\ell} ) \left( 1 - e^{-\tau_{g-1,\ell}} \right) \nonumber \\
  &+ ( \tau_{g,\ell} - \tau_{\gamma,g,\ell} ) \bigg[ \ln\left( \frac{\tau_{g,\ell}}{\tau_{g-1,\ell}} \right) + \text{E}_1(\tau_{g,\ell}) - \text{E}_1(\tau_{g-1,\ell}) \bigg] \nonumber \\
  &+ ( \tau_{\gamma,g,\ell} \tau_{g-1,\ell} - \tau_{\gamma,g-1,\ell} \tau_{g,\ell}  ) \bigg[ \frac{1 - e^{-\tau_{g-1,\ell}}}{ \tau_{g-1,\ell} }\bigg] .
\end{align}
\end{subequations}

Using a similar process, the sensitivity coefficient for the capture rate with respect to an unresolved resonance probability table elastic scattering cross section may be obtained. The form is almost identical to the result for the capture cross section sensitivity coefficient,
\begin{align}
  S_{R_\gamma,\sigma_{s,g,k_g}} = \frac{w_g N \sigma_{s,g,k_g} L}{R_\gamma}  \sum_\ell 
    &1_{g,k_g,\ell}^-    p_{g,\ell}   \left[ \frac{ h_{s,g,\ell}^- - 2 ( \tau_{g,\ell} - \tau_{g-1,\ell} ) f_{g,\ell} }{ ( \tau_{g,\ell} - \tau_{g-1,\ell} )^2 } \right] \nonumber \\
  + &1_{g+1,k_{g+1},\ell}^+  p_{g+1,\ell}\left[ \frac{ h_{s,g+1,\ell}^+ + 2 ( \tau_{g+1,\ell} - \tau_{g,\ell} ) f_{g+1,\ell} }{ ( \tau_{g+1,\ell} - \tau_{g,\ell} )^2 } \right] ,
\end{align}
with the major difference being in the $h_{s,g,\ell}^\pm$:
\begin{subequations}
\begin{align}
  h_{s,g,\ell}^- 
  &= ( \tau_{\gamma,g,\ell}   - \tau_{\gamma,g-1,\ell} ) \left( 1 - e^{-\tau_{g,\ell}} \right) \nonumber \\
  &+ \tau_{\gamma,g-1,\ell} \bigg[ \ln\left( \frac{\tau_{g,\ell}}{\tau_{g-1,\ell}} \right) + \text{E}_1(\tau_{g,\ell}) - \text{E}_1(\tau_{g-1,\ell}) \bigg] \nonumber \\
  &+ ( \tau_{\gamma,g-1,\ell} \tau_{g,\ell} - \tau_{\gamma,g,\ell} \tau_{g-1,\ell}  ) \bigg[ \frac{1 - e^{-\tau_{g,\ell}}}{ \tau_{g,\ell} }\bigg] , \\
  h_{s,g,\ell}^+
  &= ( \tau_{\gamma,g-1,\ell}   - \tau_{\gamma,g,\ell} ) \left( 1 - e^{-\tau_{g-1,\ell}} \right) \nonumber \\
  &- \tau_{\gamma,g,\ell} \bigg[ \ln\left( \frac{\tau_{g,\ell}}{\tau_{g-1,\ell}} \right) + \text{E}_1(\tau_{g,\ell}) - \text{E}_1(\tau_{g-1,\ell}) \bigg] \nonumber \\
  &+ ( \tau_{\gamma,g,\ell} \tau_{g-1,\ell} - \tau_{\gamma,g-1,\ell} \tau_{g,\ell}  ) \bigg[ \frac{1 - e^{-\tau_{g-1,\ell}}}{ \tau_{g-1,\ell} }\bigg] .
\end{align}
\end{subequations}
These derivatives are somewhat simpler than the capture case because the only dependence on the elastic scattering cross section is in the total cross section.

Inserting the $^{90}$Zr cross section values from Table~\ref{Table_Verification_90Zr_URRCrossSectionData} into the above expressions for a width $L = 10$~cm provides a set of reference solutions that are given in Tables~\ref{Table_Verification_BenchmarkFirstCollisionCapture_SensitivityCaptureXS} and~\ref{Table_Verification_BenchmarkFirstCollisionCapture_SensitivityElasticXS} for the sensitivity coefficients with respect to the capture and elastic scattering cross sections respectively. As with the previous section, solutions are given for: each energy grid point and probability band, the sum over all probability bands at each energy grid point, and the sum over the entire unresolved resonance table.

The radiative capture sensitivity coefficients given in Table~\ref{Table_Verification_BenchmarkFirstCollisionCapture_SensitivityCaptureXS} are strictly positive. This makes sense because increasing the capture cross section will always increase the radiative capture rate, either through directly increasing the reaction probability in a collision or indirectly through raising the total cross section and decreasing the likelihood of neutrons leaking out of the slab. Note that the direct effect is dominant in this case.

Conversely, the elastic scattering sensitivity coefficients given in Table~\ref{Table_Verification_BenchmarkFirstCollisionCapture_SensitivityElasticXS} are all negative. The reason for this is because the slab, having a thickness of $L = 10$~cm, is optically thick. There is a small increase in the radiative capture rate by increasing the collision probability, but this is offset by decrease in the probability that a neutron is captured as a consequence of independently increasing the elastic scattering cross section.

The Shuriken Monte Carlo results were obtained using $10^9$ histories in the same calculation as the one used to generate the sensitivity coefficients to the leakage detailed in the previous section. The calculated sensitivity coefficients are provided for the radiative capture cross section in Table~\ref{Table_Verification_MonteCarloFirstCollisionCapture_SensitivityCaptureXS} and for the elastic scattering cross section in Table~\ref{Table_Verification_MonteCarloFirstCollisionCapture_SensitivityElasticXS}.

As with those previously presented results, uncertainties are only provided for the energy-integrated values since providing them for this large dataset in a presentable manner would be difficult. As before, the uncertainties of each individual cross section value is on the order of a few tenths of a percent for those that contribute significantly. The conclusions are essentially the same as for the sensitivity coefficient to the leakage with respect to the total cross section. There is broad agreement for the significant values. Some of the results for the lower probability bands for the elastic scattering sensitivity coefficients are positive, having the wrong sign, however, this is because the magnitude of the reference values are very small and difficult to sample sufficiently. The computed energy-integrated sensitivity coefficient for the radiative capture cross section has an uncertainty of about 0.006\% and agrees with the reference solution within 0.03$\sigma$. The computed elastic scattering analog has an uncertainty of about 0.013\% and agrees within 0.96$\sigma$. This level of agreement is within the accepted 95\% confidence interval or 2$\sigma$.

\begin{table}[t!] \small
\caption{Benchmark Sensitivity Coefficients for the First-Collision Capture Rate with Respect to the Capture Cross Section for Slab Thickness $L = 10$~cm}
\begin{center}
\begin{tabular}{|l|c|c|c|c|c|c|} \hline
  $E$ (keV)  &       53.5 &       54.0 &       59.0 &       64.0 &       69.0 &       74.0 \\ \hline
  Band = 1 &  1.930e-08 &  4.339e-05 &  6.782e-06 &  1.187e-04 &  3.784e-04 &  5.460e-05 \\
  2 &  8.817e-07 &  4.488e-04 &  1.007e-04 &  1.092e-03 &  1.677e-03 &  4.103e-04 \\
  3 &  1.384e-04 &  2.197e-03 &  2.310e-03 &  3.473e-03 &  3.168e-03 &  2.896e-03 \\
  4 &  3.190e-04 &  2.707e-03 &  4.221e-03 &  5.791e-03 &  5.427e-03 &  4.302e-03 \\
  5 &  2.336e-04 &  2.870e-03 &  5.203e-03 &  4.818e-03 &  4.002e-03 &  5.255e-03 \\
  6 &  2.873e-04 &  2.810e-03 &  4.607e-03 &  6.794e-03 &  5.494e-03 &  5.554e-03 \\
  7 &  2.992e-04 &  2.374e-03 &  6.200e-03 &  5.450e-03 &  3.644e-03 &  6.659e-03 \\
  8 &  4.340e-04 &  1.587e-03 &  8.880e-03 &  7.193e-03 &  4.262e-03 &  1.011e-02 \\
  9 &  5.553e-04 &  5.009e-03 &  1.344e-02 &  1.055e-02 &  7.392e-03 &  1.263e-02 \\
 10 &  7.848e-04 &  8.640e-03 &  1.840e-02 &  1.864e-02 &  1.422e-02 &  1.664e-02 \\
 11 &  1.318e-03 &  1.511e-02 &  2.417e-02 &  2.473e-02 &  2.839e-02 &  2.474e-02 \\
 12 &  7.621e-04 &  1.074e-02 &  1.460e-02 &  1.418e-02 &  2.020e-02 &  1.099e-02 \\
 13 &  3.373e-04 &  5.426e-03 &  6.898e-03 &  5.570e-03 &  7.531e-03 &  4.771e-03 \\
 14 &  2.210e-04 &  2.489e-03 &  4.483e-03 &  3.210e-03 &  3.891e-03 &  2.588e-03 \\
 15 &  7.259e-05 &  7.919e-04 &  9.721e-04 &  6.601e-04 &  6.351e-04 &  4.587e-04 \\
 16 &  1.663e-05 &  1.679e-04 &  2.865e-04 &  2.807e-04 &  3.727e-04 &  2.713e-04 \\ \hline
 Sum   &  5.780e-03 &  6.341e-02 &  1.148e-01 &  1.125e-01 &  1.107e-01 &  1.083e-01 \\ \hline
  $E$ (keV)  &       79.0 &       84.0 &       89.0 &       94.0 &       99.0 &      100.0 \\ \hline
  Band = 1 &  6.433e-05 &  3.672e-05 &  4.181e-04 &  1.841e-06 &  1.194e-06 &  5.507e-05 \\
  2 &  1.541e-04 &  1.532e-04 &  1.653e-03 &  2.055e-04 &  7.803e-05 &  1.212e-04 \\
  3 &  2.594e-03 &  2.854e-03 &  4.771e-03 &  2.095e-03 &  1.427e-03 &  5.423e-04 \\
  4 &  3.870e-03 &  4.420e-03 &  3.714e-03 &  3.418e-03 &  2.003e-03 &  3.839e-04 \\
  5 &  4.472e-03 &  4.286e-03 &  3.110e-03 &  4.361e-03 &  2.375e-03 &  3.585e-04 \\
  6 &  6.490e-03 &  5.904e-03 &  4.384e-03 &  4.814e-03 &  2.593e-03 &  3.313e-04 \\
  7 &  6.522e-03 &  5.749e-03 &  3.971e-03 &  5.719e-03 &  4.595e-03 &  2.591e-04 \\
  8 &  9.460e-03 &  9.243e-03 &  2.874e-03 &  6.973e-03 &  5.039e-03 &  3.284e-04 \\
  9 &  1.221e-02 &  1.366e-02 &  6.573e-03 &  8.728e-03 &  7.830e-03 &  6.957e-04 \\
 10 &  1.987e-02 &  1.645e-02 &  1.235e-02 &  1.568e-02 &  1.044e-02 &  1.564e-03 \\
 11 &  2.354e-02 &  2.218e-02 &  2.984e-02 &  2.460e-02 &  1.274e-02 &  2.864e-03 \\
 12 &  9.074e-03 &  1.015e-02 &  1.855e-02 &  1.354e-02 &  5.384e-03 &  1.444e-03 \\
 13 &  5.029e-03 &  5.330e-03 &  7.142e-03 &  6.098e-03 &  3.321e-03 &  5.562e-04 \\
 14 &  2.895e-03 &  2.477e-03 &  3.239e-03 &  3.135e-03 &  1.179e-03 &  2.377e-04 \\
 15 &  3.120e-04 &  5.984e-04 &  7.829e-04 &  4.900e-04 &  3.407e-04 &  4.077e-05 \\
 16 &  9.834e-05 &  6.305e-05 &  2.507e-04 &  3.626e-04 &  2.613e-04 &  3.566e-05 \\ \hline
 Sum   &  1.067e-01 &  1.036e-01 &  1.036e-01 &  1.002e-01 &  5.961e-02 &  9.819e-03 \\ \hline
 \multicolumn{7}{|l|}{ Energy Integrated Sensitivity = 0.999011} \\ \hline
 \multicolumn{7}{|l|}{ First-Collided Capture Probability = 7.54284e-4} \\ \hline
\end{tabular}
\end{center}
\label{Table_Verification_BenchmarkFirstCollisionCapture_SensitivityCaptureXS}
\end{table}%

\begin{table}[t!] \small
\caption{Benchmark Sensitivity Coefficients for the First-Collision Capture Rate with Respect to the Elastic Scattering Cross Section for Slab Thickness $L = 10$~cm}
\begin{center}
\begin{tabular}{|l|c|c|c|c|c|c|} \hline
  $E$ (keV)  &       53.5 &       54.0 &       59.0 &       64.0 &       69.0 &       74.0 \\ \hline
  Band = 1 & -6.242e-15 & -1.288e-11 & -2.009e-12 & -9.768e-11 & -1.502e-09 & -2.444e-11 \\
  2 & -2.719e-13 & -1.610e-07 & -2.369e-11 & -1.982e-05 & -1.034e-04 & -1.277e-08 \\
  3 & -8.250e-06 & -3.635e-04 & -1.472e-04 & -6.474e-04 & -8.185e-04 & -3.882e-04 \\
  4 & -1.107e-04 & -1.265e-03 & -1.286e-03 & -2.728e-03 & -2.824e-03 & -1.683e-03 \\
  5 & -1.433e-04 & -1.867e-03 & -2.835e-03 & -3.285e-03 & -2.740e-03 & -3.339e-03 \\
  6 & -2.215e-04 & -2.086e-03 & -3.163e-03 & -5.006e-03 & -4.133e-03 & -4.127e-03 \\
  7 & -2.413e-04 & -1.928e-03 & -4.615e-03 & -4.314e-03 & -2.914e-03 & -5.293e-03 \\
  8 & -3.480e-04 & -1.328e-03 & -7.136e-03 & -5.888e-03 & -3.436e-03 & -8.251e-03 \\
  9 & -4.742e-04 & -4.176e-03 & -1.098e-02 & -8.828e-03 & -5.970e-03 & -1.085e-02 \\
 10 & -7.109e-04 & -7.591e-03 & -1.600e-02 & -1.655e-02 & -1.217e-02 & -1.510e-02 \\
 11 & -1.193e-03 & -1.389e-02 & -2.259e-02 & -2.352e-02 & -2.587e-02 & -2.356e-02 \\
 12 & -7.693e-04 & -1.010e-02 & -1.490e-02 & -1.408e-02 & -1.984e-02 & -1.095e-02 \\
 13 & -3.440e-04 & -5.244e-03 & -7.013e-03 & -5.725e-03 & -7.723e-03 & -4.817e-03 \\
 14 & -2.157e-04 & -2.492e-03 & -4.364e-03 & -3.129e-03 & -3.791e-03 & -2.552e-03 \\
 15 & -7.129e-05 & -7.640e-04 & -9.833e-04 & -6.763e-04 & -6.389e-04 & -4.772e-04 \\
 16 & -1.689e-05 & -1.647e-04 & -2.904e-04 & -2.763e-04 & -3.818e-04 & -2.675e-04 \\ \hline
Sum    & -4.869e-03 & -5.326e-02 & -9.630e-02 & -9.467e-02 & -9.336e-02 & -9.166e-02 \\ \hline
  $E$ (keV)  &       79.0 &       84.0 &       89.0 &       94.0 &       99.0 &      100.0 \\ \hline
  Band = 1 & -3.525e-11 & -1.151e-11 & -2.329e-08 & -6.843e-13 & -4.459e-13 & -1.899e-07 \\
  2 & -1.517e-10 & -8.658e-11 & -1.815e-04 & -2.352e-10 & -2.161e-11 & -1.396e-05 \\
  3 & -3.144e-04 & -3.314e-04 & -1.797e-03 & -2.594e-04 & -1.760e-04 & -2.172e-04 \\
  4 & -1.390e-03 & -1.901e-03 & -2.241e-03 & -1.367e-03 & -7.239e-04 & -2.422e-04 \\
  5 & -2.678e-03 & -2.688e-03 & -2.128e-03 & -2.642e-03 & -1.398e-03 & -2.533e-04 \\
  6 & -4.676e-03 & -4.369e-03 & -3.253e-03 & -3.479e-03 & -1.892e-03 & -2.549e-04 \\
  7 & -4.968e-03 & -4.620e-03 & -3.035e-03 & -4.501e-03 & -3.507e-03 & -2.074e-04 \\
  8 & -7.604e-03 & -7.643e-03 & -2.338e-03 & -5.735e-03 & -3.975e-03 & -2.672e-04 \\
  9 & -1.033e-02 & -1.184e-02 & -5.266e-03 & -7.446e-03 & -6.622e-03 & -5.659e-04 \\
 10 & -1.754e-02 & -1.505e-02 & -1.053e-02 & -1.386e-02 & -9.410e-03 & -1.380e-03 \\
 11 & -2.301e-02 & -2.120e-02 & -2.673e-02 & -2.273e-02 & -1.233e-02 & -2.660e-03 \\
 12 & -9.168e-03 & -1.015e-02 & -1.837e-02 & -1.346e-02 & -5.546e-03 & -1.441e-03 \\
 13 & -5.048e-03 & -5.272e-03 & -7.386e-03 & -6.089e-03 & -3.330e-03 & -5.549e-04 \\
 14 & -2.922e-03 & -2.426e-03 & -3.179e-03 & -3.112e-03 & -1.207e-03 & -2.351e-04 \\
 15 & -3.274e-04 & -6.094e-04 & -8.035e-04 & -5.003e-04 & -3.543e-04 & -4.206e-05 \\
 16 & -9.660e-05 & -6.275e-05 & -2.606e-04 & -3.561e-04 & -2.702e-04 & -3.432e-05 \\ \hline
Sum    & -9.007e-02 & -8.816e-02 & -8.751e-02 & -8.554e-02 & -5.074e-02 & -8.369e-03 \\ \hline
 \multicolumn{7}{|l|}{ Energy Integrated Sensitivity = -0.844514} \\ \hline
 \multicolumn{7}{|l|}{ First-Collided Capture Probability = 7.54284e-4} \\ \hline
\end{tabular}
\end{center}
\label{Table_Verification_BenchmarkFirstCollisionCapture_SensitivityElasticXS}
\end{table}%

\begin{table}[t!] \small
\caption{Shuriken Monte Carlo Computed Sensitivity Coefficients for the First-Collision Capture Rate with Respect to the Capture Cross Section for Slab Thickness $L = 10$~cm}
\begin{center}
\begin{tabular}{|l|c|c|c|c|c|c|} \hline
  $E$ (keV)  &       53.5 &       54.0 &       59.0 &       64.0 &       69.0 &       74.0 \\ \hline
  Band = 1 &  1.916e-08 &  4.356e-05 &  6.800e-06 &  1.187e-04 &  3.777e-04 &  5.464e-05 \\
  2 &  8.681e-07 &  4.493e-04 &  1.007e-04 &  1.092e-03 &  1.680e-03 &  4.087e-04 \\
  3 &  1.382e-04 &  2.199e-03 &  2.310e-03 &  3.477e-03 &  3.171e-03 &  2.893e-03 \\
  4 &  3.183e-04 &  2.708e-03 &  4.223e-03 &  5.786e-03 &  5.423e-03 &  4.302e-03 \\
  5 &  2.336e-04 &  2.872e-03 &  5.206e-03 &  4.818e-03 &  4.002e-03 &  5.253e-03 \\
  6 &  2.869e-04 &  2.809e-03 &  4.607e-03 &  6.791e-03 &  5.490e-03 &  5.555e-03 \\
  7 &  2.990e-04 &  2.374e-03 &  6.200e-03 &  5.450e-03 &  3.644e-03 &  6.660e-03 \\
  8 &  4.340e-04 &  1.589e-03 &  8.880e-03 &  7.192e-03 &  4.260e-03 &  1.011e-02 \\
  9 &  5.542e-04 &  5.013e-03 &  1.345e-02 &  1.055e-02 &  7.390e-03 &  1.263e-02 \\
 10 &  7.838e-04 &  8.641e-03 &  1.840e-02 &  1.865e-02 &  1.422e-02 &  1.665e-02 \\
 11 &  1.321e-03 &  1.511e-02 &  2.417e-02 &  2.474e-02 &  2.839e-02 &  2.473e-02 \\
 12 &  7.610e-04 &  1.073e-02 &  1.461e-02 &  1.418e-02 &  2.019e-02 &  1.098e-02 \\
 13 &  3.373e-04 &  5.431e-03 &  6.903e-03 &  5.570e-03 &  7.525e-03 &  4.772e-03 \\
 14 &  2.222e-04 &  2.491e-03 &  4.481e-03 &  3.204e-03 &  3.891e-03 &  2.588e-03 \\
 15 &  7.283e-05 &  7.901e-04 &  9.738e-04 &  6.627e-04 &  6.369e-04 &  4.593e-04 \\
 16 &  1.671e-05 &  1.675e-04 &  2.860e-04 &  2.818e-04 &  3.712e-04 &  2.716e-04 \\ \hline
Sum &  5.780e-03 &  6.342e-02 &  1.148e-01 &  1.126e-01 &  1.107e-01 &  1.083e-01 \\ \hline
  $E$ (keV)  &       79.0 &       84.0 &       89.0 &       94.0 &       99.0 &      100.0 \\ \hline
  Band = 1 &  6.443e-05 &  3.671e-05 &  4.181e-04 &  1.859e-06 &  1.187e-06 &  5.541e-05 \\
  2 &  1.539e-04 &  1.525e-04 &  1.652e-03 &  2.053e-04 &  7.808e-05 &  1.206e-04 \\
  3 &  2.592e-03 &  2.853e-03 &  4.772e-03 &  2.095e-03 &  1.427e-03 &  5.430e-04 \\
  4 &  3.870e-03 &  4.421e-03 &  3.711e-03 &  3.419e-03 &  2.003e-03 &  3.840e-04 \\
  5 &  4.472e-03 &  4.285e-03 &  3.110e-03 &  4.363e-03 &  2.375e-03 &  3.590e-04 \\
  6 &  6.491e-03 &  5.905e-03 &  4.383e-03 &  4.813e-03 &  2.592e-03 &  3.313e-04 \\
  7 &  6.520e-03 &  5.747e-03 &  3.972e-03 &  5.718e-03 &  4.595e-03 &  2.592e-04 \\
  8 &  9.457e-03 &  9.240e-03 &  2.875e-03 &  6.976e-03 &  5.041e-03 &  3.286e-04 \\
  9 &  1.221e-02 &  1.367e-02 &  6.573e-03 &  8.729e-03 &  7.831e-03 &  6.964e-04 \\
 10 &  1.986e-02 &  1.644e-02 &  1.235e-02 &  1.567e-02 &  1.043e-02 &  1.562e-03 \\
 11 &  2.355e-02 &  2.219e-02 &  2.985e-02 &  2.459e-02 &  1.274e-02 &  2.864e-03 \\
 12 &  9.082e-03 &  1.015e-02 &  1.855e-02 &  1.354e-02 &  5.384e-03 &  1.449e-03 \\
 13 &  5.031e-03 &  5.323e-03 &  7.142e-03 &  6.093e-03 &  3.326e-03 &  5.557e-04 \\
 14 &  2.901e-03 &  2.477e-03 &  3.234e-03 &  3.137e-03 &  1.178e-03 &  2.382e-04 \\
 15 &  3.117e-04 &  5.977e-04 &  7.831e-04 &  4.890e-04 &  3.412e-04 &  4.021e-05 \\
 16 &  9.850e-05 &  6.254e-05 &  2.501e-04 &  3.612e-04 &  2.614e-04 &  3.565e-05 \\ \hline
Sum &  1.067e-01 &  1.035e-01 &  1.036e-01 &  1.002e-01 &  5.961e-02 &  9.822e-03 \\ \hline
 \multicolumn{7}{|l|}{ Energy Integrated Sensitivity = 9.990123e-01 $\pm$ 5.685807e-05} \\ \hline
 \multicolumn{7}{|l|}{ First-Collided Capture Probability = 7.542643e-04 $\pm$ 3.037407e-08} \\ \hline
\end{tabular}
\end{center}
\label{Table_Verification_MonteCarloFirstCollisionCapture_SensitivityCaptureXS}
\end{table}%

\begin{table}[t!] \small
\caption{Shuriken Monte Carlo Computed Sensitivity Coefficients for the First-Collision Capture Rate with Respect to the Elastic Scattering Cross Section for Slab Thickness $L = 10$~cm}
\begin{center}
\begin{tabular}{|l|c|c|c|c|c|c|} \hline
  $E$ (keV)  &       53.5 &       54.0 &       59.0 &       64.0 &       69.0 &       74.0 \\ \hline
  Band = 1 & -1.225e-14 &  1.271e-07 & -4.030e-12 &  1.895e-07 &  7.423e-07 &  6.756e-11 \\
  2 & -5.359e-13 &  8.148e-06 &  1.821e-09 & -5.604e-06 & -7.695e-05 &  4.208e-06 \\
  3 & -5.264e-06 & -3.353e-04 & -1.068e-04 & -6.120e-04 & -8.165e-04 & -3.722e-04 \\
  4 & -1.092e-04 & -1.268e-03 & -1.257e-03 & -2.709e-03 & -2.857e-03 & -1.656e-03 \\
  5 & -1.445e-04 & -1.885e-03 & -2.870e-03 & -3.311e-03 & -2.762e-03 & -3.370e-03 \\
  6 & -2.230e-04 & -2.086e-03 & -3.175e-03 & -5.006e-03 & -4.126e-03 & -4.156e-03 \\
  7 & -2.415e-04 & -1.927e-03 & -4.619e-03 & -4.297e-03 & -2.915e-03 & -5.307e-03 \\
  8 & -3.472e-04 & -1.331e-03 & -7.109e-03 & -5.875e-03 & -3.427e-03 & -8.260e-03 \\
  9 & -4.703e-04 & -4.167e-03 & -1.095e-02 & -8.825e-03 & -5.976e-03 & -1.083e-02 \\
 10 & -6.904e-04 & -7.541e-03 & -1.587e-02 & -1.637e-02 & -1.210e-02 & -1.495e-02 \\
 11 & -1.191e-03 & -1.367e-02 & -2.250e-02 & -2.333e-02 & -2.529e-02 & -2.334e-02 \\
 12 & -7.834e-04 & -9.999e-03 & -1.506e-02 & -1.442e-02 & -2.014e-02 & -1.120e-02 \\
 13 & -3.557e-04 & -5.405e-03 & -7.177e-03 & -5.760e-03 & -7.966e-03 & -4.840e-03 \\
 14 & -2.212e-04 & -2.511e-03 & -4.368e-03 & -3.194e-03 & -3.846e-03 & -2.583e-03 \\
 15 & -7.254e-05 & -7.891e-04 & -9.906e-04 & -6.783e-04 & -6.519e-04 & -4.854e-04 \\
 16 & -1.675e-05 & -1.663e-04 & -2.887e-04 & -2.789e-04 & -3.817e-04 & -2.680e-04 \\ \hline
Sum & -4.872e-03 & -5.307e-02 & -9.634e-02 & -9.467e-02 & -9.333e-02 & -9.161e-02 \\ \hline
  $E$ (keV)  &       79.0 &       84.0 &       89.0 &       94.0 &       99.0 &      100.0 \\ \hline
  Band = 1 &  6.749e-11 &  5.790e-10 &  5.685e-06 &  1.164e-10 &  1.546e-09 &  1.036e-06 \\
  2 &  1.038e-06 &  1.349e-09 & -1.713e-04 &  9.929e-07 &  2.325e-08 & -1.302e-05 \\
  3 & -3.018e-04 & -3.109e-04 & -1.774e-03 & -2.584e-04 & -1.676e-04 & -2.174e-04 \\
  4 & -1.340e-03 & -1.918e-03 & -2.234e-03 & -1.372e-03 & -7.117e-04 & -2.452e-04 \\
  5 & -2.669e-03 & -2.706e-03 & -2.105e-03 & -2.688e-03 & -1.395e-03 & -2.570e-04 \\
  6 & -4.676e-03 & -4.399e-03 & -3.255e-03 & -3.486e-03 & -1.888e-03 & -2.532e-04 \\
  7 & -4.945e-03 & -4.637e-03 & -3.032e-03 & -4.517e-03 & -3.497e-03 & -2.096e-04 \\
  8 & -7.574e-03 & -7.647e-03 & -2.322e-03 & -5.745e-03 & -3.966e-03 & -2.682e-04 \\
  9 & -1.032e-02 & -1.176e-02 & -5.254e-03 & -7.415e-03 & -6.583e-03 & -5.691e-04 \\
 10 & -1.752e-02 & -1.491e-02 & -1.047e-02 & -1.363e-02 & -9.250e-03 & -1.364e-03 \\
 11 & -2.315e-02 & -2.093e-02 & -2.643e-02 & -2.249e-02 & -1.250e-02 & -2.622e-03 \\
 12 & -9.230e-03 & -1.039e-02 & -1.876e-02 & -1.377e-02 & -5.724e-03 & -1.489e-03 \\
 13 & -5.083e-03 & -5.291e-03 & -7.526e-03 & -6.121e-03 & -3.340e-03 & -5.545e-04 \\
 14 & -2.955e-03 & -2.444e-03 & -3.180e-03 & -3.150e-03 & -1.210e-03 & -2.382e-04 \\
 15 & -3.236e-04 & -6.165e-04 & -8.134e-04 & -4.970e-04 & -3.587e-04 & -4.039e-05 \\
 16 & -9.802e-05 & -6.387e-05 & -2.619e-04 & -3.526e-04 & -2.654e-04 & -3.411e-05 \\ \hline
Sum & -9.018e-02 & -8.802e-02 & -8.758e-02 & -8.548e-02 & -5.086e-02 & -8.374e-03 \\ \hline
 \multicolumn{7}{|l|}{ Energy Integrated Sensitivity = -8.444024e-01 $\pm$ 1.163467e-04} \\ \hline
 \multicolumn{7}{|l|}{ First-Collided Capture Probability = 7.542643e-04 $\pm$ 3.037407e-08} \\ \hline
\end{tabular}
\end{center}
\label{Table_Verification_MonteCarloFirstCollisionCapture_SensitivityElasticXS}
\end{table}%

\clearpage

\subsection{Once-Collided Transmittance Sensitivity} \label{Sec:Verification_OnceCollidedTransmittance}
The previous two analytic benchmarks involve uncollided monodirectional neutron beams, which does not allow for any testing that the sensitivity methods can adequately handle collision physics and the emergence of secondaries. Unfortunately, analytic solutions to the transport problem involving scattering are few in number.

A simple extension is presented to calculate the sensitivity coefficients for the transmission probability of neutrons that have undergone exactly one collision with respect to the elastic scattering cross sections in the unresolved resonance tables. The same slab geometry in the previous sections with $^{90}$Zr is used, but some modifications are made to the problem to permit an analytic solution that is not terribly unwieldy. First, the incident neutrons are monoenergetic, having an energy $E_{g-1}$ corresponding to the a point on the unresolved resonance table energy grid. 

The most significant change is that the elastic scattering physics is fictitious in this benchmark problem. The outgoing energy is distributed uniformly within the energy range on the grid directly below the incident energy grid point, $E_g \le E \le E_{g-1}$, as opposed to the normal physics of requiring the outgoing energy to be between $\alpha E$ and $E$. Additionally, the scattering is isotropic in the lab frame on the half-range with direct cosine $0 \le \mu \le 1$ and perfectly correlated with the outgoing energy such that $\mu = 1$ corresponds to an outgoing energy at the top of the energy range and $\mu = 0$ corresponds to the bottom of the energy range. This results in the interpolation parameter $\rho$ and the direction cosine $\mu$ being identical, which simplifies the solution considerably. To use this modified physics in a transport code, a fictitious thermal scattering law or $S(\alpha,\beta)$ table could be supplied, which would override the free-gas elastic scattering. (This depends on the code supporting unresolved resonance cross sampling and thermal scattering simultaneously, which may not have been considered by the developers as a practical case worth consideration.) Note that Shuriken does not currently support thermal scattering data, so some chicanery was performed for the purposes of this study by temporarily inserting a line of code that to overwrites the outgoing energy and direction after the normal collision physics.

When neutrons with energy $E_{g-1}$ enter the slab, cross sections are randomly sampled from the unresolved resonance probability table and whether or not the neutron scatters is determined. If the neutron scatters, an outgoing energy and direction are randomly sampled along with a new set of cross sections from the unresolved resonance probability table. This second set of cross sections is used to determine if the particle exits the slab before undergoing another collision. If it does, then it is counted toward the once-collided transmission probability. For clarity, neutrons are not counted if they pass through the slab without experiencing their first collision nor if they undergo a second collision.

The once-collided transmission probability can be expressed with the following double integral for the first-collided source:
\begin{align}
  J_1 = \sum_\ell \sum_m p_{g,\ell} p_{g,m} \int_0^L \int_0^1 \exp\left( -\frac{\Sigma_{t,\ell}(E) (L - x)}{\mu} \right) \Sigma_{s,m}(E_{g-1}) \exp\left( -\Sigma_{t,m}(E_{g-1}) x \right) d\mu dx .
\end{align}
The double summation accounts for two different random cross section samplings from the unresolved resonance probability table. The index $m$ corresponds to the cross section sampled for the incident neutrons, where the total and scattering cross sections are
\begin{subequations}
\begin{align}
  \Sigma_{t,m}(E_{g-1}) &= N \sigma_{t,g-1,m}, \\
  \Sigma_{s,m}(E_{g-1}) &= N \sigma_{s,g-1,m},  
\end{align}
\end{subequations}
and the index $\ell$ corresponds to the probability table band sampled for the collided neutron. Both $m$ and $\ell$ are on the unionized energy grid. Note that for the source neutrons specifically, there is no need for a linear interpolation since they are monoenergetic on a grid point. The term
\begin{align}
  \Sigma_{s,m}(E_{g-1}) \exp\left( -\Sigma_{t,m}(E_{g-1}) x \right) dx \nonumber
\end{align}
gives the probability of scattering within some $dx$ about $x$ between $0$ and $L$. The term
\begin{align}
  \exp\left( -\frac{\Sigma_{t,\ell}(E) (L - x)}{\mu} \right) \nonumber
\end{align}
gives the transmission probability for a particle traveling with direction cosine $\mu$ with respect to the $x$ axis across the remaining distance to the end of the slab. Recall the artificial physics of this benchmark problem causes the interpolation factor $\rho$ and the direction cosine $\mu$ to be identical. This implies that the cross section for the collided neutron energy is
\begin{align}
  \Sigma_{t,\ell}(E) = N [ \sigma_{t,g,\ell} + \mu ( \sigma_{t,g-1,\ell} - \sigma_{t,g,\ell} ) ] .
\end{align}
Note that it may seem that a factor of $\mu$ is missing from this integral, as one would expect from the definition of neutron current. This factor is absent in this case because the first-collision source (as opposed to the angular flux) is isotropic on the forward angular half-space, leading to a factor of $1/\mu$ canceling out the factor.

To proceed, the integral over the cosine can be evaluated to obtain
\begin{align} \label{Eqn:Verification_OnceCollidedTransmittance_Leakage}
  J_1 = \sum_\ell \sum_m p_{g,\ell} p_{g,m} \int_0^L &\exp\left[ N ( \sigma_{t,g-1,\ell} - \sigma_{t,g,\ell} ) (L - x) \right] \text{E}_2\left[ N \sigma_{t,g,\ell} (L - x) \right]  \nonumber \\
  \times &N \sigma_{s,g-1,m} \exp\left( -N \sigma_{t,g-1,m} x \right) dx .
\end{align}
This expression can be recast into the form
\begin{align} \label{Eqn:Verification_OnceCollidedTransmittance_Leakage_Simplified}
  J_1 = \sum_\ell \sum_m p_{g,\ell} p_{g,m} N \sigma_{s,g-1,m} \exp\left( -N \sigma_{t,g-1,m} L \right) f_{g,\ell,m},
\end{align}
where
\begin{align}
  f_{g,\ell,m} =  \int_0^L e^{-(b-a)x} \text{E}_2(ax) dx
\end{align}
with
\begin{subequations}
\begin{align}
  a &= N \sigma_{t,g,\ell}, \\
  b &= N ( \sigma_{t,g-1,\ell} - \sigma_{t,g-1,m} ) .
\end{align}
\end{subequations}
Subscripts on $a$ and $b$ are omitted since they can be trivially applied based on the subscripts of $f_{g,\ell,m}$ and this declutters the subsequent equations. An important point here is that $f_{g,\ell,m}$, unlike for the previous two benchmarks where it is dimensionless, now carries units of length.

Next the integral must be evaluated. Note that for certain values of $\ell$ and $m$ on the unionized grid the upper bound cross sections are the same, meaning the same probability band is sampled for both the incident and scattered neutron, and $b = 0$. It turns out that this leads to a non-trivial special case that needs to be considered. There is another special case when $b = a$ that could arise, but this can only occur in the very peculiar instance when the total cross section specified on the lower grid point for the outgoing neutron precisely equals the difference between the specified cross sections between the outgoing and incident neutrons. This would only happen if somehow the nuclear data were either engineered in this manner or as the result of an incredible coincidence.

The function $f_{g,\ell,m}$ for the two cases of concern can be expressed as
\begin{align}
  f_{g,\ell,m} = \frac{h}{b-a} ,
\end{align}
where
\begin{align}
  h &= \left\{ \begin{array}{l l}
  1 - e^{-bL} + \dfrac{ac}{b-a} , & \quad b \ne 0, \\
  -( 1 - a L ) e^{aL} \text{E}_1(aL) - \ln(aL) - \gamma, & \quad b = 0 . \end{array} \right. 
\end{align}
with
\begin{align}
  c = ( 1 + (b-a)L ) e^{-(b-a)L} \text{E}_1(aL) + \ln\left( \dfrac{a}{|b|} \right) + \text{Ei}(-bL) .
\end{align}
Here an alternative form of the exponential integral
\begin{align}
  \text{Ei}(x) = \int_{-\infty}^x \frac{e^{t}}{t} dt
\end{align}
is used. The reason for this form as opposed to $\text{E}_1(x)$ function is that $\text{Ei}(x)$ naturally handles a negative argument, whereas $\text{E}_1(x)$ would require special cases involving complex numbers. The constant $\gamma$ is the Euler-Mascheroni constant that arises when taking the difference of the $\text{Ei}(-bL) - \ln(bL)$ in the limit as $b \rightarrow 0$. (This can be shown by the power series expansion $\text{Ei}(x) = \gamma + \ln(x) + x^2/4 + \ldots$) For completeness, the special case when $b = a$ is
\begin{align}
  f_{g,\ell,m} = \int_0^L \text{E}_2(ax) dx = \frac{1}{2a} \left[ 1 - 2 \text{E}_3(aL) \right] .
\end{align}

%

The sensitivity coefficients can be computed by taking the appropriate derivatives. Since the energy index $g$ is fixed in this benchmark versus the previous ones presented, it makes sense to take derivatives with respect to the cross sections on the lower edge at $E = E_g$ and at the upper edge at $E = E_{g-1}$. The sensitivity coefficient for the elastic scattering cross section for an unresolved resonance table band $k_g$ for the lower energy edge of the range is
\begin{align} \label{Eqn:Verification_OnceCollidedTransmittance_ScatterXSSensitivity_Lower}
  S_{J_1,\sigma_{s,g,k_g}} = \frac{1}{J_1} \sum_\ell \sum_m  1_{g,k_g,\ell} p_{g,\ell} p_{g,m} N^2 \sigma_{s,g-1,m} \sigma_{s,g,\ell} \exp\left( -N \sigma_{t,g-1,m} L \right) \dho{f_{g,\ell,m}}{a},
\end{align}
and for the upper edge is
\begin{align} \label{Eqn:Verification_OnceCollidedTransmittance_ScatterXSSensitivity_Upper}
  S_{J_1,\sigma_{s,g-1,k_g}} &= \frac{1}{J_1} \sum_\ell \sum_m  p_{g,\ell} p_{g,m} \exp\left( -N \sigma_{t,g-1,m} L \right) \nonumber \\ 
  &\times \bigg[ 1_{g-1,k_{g-1},m} N \sigma_{s,g-1,m} \left( (1 - N \sigma_{s,g-1,m} L )  f_{g,\ell,m} - N \sigma_{s,g-1,m} \dho{f_{g,\ell,m}}{b} \right)  \nonumber \\
  &+ 1_{g-1,k_{g-1},\ell} N^2 \sigma_{s,g-1,m} \sigma_{s,g-1,\ell} \dho{f_{g,\ell,m}}{b} \bigg] .
\end{align}
The function $1_{g,k_g,\ell}$ is an indicator function that is one if the first index on the unionized grid $\ell$ corresponds to the index $k_g$ on the energy grid point $g$ and zero otherwise. The derivatives (omitting the redundant subscripts) are
\begin{subequations}
\begin{align}
  \dho{f}{a} &= \frac{1}{b-a} \frac{dh}{da} + \frac{h}{(b-a)^2}, \\
  \dho{f}{b} &= \frac{1}{b-a} \frac{dh}{db} - \frac{h}{(b-a)^2}. 
\end{align}
\end{subequations}
The derivatives of $h$ are
\begin{subequations}
\begin{align}
  \frac{dh}{da} &= \left\{ \begin{array}{l l}
  \dfrac{bc}{(b-a)^2} + \dfrac{a}{b-a} \dfrac{dc}{da} , & \quad b \ne 0,  \\  
  -L ( 1 - a L e^{aL} \text{E}_1( aL ) ), & \quad b = 0, \\ \end{array} \right. \\
  \frac{dh}{db} &= L e^{-bL} - \frac{ac}{(b-a)^2} + \frac{a}{b-a} \frac{dc}{db} .  
\end{align}
\end{subequations}
And the derivatives of $c$ are
\begin{subequations}
\begin{align}
  \frac{dc}{da} &= L^2 ( b - a ) e^{-(b-a) L } \text{E}_1(aL) + \frac{1}{a} \left( 1 - ( 1 + (b-a) L ) e^{-bL} \right) , \\
  \frac{dc}{db} &= L^2 ( a - b ) e^{-(b-a) L } \text{E}_1(aL) - \frac{1}{b} \left( 1 - e^{-bL} \right) .
\end{align}
\end{subequations}
Note that in Eq.~\eqref{Eqn:Verification_OnceCollidedTransmittance_ScatterXSSensitivity_Upper}, the $b = 0$ case is handled implicitly because both indicator functions evaluate to one and the terms with derivatives of $f$ with respect to $b$ cancel.

%

The sensitivity coefficient for the elastic scattering cross section at a particular energy grid point may be obtained by adding the sensitivity coefficients for each band $k_g$ at that corresponding point.

Reference numerical values are provided for the energy range where $E_g = 54$~keV and $E_{g-1} = 59$~keV.  The numerical evaluations of the analytical expressions are provided in Table~\ref{Table_Verification_OnceCollidedTransmittance_Sensitivity} for various slab thicknesses. The reference leakage or transmittance of only once-collided particles is computed from Eq.~\eqref{Eqn:Verification_OnceCollidedTransmittance_Leakage} following analytical integration as described previously. Next, the corresponding sensitivity coefficients to the lower and upper edge elastic scattering cross sections are given for each band on the probability tables. These are computed using Eqs.~\eqref{Eqn:Verification_OnceCollidedTransmittance_ScatterXSSensitivity_Lower} and \eqref{Eqn:Verification_OnceCollidedTransmittance_ScatterXSSensitivity_Upper} respectively. Finally, the sum of the sensitivity coefficients over all the probability table bands for each energy grid point is reported.

The range of slab thicknesses was selected to demonstrate the trends in the once-collided leakage. For an optically thin slab, increasing the cross sections on the upper edge (impacting only the incident neutrons) increases the rate that once-collided particles leak out of the slab because it enhances the rate that the first scattering event occurs. Recall that uncollided particles that leak out are not counted in this specific quantity of interest. For optically thick slabs, the sensitivity coefficients on the upper edge are entirely negative because increases in the scattering cross section reduce the ability of neutrons to leak out after their first collision more than the effect of increasing the production of those once-collided neutrons. The breakeven point occurs at around a thickness of $L = 3.3$~cm where the once-collided leakage is near its maximum value and the sum of the sensitivity coefficients on the upper energy grid point is small in magnitude. The sensitivity coefficients of the scattering cross section on the lower range are always negative and become increasingly so as the thickness increases. This is because the lower edge cross section does not participate at all in the scattering of the uncollided neutrons (the production of once-collided neutrons), but does hamper the ability of the once-collided neutrons from leaking out of the slab.

The process is simulated in Shuriken and the sensitivity coefficients are estimated using differential operator sampling using $10^9$ samples (in a different calculation than in the previous two sections because of the modified collision physics). The results for three different slab thicknesses are reported in Table~\ref{Table_Verification_OnceCollidedTransmittance_MonteCarlo}. Since the size data set is more manageable in this case, only being for two energy grid points, statistical uncertainties are provided for all computed values.

The once-collided leakage results agree within 2$\sigma$ of the reference analytic solutions in all cases. The sum of the sensitivity coefficients across the probability band show similar agreement, although the $L = 3.3$~cm case is right on the edge of 2-$\sigma$ error. This value is actually difficult to estimate precisely because it involves the sum of positive and negative contributions of almost equal magnitude. The individual sensitivity coefficients for the different probability bands show general agreement, except for some of the first few bands, which have both a very small cross section and low probability of being sampled, leading to a very small reference sensitivity coefficient.

\clearpage

\begin{table}[t!] \small
\caption{Benchmark First-Collided Leakage and Sensitivity Coefficients to Elastic Scattering Cross Sections}
\begin{center}
\begin{tabular}{|l|c|c|c|c|c|c|} \hline
 			& \multicolumn{2}{c|}{$L = 0.5$~cm}	& \multicolumn{2}{c|}{$L = 1$~cm}		& \multicolumn{2}{c|}{$L = 3.3$~cm}			\\ \hline
Leakage		& \multicolumn{2}{c|}{0.12400}		& \multicolumn{2}{c|}{0.17971}      	& \multicolumn{2}{c|}{0.20648}				\\ \hline
Sensitivity & 54~keV		& 59~keV		& 54~keV		& 59~keV			& 54~keV		& 59~keV			\\ \hline
Band $= 1$	& -2.359e-10	&  2.537e-11	& -5.449e-10	& 2.554e-11			& -3.237e-09	& 3.460e-12			\\ 
 2			& -5.443e-07	&  1.628e-10	& -1.227e-06	& 1.639e-10			& -6.820e-06 	& 2.241e-11			\\
 3			& -6.227e-04	&  3.214e-04	& -1.202e-03	& 3.251e-04			& -4.436e-03	& 1.172e-04			\\ 
 4			& -2.763e-03	&  4.206e-03	& -4.794e-03	& 4.163e-03			& -1.290e-02	& 1.642e-03			\\ 
 5			& -6.803e-03 	&  1.699e-02	& -1.106e-02	& 1.622e-02			& -2.437e-02	& 5.081e-03			\\ 
 6			& -1.212e-02 	&  3.800e-02	& -1.903e-02	& 3.528e-02			& -3.761e-02	& 8.104e-03			\\ 
 7			& -1.338e-02 	&  8.194e-02	& -2.064e-02 	& 7.472e-02			& -3.864e-02	& 1.325e-02			\\ 
 8			& -9.284e-03 	&  1.327e-01	& -1.419e-02	& 1.193e-01			& -2.585e-02	& 1.638e-02			\\ 
 9			& -1.758e-02 	&  1.100e-01	& -2.657e-02	& 9.738e-02			& -4.672e-02 	& 9.165e-03			\\ 
10			& -1.800e-02 	&  9.429e-02	& -2.658e-02	& 8.141e-02			& -4.357e-02	& 2.365e-03			\\ 
11			& -1.788e-02 	&  1.043e-01	& -2.491e-02	& 8.385e-02			& -3.423e-02	& -1.177e-02		\\ 
12			& -1.021e-02  	&  9.000e-02	& -1.208e-02	& 5.578e-02			& -1.074e-02	& -2.530e-02		\\ 
13			& -5.419e-03  	&  5.493e-02	& -4.924e-03	& 1.753e-02			& -2.638e-03	& -1.195e-02		\\ 
14			& -1.614e-03  	&  2.184e-02	& -1.158e-03	& 1.421e-03			& -4.713e-04	& -3.105e-03		\\ 
15			& -4.000e-04  	&  3.865e-03	& -2.483e-04	& -7.802e-04		& -9.582e-05	& -4.815e-04		\\ 
16			& -1.151e-04  	&  1.276e-03	& -6.714e-05 	& -4.053e-04		& -2.565e-05	& -1.714e-04		\\ \hline
Total		& -1.162e-01  	&  7.547e-01	& -1.674e-01	& 5.862e-01			& -2.823e-01	& 3.329e-03			\\ \hline
			& \multicolumn{2}{c|}{$L = 5$~cm}	& \multicolumn{2}{c|}{$L = 10$~cm}	& \multicolumn{2}{c|}{$L = 20$~cm}	\\ \hline
Leakage		& \multicolumn{2}{c|}{0.16911}		& \multicolumn{2}{c|}{0.07521}		& \multicolumn{2}{c|}{0.01748}		\\ \hline			
Sensitivity & 54~keV		& 59~keV		& 54~keV		& 59~keV			& 54~keV		& 59~keV				\\ \hline
Band $= 1$	& -7.218e-09 	& -4.685e-11	& -3.997e-08	& -5.876e-10		& -3.686e-07	& -7.439e-09				\\ 
 2			& -1.466e-05 	& -2.999e-10	& -7.394e-05	& -3.761e-09		& -5.782e-04	& -4.754e-08				\\
 3			& -7.714e-03 	& -3.292e-04	& -2.391e-02	& -4.529e-03		& -9.118e-02	& -4.593e-02				\\ 
 4			& -1.876e-02 	& -2.560e-03	& -3.712e-02	& -3.076e-02		& -6.465e-02	& -1.760e-01				\\ 
 5			& -3.156e-02 	& -8.815e-03	& -4.599e-02	& -7.262e-02		& -4.425e-02	& -2.208e-01				\\ 
 6			& -4.568e-02 	& -2.031e-02	& -5.645e-02	& -1.216e-01		& -4.073e-02	& -2.426e-01				\\ 
 7			& -4.552e-02 	& -4.478e-02	& -5.222e-02	& -2.212e-01		& -3.376e-02	& -3.437e-01				\\ 
 8			& -2.997e-02 	& -7.384e-02	& -3.305e-02	& -3.173e-01		& -2.018e-02	& -4.121e-01				\\ 
 9			& -5.312e-02 	& -6.230e-02	& -5.604e-02	& -2.312e-01		& -3.250e-02	& -2.495e-01				\\ 
10			& -4.764e-02 	& -5.421e-02	& -4.604e-02	& -1.616e-01		& -2.432e-02	& -1.333e-01				\\ 
11			& -3.415e-02 	& -6.115e-02	& -2.758e-02	& -1.091e-01		& -1.292e-02	& -5.355e-02				\\ 
12			& -9.011e-03 	& -3.591e-02	& -5.833e-03	& -2.292e-02		& -2.676e-03	& -8.128e-03				\\ 
13			& -1.978e-03 	& -8.345e-03	& -1.193e-03	& -3.643e-03		& -5.788e-04	& -1.525e-03				\\ 
14			& -3.490e-04 	& -1.878e-03	& -2.133e-04	& -8.567e-04		& -1.067e-04	& -3.720e-04				\\ 
15			& -7.131e-05 	& -2.952e-04	& -4.384e-05	& -1.390e-04		& -2.212e-05	& -6.148e-05				\\ 
16			& -1.912e-05 	& -1.059e-04	& -1.178e-05	& -5.024e-05		& -5.968e-06	& -2.250e-05				\\ \hline
Total		& -3.256e-01 	& -3.748e-01	& -3.858e-01	& -1.297e+00		& -3.684e-01	& -1.888e+00				\\ \hline
\end{tabular}
\end{center}
\label{Table_Verification_OnceCollidedTransmittance_Sensitivity}
\end{table}%

\clearpage

\begin{table}[t!] \small
\caption{Shuriken Monte Carlo Results of First-Collided Leakage and Sensitivity Coefficients to Elastic Scattering Cross Sections}
\begin{center}
\begin{tabular}{|l|c|c|c|} \hline
 			& $L = 0.5$~cm	& $L = 3.3$~cm		& $L = 10$~cm	 \\ \hline
 Leakage		
 			& 1.2401e-01 $\pm$ 1.0423e-05		& 2.0649e-01 $\pm$ 1.2800e-05      	& 7.5198e-02 $\pm$ 8.3392e-06 \\ \hline
 Sensitivity      
 			& 54~keV					  & 54~keV  & 54~keV	\\ \hline
Band $= 1$	& -2.509e-10 $\pm$ 1.552e-11  & -3.188e-09 $\pm$ 4.378e-11    	& -4.008e-08 $\pm$ 3.290e-10			\\ 
 2			& -5.347e-07 $\pm$ 9.535e-09  & -6.796e-06 $\pm$ 3.856e-08 		& -7.366e-05 $\pm$ 2.293e-07			\\
 3			& -6.240e-04 $\pm$ 1.232e-06  & -4.436e-03 $\pm$ 3.324e-06  	& -2.392e-02 $\pm$ 1.531e-05			\\ 
 4			& -2.763e-03 $\pm$ 3.004e-06  & -1.290e-02 $\pm$ 6.647e-06 		& -3.708e-02 $\pm$ 2.208e-05  		\\ 
 5			& -6.807e-03 $\pm$ 5.038e-06  & -2.436e-02 $\pm$ 9.698e-06  	& -4.597e-02 $\pm$ 2.571e-05			\\ 
 6			& -1.214e-02 $\pm$ 6.920e-06  & -3.760e-02 $\pm$ 1.233e-05  	& -5.647e-02 $\pm$ 2.890e-05			\\ 
 7			& -1.337e-02 $\pm$ 7.355e-06  & -3.864e-02 $\pm$ 1.265e-05  	& -5.221e-02 $\pm$ 2.790e-05			\\ 
 8			& -9.281e-03 $\pm$ 6.176e-06  & -2.585e-02 $\pm$ 1.041e-05  	& -3.303e-02 $\pm$ 2.223e-05			\\ 
 9			& -1.759e-02 $\pm$ 8.556e-06  & -4.671e-02 $\pm$ 1.406e-05  	& -5.601e-02 $\pm$ 2.900e-05			\\ 
10			& -1.799e-02 $\pm$ 8.779e-06  & -4.358e-02 $\pm$ 1.374e-05  	& -4.604e-02 $\pm$ 2.635e-05			\\ 
11			& -1.787e-02 $\pm$ 9.050e-06  & -3.423e-02 $\pm$ 1.246e-05  	& -2.757e-02 $\pm$ 2.038e-05 			\\ 
12			& -1.021e-02 $\pm$ 7.403e-06  & -1.075e-02 $\pm$ 7.294e-06  	& -5.826e-03 $\pm$ 9.303e-06			\\ 
13			& -5.414e-03 $\pm$ 5.956e-06  & -2.641e-03 $\pm$ 3.720e-06  	& -1.195e-03 $\pm$ 4.159e-06			\\ 
14			& -1.616e-03 $\pm$ 3.430e-06  & -4.737e-04 $\pm$ 1.543e-06  	& -2.128e-04 $\pm$ 1.704e-06			\\ 
15			& -3.992e-04 $\pm$ 1.769e-06  & -9.566e-05 $\pm$ 6.950e-07  	& -4.616e-05 $\pm$ 8.196e-07			\\ 
16			& -1.163e-04 $\pm$ 9.721e-07  & -2.570e-05 $\pm$ 3.592e-07  	& -1.243e-05 $\pm$ 4.114e-07			\\ \hline
Total		& -1.162e-01 $\pm$ 2.241e-05  & -2.823e-01 $\pm$ 3.334e-05  	& -3.857e-01 $\pm$ 7.360e-05			\\ \hline
 Sensitivity      
 			& 59~keV					  & 59~keV	& 59~keV	\\ \hline
Band $= 1$	& -3.213e-12 $\pm$ 1.967e-14  & -5.521e-11 $\pm$ 1.683e-13 	& -8.084e-10 $\pm$ 1.939e-12			\\ 
 2			& -2.043e-11 $\pm$ 2.043e-11  &  4.488e-09 $\pm$ 4.843e-09	& -5.192e-09 $\pm$ 5.192e-09			\\
 3			&  3.206e-04 $\pm$ 1.701e-06  &  1.177e-04 $\pm$ 1.848e-06	& -4.531e-03 $\pm$ 6.074e-06			\\ 
 4			&  4.209e-03 $\pm$ 6.105e-06  &  1.639e-03 $\pm$ 5.956e-06	& -3.078e-02 $\pm$ 1.936e-05			\\ 
 5			&  1.699e-02 $\pm$ 1.212e-05  &  5.080e-03 $\pm$ 1.024e-05	& -7.259e-02 $\pm$ 3.354e-05			\\ 
 6			&  3.799e-02 $\pm$ 1.795e-05  &  8.116e-03 $\pm$ 1.368e-05	& -1.215e-01 $\pm$ 4.733e-05			\\ 
 7			&  8.190e-02 $\pm$ 2.611e-05  &  1.326e-02 $\pm$ 1.835e-05	& -2.213e-01 $\pm$ 6.757e-05			\\ 
 8			&  1.327e-01 $\pm$ 3.292e-05  &  1.640e-02 $\pm$ 2.162e-05	& -3.172e-01 $\pm$ 8.474e-05			\\ 
 9			&  1.100e-01 $\pm$ 2.996e-05  &  9.157e-03 $\pm$ 1.956e-05	& -2.313e-01 $\pm$ 7.570e-05			\\ 
10			&  9.428e-02 $\pm$ 2.762e-05  &  2.371e-03 $\pm$ 1.768e-05	& -1.616e-01 $\pm$ 6.733e-05			\\ 
11			&  1.043e-01 $\pm$ 2.861e-05  & -1.176e-02 $\pm$ 1.794e-05 	& -1.091e-01 $\pm$ 6.281e-05			\\ 
12			&  8.999e-02 $\pm$ 2.547e-05  & -2.529e-02 $\pm$ 1.868e-05 	& -2.291e-02 $\pm$ 3.494e-05			\\ 
13			&  5.493e-02 $\pm$ 1.894e-05  & -1.195e-02 $\pm$ 1.574e-05 	& -3.654e-03 $\pm$ 1.664e-05			\\ 
14			&  2.185e-02 $\pm$ 1.261e-05  & -3.095e-03 $\pm$ 9.619e-06 	& -8.473e-04 $\pm$ 9.640e-06			\\ 
15			&  3.860e-03 $\pm$ 7.176e-06  & -4.837e-04 $\pm$ 4.466e-06 	& -1.374e-04 $\pm$ 4.671e-06			\\ 
16			&  1.280e-03 $\pm$ 4.947e-06  & -1.719e-04 $\pm$ 2.808e-06 	& -5.502e-05 $\pm$ 2.997e-06			\\ \hline
Total		&  7.547e-01 $\pm$ 6.658e-05  &  3.395e-03 $\pm$ 3.274e-05 	& -1.297e+00 $\pm$ 1.746e-04			\\ \hline
\end{tabular}
\end{center}
\label{Table_Verification_OnceCollidedTransmittance_MonteCarlo}
\end{table}%

\clearpage

\subsection{Discussion} \label{Sec:Verification_Discussion}
To summarize, three analytic benchmark solutions for cross section sensitivity coefficients using real continuous-energy unresolved resonance data were derived. Monte Carlo calculations were performed and the results show general agreement within the accepted 2-$\sigma$ confidence band. The benchmark solutions were obtained by solving specific transport problems for a quantity of interest and then sensitivity coefficients were computed by taking the appropriate derivatives. On the other hand, the Monte Carlo estimators for the sensitivity coefficients using differential operator sampling were derived for a general random process. While these two are obviously related, the approaches for obtaining the end results are quite different, so it should hopefully provide some confidence that the equations for the estimators that were derived in Sec.~\ref{Sec:Methodology} are correct and consistent with the transport process. Additionally, the agreement should provide confidence that the specific implementation used to generate the results in this paper is also correct.

Admittedly, these analytic benchmarks only cover a subset of the cases that this paper addresses. To review, the processed $^{90}$Zr unresolved resonance data from ENDF/B-VII.1 represent their cross sections as values as opposed to multiplicative factors and linear-linear interpolation is used. Stated another way, the results for multiplicative factors nor those for the seldom-used logarithmic interpolation are tested by these analytical benchmark solutions. Additionally, the solutions were obtained assuming that the cross section values (or factors) are not interpolated across probability bands with the only interpolation being in energy. This may limit their applicability to some codes depending on the implementation. Future work could attempt to extend these solutions to bilinear interpolation across both energy and probability band.

Constructing analytical solutions using multiplicative factors is possible, but the results would be more complicated by the fact that the energy grids for the unresolved resonance data and the mean-value cross sections that they are applied to are different and do not usually align. Furthermore, the implementation of the estimators for the case of provided multiplicative factors is identical to the case of provided values with the difference being in the routines for the computation of the actual cross sections. In other words, only the final cross sections are returned for or cached within the routines sampling the unresolved resonance cross sections, whereas the multiplicative factors are only used in the intermediate processing steps. Additionally, testing done for eigenvalue problems with $^{235/238}$U, which uses multiplicative factors, shows favorable agreement. So while there is certainly value in obtaining an analytic benchmark solution for the case of multiplicative factors, the fact they are so similar suggests that the results should be correct regardless.

Unfortunately, these analytic benchmarks are only for fixed-source problems and not eigenvalue or criticality problems. As such, the derived equations specific to eigenvalue problems are not covered. Additional testing that is not given in this paper was performed to test these cases. In particular, the sensitivity coefficient to the total cross sections were tested using direct perturbations on the constituent atomic density with the central difference method. Additionally, the computation of sensitivity coefficients for $k$ with respect to the individual reaction cross sections was tested by comparing the energy-integrated values to the equivalent quantities obtained from the adjoint-based method in MCNP6.2. (Unfortunately, the currently available version of MCNP does not support sensitivity coefficients of reaction rate ratios.) The results show agreement in all the cases that were tested, so this provides confidence that the equations specific to eigenvalue problems have been correctly derived and the corresponding estimators have been correctly implemented.

\section{Numerical Results} \label{Sec:Results}
Given that the method shows agreement with analytical benchmarks, the next step is to demonstrate it on more realistic calculations. This section presents two systems: the Big Ten benchmark (Sec.~\ref{Sec:Results_BigTen}) and a simplified Molten Chloride Fast Reactor model (Sec.~\ref{Sec:Results_MCFR}). Both calculations use cross sections from the ENDF/B-VII.1 nuclear data library. 

\subsection{Big Ten Critical Experiment} \label{Sec:Results_BigTen}
Big Ten was a cylindrical critical assembly at the Los Alamos Critical Experiments Facility that was used for experiments in the 1970s and the early 1990s. The primary purpose of the experiment was to assess the quality of uranium cross sections in the fast spectrum range that was representative of the liquid metal cooled reactor designs of the time. The assembly had about 10 metric tons of metallic uranium having a core with an average enrichment of 10\%. The core consisted of various metallic uranium plates of different enrichments and was surrounded by a depleted uranium reflector. The Big Ten experiment is documented in the International Criticality Safety Benchmark Evaluation Project (ICSBEP) Handbook with the identifier IEU-MET-FAST-007. 

Big Ten was selected as a test case for this paper because its geometric created a neutron spectrum such that a significant portion is in the unresolved resonance range of $^{238}$U and less so for $^{235}$U. The simplified benchmark model consisting of a homogenized central core region surrounded by the depleted uranium reflector. The homogenized cylindrical core has a height of 57.6338~cm and a radius of 26.67~cm. The reflector surrounding the core is described as another cylinder with a height of 96.5200~cm and a radius of 41.91~cm. The atomic densities are given in Table~\ref{Table_Results_BigTen_Isotopics}.

The two dominant reaction cross sections are fission in $^{235}$U and capture in $^{238}$U, and therefore the sensitivity coefficients are computed with respect to those values. Three responses are considered: the effective multiplication factor $k$, the ratio of leakage to fission production (i.e. the leakage probability), and the ratio of the volume-integrated $^{238}$U capture to $^{235}$U fission within the core region (excluding the reflector). 

These sensitivity coefficients were calculated by Shuriken. The calculation used $10^5$ source neutrons (on average) per batch, 20 inactive batches, and 1000 active batches. The responses and sensitivity coefficients of $k$ were compared to an analogous calculation with MCNP6.2 to assess the level of agreement. The eigenvalue sensitivity coefficients are computing using adjoint-weighting with the iterated fission probability method using a block size of 5 generations. As mentioned, MCNP6.2 cannot compute reaction rate sensitivity coefficients in eigenvalue problems. 

The results are displayed in Table~\ref{Table_Results_BigTen_EnergyIntegratedResults}, where the top value is the one obtained from the Shuriken and the bottom is the one from MCNP6.2. Note that the sensitivity coefficients in this table are over the unresolved energy range of each respective nuclide. Based on the calculated magnitudes of the sensitivity coefficients, the unresolved resonance data of $^{238}$U(n,$\gamma$) is generally more important. This is perhaps not too surprising given that the unresolved resonance range of $^{238}$U occurs at a higher energy range than $^{235}$U and that the neutron spectrum in the Big Ten assembly is quite fast.

The calculated integral values all agree within the 2-$\sigma$ confidence band. The eigenvalue sensitivity coefficients for $^{235}$U(n,f) agree to about 1$\sigma$. On the other hand, the agreement for the eigenvalue sensitivity coefficients for $^{238}$U(n,$\gamma$) are just outside 2$\sigma$ where the reported statistical uncertainties in the calculated and MCNP6.2 results are just under 1\% and around 0.15\% respectively. Ideally the difference would be somewhat smaller, but it is not too far outside of reasonable confidence bounds considering the level of agreement in the other results given in this paper and those from other testing not presented herein. Also, the uncertainty estimates in eigenvalue correlations are reported lower than they actually are because autocorrelation between batches is neglected. Moreover, even excluding the possibility that minor software errors exist in Shuriken, some differences are to be expected when performing code-to-code comparisons. Even though nominally both Shuriken and MCNP6.2 are doing very similar operations, there are some small differences in, for example, how the continuous-energy nuclear data is handled that can lead to minor differences in results that only become apparent when solutions are tightly resolved.

Detailed numerical results for the sensitivity coefficients with respect to all of the $^{238}$U(n,$\gamma$) unresolved resonance data for the effective multiplication factor $k$ and the $^{238}$U capture to $^{235}$U core fission ratios are provided in Tables~\ref{Table_Results_BigTen_Sensitivity_k_238UCapture} and~\ref{Table_Results_BigTen_Sensitivity_238Capture235FissionRatio_238UCapture} respectively. (The sensitivity coefficients for the leakage to neutron production ratio do not show anything significantly different in shape, so they are not reported.)


\begin{table}[h!] \small
\caption{Atomic Densities (b$^{-1}$cm$^{-1}$) of the Simplified Big Ten Model}
\begin{center}
\begin{tabular}{|l|c|c|} \hline
			& Homogenized Core			& Reflector					\\ \hline
$^{234}$U	& $4.8416 \times 10^{-5}$	& $2.8672 \times 10^{-7}$	\\
$^{235}$U	& $4.8151 \times 10^{-3}$	& $1.0058 \times 10^{-4}$	\\
$^{236}$U	& $1.7407 \times 10^{-5}$	& $1.1468 \times 10^{-6}$	\\
$^{238}$U	& $4.3181 \times 10^{-2}$	& $4.7677 \times 10^{-2}$	\\ \hline
\end{tabular}
\end{center}
\label{Table_Results_BigTen_Isotopics}
\end{table}%

\begin{table}[h!] \small
\caption{Energy-Integrated Response and Unresolved Resonance Range Sensitivity Coefficients of the Simplified Big Ten Benchmark Model (Top: Shuriken, Bottom: MCNP6.2)}
\begin{center}
\begin{tabular}{|l|c|c|c|} \hline
Response			& Value						& Sensitivity $^{235}$U(n,f)	& Sensitivity $^{238}$U(n,$\gamma$)		\\ \hline
\multirow{2}{*}{$k$}& 0.99475 $\pm$ 0.00009		&  2.0092e-02 $\pm$ 4.0312e-04	& -1.0758e-01 $\pm$ 9.8814e-04			\\
					& 0.99470 $\pm$ 0.00005		&  2.0499e-02 $\pm$ 1.0659e-04	& -1.0552e-01 $\pm$ 1.4773e-04			\\ \hline
\multirow{2}{*}{Leakage}
					& 0.10881 $\pm$ 0.00003		& -2.0142e-02 $\pm$ 9.9846e-04 	& -1.6554e-01 $\pm$ 2.4480e-03			\\
					& 0.10885 $\pm$	0.00002		&  ---							& ---									\\ \hline
\multirow{2}{*}{$^{238}$U(n,$\gamma$) / $^{235}$U(n,f)}
					& 0.97209 $\pm$ 0.00017		& -4.9678e-02 $\pm$ 6.7460e-04	&  3.7862e-01 $\pm$ 1.6498e-03			\\
					& 0.97190 $\pm$ 0.00014		&  ---							& ---									\\ \hline
\end{tabular}
\end{center}
\label{Table_Results_BigTen_EnergyIntegratedResults}
\end{table}

\begin{table}[h!] \tiny
\caption{Computed Unresolved Resonance Sensitivity Coefficients for $k$ With Respect to $^{238}$U(n,$\gamma$) of the Simplified Big Ten Benchmark Model}
\begin{center}
\begin{tabular}{|l|c|c|c|c|c|c|c|c|c|} \hline
  $E$ (keV)  &  20		& 23		& 26		& 30		& 35		& 40		& 45		& 45.09		& 50			\\ \hline
Band = 1	& -6.358e-07 & -5.690e-06 & -6.994e-06 & -2.360e-06 & -1.760e-05 & -1.429e-05 & -7.829e-06 & -2.061e-06 & -1.434e-05	\\
2	& -1.717e-06 & -1.413e-05 & -2.021e-05 & -2.523e-05 & -4.489e-05 & -1.945e-05 & -6.461e-06 & -6.541e-06 & -4.012e-05	\\
3	& -2.254e-05 & -5.428e-05 & -6.324e-05 & -1.249e-04 & -7.386e-05 & -8.967e-05 & -4.973e-05 & -3.620e-05 & -8.757e-05	\\
4	& -4.148e-05 & -8.667e-05 & -1.264e-04 & -1.728e-04 & -1.705e-04 & -2.682e-04 & -1.102e-04 & -6.879e-05 & -1.653e-04	\\
5	& -6.630e-05 & -1.728e-04 & -2.049e-04 & -2.688e-04 & -2.652e-04 & -3.648e-04 & -1.830e-04 & -1.753e-04 & -3.776e-04	\\
6	& -1.310e-04 & -2.740e-04 & -3.414e-04 & -5.742e-04 & -6.963e-04 & -6.527e-04 & -3.524e-04 & -3.412e-04 & -5.595e-04	\\
7	& -1.384e-04 & -2.983e-04 & -3.840e-04 & -5.914e-04 & -6.940e-04 & -8.045e-04 & -4.312e-04 & -3.671e-04 & -6.756e-04	\\
8	& -1.920e-04 & -3.770e-04 & -4.908e-04 & -6.723e-04 & -6.984e-04 & -8.825e-04 & -4.111e-04 & -4.013e-04 & -6.945e-04	\\
9	& -2.776e-04 & -5.317e-04 & -5.935e-04 & -7.985e-04 & -9.402e-04 & -9.256e-04 & -4.449e-04 & -3.650e-04 & -6.687e-04	\\
10	& -2.857e-04 & -5.710e-04 & -6.357e-04 & -8.509e-04 & -1.002e-03 & -1.036e-03 & -4.846e-04 & -4.126e-04 & -9.290e-04	\\
11	& -3.063e-04 & -5.783e-04 & -6.450e-04 & -7.930e-04 & -1.014e-03 & -8.408e-04 & -4.595e-04 & -4.552e-04 & -9.092e-04	\\
12	& -1.486e-04 & -2.135e-04 & -2.814e-04 & -3.906e-04 & -4.581e-04 & -4.130e-04 & -2.143e-04 & -2.484e-04 & -3.819e-04	\\
13	& -5.888e-05 & -1.080e-04 & -1.152e-04 & -1.741e-04 & -1.911e-04 & -1.776e-04 & -1.150e-04 & -7.922e-05 & -1.642e-04	\\
14	& -8.864e-06 & -3.660e-05 & -3.911e-05 & -7.780e-05 & -7.620e-05 & -4.357e-05 & -3.899e-05 & -1.487e-05 & -6.314e-05	\\
15	& -1.679e-06 & -6.348e-06 & -6.755e-06 & -9.447e-06 & -1.399e-05 & -7.456e-06 & -9.310e-06 & -4.692e-06 & -1.215e-05	\\
16	& -1.244e-06 & -3.463e-06 & -5.797e-06 & -6.983e-06 & -1.612e-05 & -5.827e-06 & -5.974e-06 & -6.684e-06 & -1.220e-05	\\ \hline
Sum	& -1.683e-03 & -3.332e-03 & -3.961e-03 & -5.533e-03 & -6.373e-03 & -6.546e-03 & -3.325e-03 & -2.985e-03 & -5.755e-03	\\ \hline
  $E$ (keV)  &  55		& 60		& 70		& 80		& 90		& 100		& 120		& 140		& 149			\\ \hline
Band = 1	& -5.494e-07 & -6.773e-06 & -4.129e-06 & -1.845e-05 & -2.810e-05 & -2.879e-05 & -5.872e-07 & -2.158e-05 &  7.323e-08	\\
2	& -1.467e-05 & -2.336e-05 & -3.318e-05 & -8.174e-05 & -3.736e-05 & -4.910e-05 & -4.984e-05 & -1.456e-05 & -8.836e-06	\\
3	& -5.733e-05 & -1.058e-04 & -8.521e-05 & -1.551e-04 & -1.642e-04 & -1.525e-04 & -1.849e-04 & -8.006e-05 & -2.339e-05	\\
4	& -1.362e-04 & -2.081e-04 & -2.679e-04 & -2.638e-04 & -2.622e-04 & -4.141e-04 & -5.054e-04 & -3.106e-04 & -7.884e-05	\\
5	& -2.759e-04 & -3.968e-04 & -4.353e-04 & -4.466e-04 & -4.412e-04 & -6.778e-04 & -8.369e-04 & -4.765e-04 & -1.153e-04	\\
6	& -6.019e-04 & -7.580e-04 & -9.821e-04 & -8.125e-04 & -8.001e-04 & -1.085e-03 & -1.283e-03 & -8.296e-04 & -3.302e-04	\\
7	& -6.930e-04 & -8.236e-04 & -1.135e-03 & -8.107e-04 & -8.325e-04 & -1.139e-03 & -1.417e-03 & -9.099e-04 & -3.422e-04	\\
8	& -7.337e-04 & -9.836e-04 & -1.338e-03 & -9.185e-04 & -8.337e-04 & -1.076e-03 & -1.533e-03 & -1.041e-03 & -2.651e-04	\\
9	& -8.320e-04 & -1.005e-03 & -1.404e-03 & -9.869e-04 & -9.514e-04 & -1.305e-03 & -1.519e-03 & -1.014e-03 & -2.728e-04	\\
10	& -7.304e-04 & -1.037e-03 & -1.298e-03 & -1.370e-03 & -8.850e-04 & -1.123e-03 & -1.264e-03 & -9.388e-04 & -2.578e-04	\\
11	& -8.349e-04 & -1.037e-03 & -1.209e-03 & -1.241e-03 & -9.586e-04 & -1.092e-03 & -1.509e-03 & -1.007e-03 & -2.661e-04	\\
12	& -3.552e-04 & -5.111e-04 & -6.467e-04 & -5.118e-04 & -4.914e-04 & -6.221e-04 & -8.069e-04 & -4.939e-04 & -1.566e-04	\\
13	& -1.170e-04 & -2.881e-04 & -2.405e-04 & -3.576e-04 & -2.527e-04 & -3.833e-04 & -3.926e-04 & -2.953e-04 & -5.504e-05	\\
14	& -4.667e-05 & -1.034e-04 & -5.339e-05 & -1.605e-04 & -1.583e-04 & -1.663e-04 & -1.751e-04 & -1.059e-04 & -2.650e-05	\\
15	& -1.100e-05 & -2.223e-05 & -2.479e-05 & -2.985e-05 & -3.401e-05 & -1.143e-04 & -4.920e-05 & -2.041e-05 & -1.177e-06	\\
16	& -1.411e-06 & -2.532e-05 & -2.753e-05 & -3.655e-05 & -2.707e-05 & -2.372e-05 & -2.808e-05 & -3.249e-06 & -5.829e-07	\\ \hline
Sum	& -5.442e-03 & -7.335e-03 & -9.184e-03 & -8.201e-03 & -7.158e-03 & -9.452e-03 & -1.155e-02 & -7.562e-03 & -2.200e-03	\\ \hline
\end{tabular}
\end{center}
\label{Table_Results_BigTen_Sensitivity_k_238UCapture}
\end{table}%

\begin{table}[t!] \tiny
\caption{Computed Unresolved Resonance Sensitivity Coefficients for the $^{238}$U(n,$\gamma$) to $^{235}$U(n,f) Ratio With Respect to $^{238}$U(n,$\gamma$) of the Simplified Big Ten Benchmark Model}
\begin{center}
\begin{tabular}{|l|c|c|c|c|c|c|c|c|c|} \hline
  $E$ (keV)  &  20		& 23		& 26		& 30		& 35		& 40		& 45		& 45.09		& 50			\\ \hline
Band $ = 1$   & 5.486e-06 & 2.159e-05 & 2.878e-05 & 1.063e-05 & 6.948e-05 & 5.308e-05 & 2.282e-05 & 9.190e-06 & 4.778e-05	\\
2   & 1.303e-05 & 5.419e-05 & 6.788e-05 & 7.693e-05 & 1.411e-04 & 6.423e-05 & 3.012e-05 & 4.068e-05 & 1.188e-04	\\
3   & 6.774e-05 & 1.980e-04 & 2.292e-04 & 4.201e-04 & 3.027e-04 & 2.858e-04 & 1.659e-04 & 1.352e-04 & 3.004e-04	\\
4   & 1.547e-04 & 3.096e-04 & 4.519e-04 & 5.400e-04 & 6.590e-04 & 9.051e-04 & 3.908e-04 & 2.698e-04 & 5.698e-04	\\
5   & 2.667e-04 & 5.437e-04 & 7.415e-04 & 9.678e-04 & 9.998e-04 & 1.271e-03 & 5.449e-04 & 5.906e-04 & 1.279e-03	\\
6   & 5.116e-04 & 1.019e-03 & 1.278e-03 & 1.944e-03 & 2.406e-03 & 2.265e-03 & 1.183e-03 & 1.232e-03 & 1.966e-03	\\
7   & 5.173e-04 & 1.007e-03 & 1.444e-03 & 2.043e-03 & 2.533e-03 & 2.762e-03 & 1.468e-03 & 1.264e-03 & 2.352e-03	\\
8   & 7.223e-04 & 1.296e-03 & 1.733e-03 & 2.368e-03 & 2.542e-03 & 3.097e-03 & 1.419e-03 & 1.374e-03 & 2.420e-03	\\
9   & 9.491e-04 & 1.782e-03 & 2.002e-03 & 2.832e-03 & 3.369e-03 & 3.272e-03 & 1.565e-03 & 1.327e-03 & 2.374e-03	\\
10  & 9.932e-04 & 1.920e-03 & 2.155e-03 & 2.913e-03 & 3.409e-03 & 3.551e-03 & 1.749e-03 & 1.444e-03 & 3.140e-03	\\
11  & 9.963e-04 & 2.098e-03 & 2.234e-03 & 2.794e-03 & 3.496e-03 & 2.939e-03 & 1.674e-03 & 1.576e-03 & 3.257e-03	\\
12  & 4.856e-04 & 7.798e-04 & 9.604e-04 & 1.316e-03 & 1.543e-03 & 1.405e-03 & 7.617e-04 & 8.485e-04 & 1.277e-03	\\
13  & 1.882e-04 & 4.138e-04 & 3.513e-04 & 5.994e-04 & 6.723e-04 & 6.037e-04 & 4.172e-04 & 2.761e-04 & 5.452e-04	\\
14  & 2.662e-05 & 1.248e-04 & 1.383e-04 & 2.700e-04 & 2.605e-04 & 1.720e-04 & 1.386e-04 & 5.538e-05 & 2.274e-04	\\
15  & 6.604e-06 & 2.029e-05 & 3.363e-05 & 4.209e-05 & 4.436e-05 & 3.654e-05 & 2.947e-05 & 2.071e-05 & 4.071e-05	\\
16  & 2.508e-06 & 1.748e-05 & 2.303e-05 & 1.894e-05 & 4.877e-05 & 1.651e-05 & 1.680e-05 & 1.719e-05 & 4.381e-05	\\ \hline
Sum & 5.907e-03 & 1.161e-02 & 1.387e-02 & 1.916e-02 & 2.250e-02 & 2.270e-02 & 1.158e-02 & 1.048e-02 & 1.996e-02	\\ \hline
  $E$ (keV)  &  55		& 60		& 70		& 80		& 90		& 100		& 120		& 140		& 149			\\ \hline
Band $ = 1$    & 3.392e-06 & 3.190e-05 & 2.156e-05 & 6.977e-05 & 1.060e-04 & 1.104e-04 & 2.657e-06 & 7.721e-05 & 7.482e-07	\\
2   & 5.774e-05 & 1.019e-04 & 1.093e-04 & 3.171e-04 & 1.276e-04 & 1.899e-04 & 1.703e-04 & 4.702e-05 & 2.945e-05	\\
3   & 1.911e-04 & 4.013e-04 & 3.264e-04 & 5.541e-04 & 5.895e-04 & 5.308e-04 & 6.351e-04 & 2.853e-04 & 8.195e-05	\\
4   & 4.493e-04 & 7.629e-04 & 9.138e-04 & 9.605e-04 & 9.964e-04 & 1.523e-03 & 1.804e-03 & 1.105e-03 & 2.839e-04	\\
5   & 9.730e-04 & 1.406e-03 & 1.517e-03 & 1.560e-03 & 1.658e-03 & 2.466e-03 & 2.988e-03 & 1.639e-03 & 4.088e-04	\\
6   & 2.081e-03 & 2.732e-03 & 3.492e-03 & 2.807e-03 & 2.918e-03 & 3.985e-03 & 4.561e-03 & 3.016e-03 & 1.160e-03	\\
7   & 2.366e-03 & 2.890e-03 & 3.818e-03 & 2.827e-03 & 3.026e-03 & 4.102e-03 & 5.073e-03 & 3.286e-03 & 1.219e-03	\\
8   & 2.547e-03 & 3.375e-03 & 4.537e-03 & 3.184e-03 & 3.080e-03 & 3.869e-03 & 5.404e-03 & 3.716e-03 & 9.780e-04	\\
9   & 2.832e-03 & 3.418e-03 & 4.912e-03 & 3.442e-03 & 3.362e-03 & 4.702e-03 & 5.379e-03 & 3.569e-03 & 9.686e-04	\\
10  & 2.556e-03 & 3.553e-03 & 4.486e-03 & 4.762e-03 & 3.195e-03 & 4.101e-03 & 4.521e-03 & 3.371e-03 & 9.197e-04	\\
11  & 2.890e-03 & 3.590e-03 & 4.136e-03 & 4.331e-03 & 3.345e-03 & 3.983e-03 & 5.489e-03 & 3.595e-03 & 9.391e-04	\\
12  & 1.194e-03 & 1.770e-03 & 2.236e-03 & 1.774e-03 & 1.737e-03 & 2.233e-03 & 2.880e-03 & 1.805e-03 & 5.570e-04	\\
13  & 3.964e-04 & 9.645e-04 & 8.756e-04 & 1.238e-03 & 9.174e-04 & 1.382e-03 & 1.418e-03 & 1.058e-03 & 2.131e-04	\\
14  & 1.544e-04 & 4.111e-04 & 1.736e-04 & 5.330e-04 & 5.718e-04 & 5.940e-04 & 6.438e-04 & 3.862e-04 & 1.052e-04	\\
15  & 4.023e-05 & 7.358e-05 & 8.814e-05 & 8.953e-05 & 1.158e-04 & 3.826e-04 & 1.606e-04 & 7.910e-05 & 7.607e-06	\\
16  & 5.024e-06 & 9.399e-05 & 9.363e-05 & 1.360e-04 & 1.103e-04 & 8.666e-05 & 9.906e-05 & 8.846e-06 & 6.651e-07	\\ \hline
Sum & 1.874e-02 & 2.558e-02 & 3.174e-02 & 2.858e-02 & 2.586e-02 & 3.424e-02 & 4.123e-02 & 2.704e-02 & 7.874e-03	\\ \hline
\end{tabular}
\end{center}
\label{Table_Results_BigTen_Sensitivity_238Capture235FissionRatio_238UCapture}
\end{table}%

\clearpage
\subsection{Molten Chloride Fast Reactor} \label{Sec:Results_MCFR}
The Molten Chloride Fast Reactor (MCFR) is an advanced reactor design that has attracted commercial interest. The design typically involves dissolving fissile material into a chloride-based salt such as NaCl and running the molten salt through a reactor vessel where fission generates heat. The MCFR can operate at much high temperatures while remaining at atmospheric pressure without boiling the fuel and coolant. This allows for high thermal and operational efficiency. For the purposes of this paper, as the name implies, the MCFR has a fast spectrum and is an ideal candidate to assess the importance of unresolved resonances.


A simplified model for an MCFR with was constructed using parameters from Ref.~\cite{Mausloff_MCFR_NED379_2021}. The geometric model is qualitatively similar to that of the Big Ten model in the previous section, consisting of two axisymmetric cylinders: a central core and a reflector. The central core is a cylinder with a diameter and height of 4~m. The core contains a molten chloride salt that consists of 33.3\%~UCl$_3$ and 66.7\%~NaCl. The uranium has an enrichment of 15.5\%. The temperature of the core is 900~K, giving an estimated mass density of 3.112~g/cm$^2$. The reflector is a cylinder surrounding the core region with a thickness of 20~cm in both the axial (top and bottom) and radial directions. The reflector made of 316-stainless steel with material data from Ref.~\cite{Detwiler_MaterialCompendium_PNNL-15870_2021}. The temperature of the reflector is taken to be room temperature for simplicity. The isotopic atom densities are provided in Table~\ref{Table_Results_MCFR_Isotopics}. Note that the model is purely conceptual and lacks important design features such as channels for circulating the salt into the reactor vessel or cooling the reflector.

As with the Big Ten model, the most important reactions are $^{235}$U(n,f) and $^{238}$U(n,$\gamma$). The responses considered are as in the previous case: the eigenvalue $k$, the ratio of leakage to neutron production, and the ratio of $^{238}$U(n,$\gamma$) to $^{235}$U(n,f) in the core. Calculations were done in both the research code Shuriken and MCNP6.2 using $10^5$ particles (on average) per batch, 50 inactive batches, and 1000 active batches. 

The energy-integrated results for the responses and unresolved resonance sensitivity coefficients are given in Table~\ref{Table_Results_MCFR_EnergyIntegratedResults}. All three responses considered are quite sensitive to the unresolved resonance cross sections of $^{235}$U(n,f) and $^{238}$U(n,$\gamma$). While still a fast spectrum system, the neutron spectrum in this MCFR model is considerably softer than the Big Ten model. For reference, about 50\% of the fissions in the MCFR model are from neutrons with a kinetic energy $< 100$~keV, whereas in the Big Ten model this is about 20\%. Because the spectrum is softer, there is a larger overlap with the unresolved resonance range of $^{235}$U than in the Big Ten case.

The obtained results largely agree within 2$\sigma$ of their MCNP6.2 counterparts. The ratio of leakage to neutron production ratio is slightly outside 2$\sigma$, but the agreement is within 0.1\%, so this could be the result of either a statistical anomaly or perhaps from minor differences in the implementation of handling and sampling the continuous-energy nuclear data.

Tables~\ref{Table_Results_MCFR_Sensitivity_k_238UCapture} and~\ref{Table_Results_MCFR_Sensitivity_238Capture235FissionRatio_238UCapture} provide unresolved resonance sensitivity coefficients for $^{235}$U(n,f) with respect to $k$ and the ratio of the core $^{238}$U(n,$\gamma$) to $^{235}$U(n,f) reaction rates. As with Big Ten, the values for leakage are not significantly different in shape.

\begin{table}[h!] \small
\caption{Atomic Densities (b$^{-1}$cm$^{-1}$) of the Molten Chloride Fast Reactor Model}
\begin{center}
\begin{tabular}{|l|c|l|c|} \hline
\multicolumn{4}{|c|}{UCl$_3$-NaCl Core}	\\ \hline
$^{23}$Na	& $8.1425 \times 10^{-3}$	& $^{235}$U	& $6.3010 \times 10^{-4}$	\\
$^{35}$Cl	& $1.5410 \times 10^{-2}$	& $^{238}$U	& $3.4350 \times 10^{-3}$	\\
$^{37}$Cl	& $4.9279 \times 10^{-3}$	&				&						\\ \hline
\multicolumn{4}{|c|}{316 SS Reflector}	\\ \hline
C    		& $3.2090 \times 10^{-4}$	& $^{58}$Ni   & $6.7052 \times 10^{-3}$	\\
$^{28}$Si   & $1.5815 \times 10^{-3}$	& $^{60}$Ni   & $2.5828 \times 10^{-3}$	\\
$^{29}$Si   & $8.0620 \times 10^{-5}$	& $^{61}$Ni   & $1.1228 \times 10^{-4}$	\\
$^{30}$Si   & $5.3175 \times 10^{-5}$	& $^{62}$Ni   & $3.5803 \times 10^{-4}$	\\
$^{50}$Cr	& $6.8436 \times 10^{-4}$	& $^{64}$Ni   & $9.1107 \times 10^{-5}$	\\
$^{52}$Cr	& $1.3197 \times 10^{-2}$	& $^{92}$Mo   & $1.8389 \times 10^{-4}$	\\
$^{53}$Cr	& $1.4965 \times 10^{-3}$   & $^{94}$Mo   & $1.1535 \times 10^{-4}$	\\
$^{54}$Cr	& $3.7250 \times 10^{-4}$	& $^{95}$Mo   & $1.9920 \times 10^{-4}$	\\
$^{55}$Mn   & $1.7538 \times 10^{-3}$	& $^{96}$Mo   & $2.0924 \times 10^{-4}$	\\
$^{54}$Fe   & $3.3044 \times 10^{-3}$	& $^{97}$Mo   & $1.2025 \times 10^{-4}$	\\
$^{56}$Fe   & $5.1825 \times 10^{-2}$	& $^{98}$Mo   & $3.0489 \times 10^{-4}$	\\
$^{57}$Fe   & $1.1975 \times 10^{-3}$	& $^{100}$Mo  & $1.2226 \times 10^{-4}$	\\
$^{58}$Fe   & $1.5816 \times 10^{-4}$	&				&						\\ \hline
\end{tabular}
\end{center}
\label{Table_Results_MCFR_Isotopics}
\end{table}%

\begin{table}[t!] \small
\caption{Energy-Integrated Response and Unresolved Resonance Range Sensitivity Coefficients of the Molten Chloride Fast Reactor Model (Top: Shuriken, Bottom: MCNP6.2)}
\begin{center}
\begin{tabular}{|l|c|c|c|} \hline
Response			& Value						& Sensitivity $^{235}$U(n,f)	& Sensitivity $^{238}$U(n,$\gamma$)		\\ \hline
\multirow{2}{*}{$k$}& 1.00007 $\pm$ 0.00010		&  1.1080e-01 $\pm$ 1.0651e-03	& -1.0849e-01 $\pm$ 4.7897e-04			\\
					& 1.00011 $\pm$ 0.00004		&  1.1147E-01 $\pm$ 2.3409e-04	& -1.0821E-01 $\pm$ 7.5747e-05			\\ \hline
\multirow{2}{*}{Leakage}
					& 0.09570 $\pm$ 0.00003		& -1.0527e-01 $\pm$ 2.6950e-03 	&  8.2752e-02 $\pm$ 1.2075e-03			\\
					& 0.09578 $\pm$	0.00003		&  ---							& ---									\\ \hline
\multirow{2}{*}{$^{238}$U(n,$\gamma$) / $^{235}$U(n,f)}
					& 0.76162 $\pm$ 0.00013		& -2.3053e-01 $\pm$ 1.6872e-03	&  3.8511e-01 $\pm$ 7.5835e-04			\\
					& 0.76145 $\pm$ 0.00011		&  ---							& ---									\\ \hline
\end{tabular}
\end{center}
\label{Table_Results_MCFR_EnergyIntegratedResults}
\end{table}
  
\begin{table}[t!] \tiny
\caption{Computed Unresolved Resonance Sensitivity Coefficients for $k$ With Respect to $^{235}$U(n,f) of the Molten Chloride Fast Reactor Model}
\begin{center}
\begin{tabular}{|l|c|c|c|c|c|c|c|c|c|} \hline
  $E$ (keV)  &  2.25	& 2.5		& 3		& 3.5		& 4		& 4.5		& 5.5		& 6.5		& 7.5			\\ \hline
Band = 1  &  6.585e-08 & 1.376e-06 & 5.469e-07 & 1.968e-06 & -1.334e-06 & 8.198e-07 & -4.987e-07 & 9.895e-06 & -3.440e-06 \\
2  & -9.520e-07 & 2.207e-06 & 2.438e-06 & 1.519e-06 & 8.887e-07 & -1.207e-06 & 3.239e-06 & 1.983e-06 & 9.263e-06 \\
3  & -6.955e-07 & 4.342e-06 & 3.701e-06 & 1.437e-05 & 3.124e-05 & 2.303e-05 & 6.068e-05 & 3.037e-05 & 1.329e-04 \\
4  & -2.384e-06 & 9.180e-06 & 2.665e-06 & 1.701e-05 & 7.944e-05 & 3.320e-05 & 1.430e-04 & 1.098e-04 & 2.276e-04 \\
5  &  2.350e-05 & 3.670e-05 & 1.074e-05 & 3.624e-05 & 7.780e-05 & 2.218e-04 & 2.950e-04 & 2.421e-04 & 2.933e-04 \\
6  &  4.203e-05 & 2.359e-05 & 4.364e-05 & 1.263e-04 & 2.398e-04 & 2.843e-04 & 6.055e-04 & 4.398e-04 & 4.580e-04 \\
7  &  3.653e-05 & 4.654e-05 & 1.917e-06 & 1.614e-04 & 1.710e-04 & 4.133e-04 & 7.057e-04 & 8.884e-04 & 4.862e-04 \\
8  &  2.723e-05 & 3.136e-05 & 9.914e-05 & 1.901e-04 & 3.220e-04 & 3.920e-04 & 7.093e-04 & 7.432e-04 & 5.577e-04 \\
9  &  4.083e-05 & 1.022e-04 & 3.044e-05 & 1.094e-04 & 2.188e-04 & 4.543e-04 & 6.263e-04 & 7.116e-04 & 9.014e-04 \\
10 &  3.399e-05 & 6.551e-05 & 8.174e-05 & 1.637e-04 & 3.026e-04 & 4.507e-04 & 5.791e-04 & 7.111e-04 & 9.447e-04 \\
11 &  8.878e-05 & 1.211e-04 & 1.078e-04 & 1.914e-04 & 3.059e-04 & 7.250e-04 & 7.849e-04 & 1.047e-03 & 1.185e-03 \\
12 &  2.846e-05 & 6.612e-05 & 1.929e-05 & 7.662e-05 & 1.024e-04 & 3.362e-04 & 6.807e-04 & 5.671e-04 & 4.548e-04 \\
13 &  1.946e-05 & 7.488e-06 & 1.287e-05 & 5.362e-05 & 1.362e-04 & 2.786e-04 & 5.135e-04 & 4.366e-04 & 3.273e-04 \\
14 &  5.856e-06 & 1.031e-05 & 1.314e-06 & 9.514e-06 & 1.758e-05 & 1.764e-04 & 1.111e-04 & 2.711e-04 & 2.511e-04 \\
15 &  2.177e-06 & 1.812e-05 & 1.222e-06 & 6.223e-07 & 1.850e-05 & 2.913e-05 & 1.365e-05 & 1.995e-05 & 6.682e-05 \\
16 & -7.800e-07 & 1.167e-05 & 1.967e-07 & -8.238e-07 & 1.042e-05 & 3.047e-05 & 3.014e-04 & 6.387e-05 & 1.312e-04 \\ \hline
Sum & 3.441e-04 & 5.578e-04 & 4.197e-04 & 1.153e-03 & 2.033e-03 & 3.848e-03 & 6.133e-03 & 6.293e-03 & 6.424e-03 \\ \hline
 8.5 	& 9.5 	& 10 		& 12.5		& 13.1	 	& 15 		& 17 		& 20 		& 24 		& 25	\\ \hline
8.496e-05 & 2.992e-06 & 1.503e-05 & 2.341e-05 & 3.327e-06 & 7.712e-05 & 6.289e-05 & 4.271e-05 & 2.521e-04 & 1.023e-05 \\
4.272e-05 & 4.841e-06 & 5.608e-06 & 6.246e-06 & 1.974e-06 & 1.748e-06 & 9.947e-06 & 2.508e-05 & 1.901e-05 & 2.391e-08 \\
6.355e-05 & 2.455e-05 & 4.961e-05 & 2.763e-05 & 4.806e-05 & 8.042e-05 & 5.979e-05 & 2.199e-04 & 4.517e-05 & 4.369e-05 \\
1.044e-04 & 6.530e-05 & 8.528e-05 & 1.537e-04 & 2.008e-04 & 1.515e-04 & 9.329e-05 & 9.388e-04 & 2.701e-04 & 1.023e-04 \\
2.439e-04 & 1.134e-04 & 3.637e-04 & 5.609e-04 & 3.567e-04 & 4.313e-04 & 3.180e-04 & 8.074e-04 & 4.690e-04 & 1.373e-04 \\
4.755e-04 & 4.363e-04 & 8.468e-04 & 7.029e-04 & 8.028e-04 & 1.079e-03 & 8.828e-04 & 2.076e-03 & 1.060e-03 & 3.002e-04 \\
4.401e-04 & 4.763e-04 & 1.528e-03 & 8.514e-04 & 7.780e-04 & 9.248e-04 & 1.338e-03 & 2.227e-03 & 1.468e-03 & 3.590e-04 \\
6.533e-04 & 5.777e-04 & 1.356e-03 & 8.712e-04 & 7.786e-04 & 1.046e-03 & 1.805e-03 & 1.644e-03 & 1.585e-03 & 3.241e-04 \\
4.881e-04 & 6.370e-04 & 9.024e-04 & 1.193e-03 & 1.024e-03 & 8.259e-04 & 1.805e-03 & 1.885e-03 & 1.771e-03 & 2.891e-04 \\
6.693e-04 & 6.641e-04 & 1.112e-03 & 1.119e-03 & 1.014e-03 & 9.611e-04 & 1.294e-03 & 2.142e-03 & 1.055e-03 & 2.579e-04 \\
6.546e-04 & 5.838e-04 & 9.619e-04 & 1.563e-03 & 9.556e-04 & 1.587e-03 & 1.722e-03 & 2.843e-03 & 1.253e-03 & 2.129e-04 \\
3.220e-04 & 5.054e-04 & 4.777e-04 & 3.945e-04 & 5.665e-04 & 8.032e-04 & 9.399e-04 & 8.486e-04 & 1.396e-03 & 5.516e-05 \\
1.776e-04 & 2.898e-04 & 2.929e-04 & 2.822e-04 & 2.686e-04 & 4.787e-04 & 8.892e-04 & 6.272e-04 & 4.134e-04 & 5.220e-05 \\
2.927e-05 & 1.095e-04 & 2.858e-04 & 1.839e-04 & 1.158e-04 & 2.150e-04 & 7.803e-05 & 3.039e-04 & 1.954e-04 & 6.479e-05 \\
4.725e-05 & 2.918e-05 & 2.031e-05 & 1.029e-05 & 2.263e-05 & 4.583e-05 & -6.709e-07 & 5.236e-05 & 2.613e-05 & 5.154e-06 \\
7.457e-05 & 3.009e-06 & 2.997e-06 & 8.148e-05 & 2.602e-04 & 2.469e-04 & 4.977e-05 & 1.045e-04 & 3.578e-04 & 3.659e-05 \\ \hline
4.571e-03 & 4.523e-03 & 8.306e-03 & 8.025e-03 & 7.197e-03 & 8.956e-03 & 1.135e-02 & 1.679e-02 & 1.164e-02 & 2.251e-03 \\ \hline
\end{tabular}
\end{center}
\label{Table_Results_MCFR_Sensitivity_k_238UCapture}
\end{table}%

\begin{table}[t!] \tiny
\caption{Computed Unresolved Resonance Sensitivity Coefficients for the $^{238}$U(n,$\gamma$) to $^{235}$U(n,f) Ratio With Respect to $^{235}$U(n,f) of the Molten Chloride Fast Reactor Model}
\begin{center}
\begin{tabular}{|l|c|c|c|c|c|c|c|c|c|} \hline
  $E$ (keV)  &  2.25	& 2.5		& 3		& 3.5		& 4		& 4.5		& 5.5		& 6.5		& 7.5			\\ \hline
Band = 1 & -1.053e-06 & -8.079e-07 & -6.827e-07 & -4.803e-06 & -4.783e-06 & -3.633e-06 & -6.197e-06 & -4.882e-05 & -4.981e-05 \\ 
2 & -1.053e-06 & -2.257e-06 & -8.491e-07 & -8.058e-06 & -3.958e-06 & -4.013e-06 & -6.223e-06 & -2.616e-06 & -1.799e-05 \\  
3 & -3.271e-06 & -1.008e-05 & -1.276e-05 & -4.003e-05 & -7.099e-05 & -3.786e-05 & -6.678e-05 & -6.332e-05 & -2.481e-04 \\
4 & -6.343e-06 & -1.405e-05 & -2.112e-05 & -5.552e-05 & -1.010e-04 & -1.456e-04 & -2.897e-04 & -2.654e-04 & -3.746e-04 \\  
5 & -4.452e-05 & -5.494e-05 & -6.973e-05 & -1.459e-04 & -2.145e-04 & -3.137e-04 & -5.091e-04 & -4.768e-04 & -6.795e-04 \\  
6 & -1.193e-04 & -9.677e-05 & -1.041e-04 & -2.440e-04 & -4.633e-04 & -6.493e-04 & -1.282e-03 & -9.484e-04 & -9.448e-04 \\  
7 & -1.126e-04 & -1.249e-04 & -1.012e-04 & -2.776e-04 & -4.269e-04 & -7.885e-04 & -1.556e-03 & -1.601e-03 & -1.024e-03 \\  
8 & -1.394e-04 & -1.266e-04 & -1.152e-04 & -3.979e-04 & -4.399e-04 & -8.874e-04 & -1.519e-03 & -1.538e-03 & -1.235e-03 \\  
9 & -1.461e-04 & -1.614e-04 & -1.419e-04 & -3.318e-04 & -4.804e-04 & -8.180e-04 & -1.332e-03 & -1.432e-03 & -1.778e-03 \\  
10 & -1.453e-04 & -1.962e-04 & -1.006e-04 & -4.071e-04 & -6.918e-04 & -1.141e-03 & -1.384e-03 & -1.509e-03 & -2.010e-03 \\ 
11 & -1.571e-04 & -2.026e-04 & -1.533e-04 & -4.522e-04 & -6.989e-04 & -1.469e-03 & -1.611e-03 & -2.005e-03 & -1.998e-03 \\ 
12 & -1.040e-04 & -1.377e-04 & -8.461e-05 & -1.913e-04 & -3.382e-04 & -8.270e-04 & -1.345e-03 & -1.031e-03 & -8.702e-04 \\ 
13 & -4.049e-05 & -5.994e-05 & -3.821e-05 & -1.010e-04 & -3.012e-04 & -8.678e-04 & -9.932e-04 & -9.276e-04 & -6.214e-04 \\ 
14 & -2.413e-05 & -6.905e-05 & -2.271e-05 & -8.098e-05 & -8.485e-05 & -2.876e-04 & -3.595e-04 & -4.651e-04 & -4.018e-04 \\ 
15 & -4.518e-06 & -1.503e-05 & -1.646e-06 & -7.509e-06 & -2.328e-05 & -5.523e-05 & -3.814e-05 & -5.590e-05 & -1.677e-04 \\ 
16 & -6.964e-06 & -1.106e-05 & -3.216e-06 & -4.749e-06 & -1.801e-05 & -7.531e-05 & -3.878e-04 & -1.368e-04 & -2.111e-04 \\ \hline
Sum & -1.056e-03 & -1.283e-03 & -9.719e-04 & -2.750e-03 & -4.362e-03 & -8.371e-03 & -1.269e-02 & -1.251e-02 & -1.263e-02 \\ \hline
 8.5 	& 9.5 	& 10 		& 12.5		& 13.1	 	& 15 		& 17 		& 20 		& 24 		& 25	\\ \hline
-1.260e-04 & -6.553e-06 & -9.832e-06 & -1.090e-04 & -7.456e-06 & -1.473e-04 & -1.255e-04 & -1.153e-04 & -4.863e-04 & -6.671e-06 \\ 
-2.338e-05 & -1.567e-06 & -1.141e-05 & -1.245e-05 & -8.285e-06 & -2.039e-05 & -1.960e-05 & -2.910e-05 & -5.837e-05 & -1.156e-06 \\ 
-1.196e-04 & -2.819e-05 & -1.105e-04 & -8.708e-05 & -9.230e-05 & -1.131e-04 & -1.654e-04 & -4.421e-04 & -9.416e-05 & -7.490e-05 \\ 
-1.922e-04 & -1.086e-04 & -2.161e-04 & -2.921e-04 & -3.149e-04 & -3.656e-04 & -1.762e-04 & -2.010e-03 & -5.898e-04 & -2.302e-04 \\ 
-5.165e-04 & -2.627e-04 & -6.610e-04 & -1.236e-03 & -9.411e-04 & -1.003e-03 & -7.096e-04 & -1.591e-03 & -1.018e-03 & -3.198e-04 \\ 
-1.063e-03 & -7.453e-04 & -1.884e-03 & -1.716e-03 & -1.633e-03 & -2.293e-03 & -1.767e-03 & -4.129e-03 & -2.461e-03 & -6.048e-04 \\ 
-9.508e-04 & -9.308e-04 & -3.163e-03 & -1.830e-03 & -1.628e-03 & -1.831e-03 & -2.422e-03 & -4.420e-03 & -2.980e-03 & -5.833e-04 \\ 
-1.062e-03 & -1.128e-03 & -2.503e-03 & -1.909e-03 & -1.735e-03 & -2.202e-03 & -3.608e-03 & -3.218e-03 & -3.155e-03 & -5.067e-04 \\ 
-1.046e-03 & -1.078e-03 & -2.270e-03 & -2.440e-03 & -2.165e-03 & -1.823e-03 & -3.809e-03 & -4.218e-03 & -3.853e-03 & -5.480e-04 \\ 
-1.393e-03 & -1.326e-03 & -2.252e-03 & -2.707e-03 & -2.124e-03 & -2.091e-03 & -2.744e-03 & -4.396e-03 & -2.045e-03 & -5.301e-04 \\ 
-1.500e-03 & -1.350e-03 & -2.168e-03 & -3.157e-03 & -1.948e-03 & -3.169e-03 & -3.273e-03 & -5.713e-03 & -2.651e-03 & -4.255e-04 \\ 
-6.346e-04 & -8.983e-04 & -1.066e-03 & -1.060e-03 & -1.123e-03 & -1.812e-03 & -2.056e-03 & -1.775e-03 & -3.261e-03 & -2.616e-04 \\ 
-2.709e-04 & -6.813e-04 & -9.181e-04 & -6.680e-04 & -6.193e-04 & -1.157e-03 & -1.667e-03 & -1.075e-03 & -8.955e-04 & -6.643e-05 \\ 
-8.855e-05 & -2.084e-04 & -4.973e-04 & -3.325e-04 & -2.336e-04 & -5.110e-04 & -1.787e-04 & -5.199e-04 & -3.457e-04 & -6.127e-05 \\ 
-7.504e-05 & -4.167e-05 & -7.747e-05 & -4.102e-05 & -5.300e-05 & -5.500e-05 & -1.795e-05 & -8.902e-05 & -4.955e-05 & -9.763e-06 \\ 
-1.316e-04 & -6.722e-05 & -2.445e-05 & -1.919e-04 & -7.083e-04 & -5.534e-04 & -1.313e-04 & -2.575e-04 & -6.162e-04 & -9.073e-05 \\ \hline
-9.192e-03 & -8.864e-03 & -1.783e-02 & -1.779e-02 & -1.533e-02 & -1.915e-02 & -2.287e-02 & -3.400e-02 & -2.456e-02 & -4.321e-03 \\ \hline
\end{tabular}
\end{center}
\label{Table_Results_MCFR_Sensitivity_238Capture235FissionRatio_238UCapture}
\end{table}%

\clearpage
\section{Conclusions \& Future Work} \label{Sec:Conclusions}
A Monte Carlo method for computing sensitivity coefficients of unresolved resonance probability table data using differential operator sampling was derived and implemented in a research code, Shuriken. Analytical benchmarks using slab geometry were created to verify the method and implementation. Numerical calculations of eigenvalue problems were performed quantifying the magnitude of the unresolved resonance sensitivity coefficients in representative application problems.

Implementing this method in a production transport code should be feasible provided it already supports differential operator sampling. The one significant modification to the codebase would be to provide the perturbation routines access to the sampled data and bands from the probability table. This may necessitate modifying or exposing data structures used in the collision physics routines to the scoring routines.

The hope of the author is that by providing a method to compute sensitivity coefficients to unresolved resonance probability table data, it provides a reason for the nuclear data community to develop a format for unresolved resonance covariances and to populate the data for important nuclides. This covariance data could then be used to obtain more accurate estimates of uncertainties of quantities of interest in fast spectrum systems.

Regardless of whether new unresolved resonance covariance table data is developed, the three analytical benchmarks developed in this paper can be applied to existing or newly developed Monte Carlo neutron transport codes to verify their probability table sampling algorithms. As stated previously, the benchmarks made some simplifying assumptions and followed the MCNP-style implementation. Their application to other codes therefore may require some adjustment to be consistent with other implementations. The benchmarks can in principle be further generalized to relax many of the assumptions, albeit increasing the complexity of the analytical solution.

%

\clearpage

\end{document}